\documentclass[twocolumn,apj,numberedappendix,twocolappendix,iop]{openjournal}
\usepackage[T1]{fontenc}
\usepackage{graphicx}	
\usepackage{amsmath}	
\usepackage{newtxtext,newtxmath}    
\usepackage[dvipsnames]{xcolor}  
\usepackage[colorlinks,linkcolor=blue,citecolor=blue,urlcolor=blue ]{hyperref}
\usepackage{amssymb}	
\usepackage{dsfont}
\usepackage{ulem}
\usepackage{soul}
\usepackage{enumitem}
\usepackage{orcidlink}
\usepackage{hyperref}

\usepackage{natbib}

\usepackage{fontawesome}
\usepackage{lineno}

\usepackage{macros} 

\begin{document}


\title{Statistical Predictions of the Accreted Stellar Halos around Milky Way-Like Galaxies}

\shorttitle{Stellar Halos of Milky Way-Like Galaxies}
\shortauthors{Monzon et al.}

\author{J. Sebastian Monzon\orcidlink{0000-0002-9986-4604}}
\author{Frank C. van den Bosch\orcidlink{0000-0003-3236-2068}}
\author{Martin P. Rey\orcidlink{0000-0002-1515-995X}}   

\affiliation{Department of Astronomy, Yale University, PO. Box 208101, New Haven, CT 06520-8101}
\affiliation{Department of Physics, University of Bath, Claverton Down, Bath, BA2 7AY, UK}

\email{s.monzon@yale.edu}

\label{firstpage}


\begin{abstract}
In the $\Lambda$CDM paradigm, stellar halos form through the accretion and disruption of satellite galaxies. We introduce new semi-analytic modeling within the SatGen framework to track the ex-situ stellar components of Milky Way-like galaxies across large ensembles of merger trees, enabling a statistical study of the stochastic nature of galaxy assembly. We find that accreted stellar halos are typically built by only a few progenitors and are highly sensitive to the fate of the most massive satellite, producing order-of-magnitude variations in accreted stellar halo mass even at fixed host halo mass. Different stellar components trace distinct phases of host halo growth: central and accreted stellar mass correlate most strongly with early assembly, while surviving satellites trace more recent accretion. Finally, using Random Forest Regression, we quantify how well observable galaxy properties can recover halo assembly histories, providing a framework for interpreting upcoming low-surface-brightness observations of stellar halos.
\end{abstract}

\keywords{
methods: analytical ---
methods: statistical ---
galaxies: satellites --- 
galaxies: stellar halos ---
cosmology: dark matter
}

\maketitle


\section{Introduction}
\label{sec:intro}

According to the standard $\Lambda$CDM cosmological paradigm, galaxies form hierarchically through the accretion of lower mass satellites. As these satellites infall, they are subjected to disruptive tidal forces that strip their mass and enrich galactic halos with stellar debris. The ages, star formation histories, and kinematic properties of accreted satellites and their tidal debris therefore provide valuable insight into the assembly histories of their parent galaxies. Particularly for Milky Way like galaxies, the accreted stellar halo (ASH) component is an invaluable laboratory for learning about the earliest generations of stellar populations \citep{Helmi.etal.20} and investigating the low-mass end of the galaxy-halo connection \citep{Bullock.Johnston.05, Wechsler.Tinker.18, Sales.etal.22, Rey.and.Starkenburg.2022, Forourhar.etal.25}. 

Due to the long dynamical timescales in the outskirts of stellar halos, merger events can be preserved in the kinematic properties and chemical signatures of accreted halo stars. Using precision astrometry from the \textit{Gaia} mission \citep[see][for recent reviews]{Deason.etal.24, Bonaca.etal.25} observers have leveraged this fact and extensively studied the stellar halo of the Milky Way (MW) \citep[e.g.,][]{Deason.etal.15, Naidu.etal.21, Malhan.etal.22, Sharpe.etal.24}. Detailed studies in this subfield of ``galactic archaeology'' have uncovered the MW's most recent major-merger -- the Gaia Sausage Enceladus (GSE) system which is inferred to have been accreted $\sim 10\Gyr$ ago \citep{Belokurov.etal.18, Helmi.etal.18, Koppelman.etal.18}. This is consistent with the general consensus \citep{Bullock.Johnston.05, Bell.etal.08, Cooper.etal.10, Deason.etal.16, Fattahi.etal.19, Cooper.etal.25} that the MW's stellar halo was built up relatively quickly via a handful of massive merger events at early times. 

Extrapolations from local measurements ($d_{\rm helio} \lesssim 5$ kpc) give \textit{total} mass estimates of $\sim 10^9 \Msun$ for the MW's stellar halo \citep{Deason.etal.16, Deason.etal.19, Malhan.etal.22}. Exactly what fraction of the stellar halo comes from stars formed in the main progenitor (``in-situ'') vs. stars formed in systems that were then accreted onto the main progenitor (``ex-situ'') is still very uncertain \citep{RodriguezGomez.etal.16, Sanderson.etal.18, Font.etal.20}. This uncertainty arises primarily because the dynamical evolution of galaxies tends to mix the two populations. For example, satellites on low-inclination orbits can deposit some of their stars in the disc \citep{Abadi.etal.03} or the bulge \citep{Zhu.etal.22}. Similarly, stars that were born in the disc can be perturbed onto more energetic halo orbits \citep{Zolotov.etal.09, Tissera.etal.13, Cooper.etal.15}. 

Recent advances in wide-field, deep-imaging programs have begun the challenging work of contextualizing the MW’s specific formation scenario by focusing on the ultra-low surface brightness outskirts of galaxies outside of the Local Group. Instruments such as the Dragonfly Telephoto Array \citep{Abraham.etal.14} have demonstrated the capability to detect diffuse structures at extremely low surface brightness for a handful of galaxies \citep{Merritt.etal.16, Gilhuly.etal.22}. Complementary surveys such as the CFHT Large Area U-band Deep Survey \cite[CLAUDS;][]{Sawicki.etal.2019} and the Hyper Suprime-Cam Subaru Strategic Program \citep[HSC-SSP,][]{Aihara.etal.18} provide the photometric depth to systematically study the morphology of galaxy outskirts across cosmic time \citep{Williams.etal.25}. Looking ahead, the upcoming ARRAKIHS mission \citep{Guzman24} is designed to push these studies to even lower surface brightness levels across larger samples of galaxies, promising a significant expansion in our knowledge of MW-mass galaxy assembly.

Cosmological hydrodynamical simulations like Illustris \citep{Elias.etal.18, Merritt.etal.20}, EAGLE \citep{Davison.etal.20, Proctor.etal.24}, FIRE \citep{Sanderson.etal.18} and FOGGIE \citep{Wright.etal.24} provide crucial insights by capturing the complex interplay between baryonic processes and hierarchical formation. \rm{while still resolving low-mass systems.} These studies show that disrupted remains of accreted satellites are ubiquitous in the stellar halos of MW-mass galaxies. These accretion events can manifest themselves as metallicity/age gradients in the density profiles of the halo \citep{Font.etal.20}, dynamically heated inner regions of the halo \citep{Zhu.etal.22, Tau.etal.25}, or as populations of stellar streams or shells \citep{Riley.etal.24, Shipp.etal.24}. However, because these analyses use a variety of subgrid prescriptions to model star formation and stellar evolution, there is no consensus on the predicted accreted stellar mass budget across simulation suites \citep{Sales.etal.22}. Furthermore, although modern simulations \cite{Monachesi.etal.19, Orkney.etal.22, Celiz.etal.25} can reach mass resolutions of $\sim 10^4 - 10^5\Msun$, they come at great computational cost, making it challenging to build up a statistically significant sample of MW-mass halos \citep[e.g.,][]{Engler.etal.21a}. Finally, these analyses also suffer from numerical artifacts such as artificial disruption \citep{vdBosch.etal.18a, vdBosch.etal.18b, Errani.Penarrubia.20} and uncertainties associated with subhalo finding algorithms \citep{Diemer.etal.24, Mansfield.etal.24, Kong.etal.25}. Together, these issues limit the use of hydrodynamical simulations in isolating and studying the impact that different mass accretion histories have on present-day stellar halos. Complementary modeling approaches are therefore needed to efficiently explore large ensembles of MW-mass systems without compromising control over the relevant physical parameters.

In this paper, we use the semi-analytic model \SatGen to investigate the correlations between the mass accretion histories (MAH) of MW-mass galaxies, the stellar mass of the ASH, and the surviving populations of satellite galaxies. Instead of modeling the orbits of star particles post accretion \citep{Dropulic.etal.25}, we develop a computationally-inexpensive and flexible framework that outputs realistic galactic properties. We apply this framework to large samples of MW-mass merger trees. Given the significant halo-to-halo variance inherent in surviving populations of satellite galaxies \citep{Monzon.etal.24}, it should come as no surprise that accounting for variance in the \textit{disrupted} populations of satellite galaxies is critical \citep{Deason.etal.16, Harmsen.etal.17, Cooper.etal.25}. The main goal of this work is to understand what shapes the ASHs of MW-mass galaxies without relying on galaxy–halo connection prescriptions that are specific to any single simulation suite, and to test the efficacy of inferring MAHs from observables.

This paper is organized as follows: Section~\ref{sec:methods} presents a detailed description of our semi-analytic framework and the assumptions that go into modeling the stellar halos. Section~\ref{sec:results} presents our fiducial results and describes how different sources of stochasticity impact the populations of surviving and disrupted satellites. Section \ref{sec:assembly_histories} investigates the correlation strengths between properties of the ASH, the surviving satellite population, and the MAH of their host halos. We finish with a discussion of our main results in Section~\ref{sec:discussion} and a summary of our conclusions in Section \ref{sec:conclusions}. Throughout, we adopt a flat $\Lambda$CDM cosmology with a present-day matter density $\Omega_\rmm = 0.3$, a baryon density $\Omega_\rmb = 0.0465$, power spectrum normalization $\sigma_8 = 0.8$,  spectral index $n_\rms = 1.0$, and a Hubble parameter  $h = (H_0/100\kmsmpc) = 0.7$, roughly in agreement with the Planck18 constraints \citep{Planck.18}.

\section{Methodology}
\label{sec:methods}

Our analysis employs the \SatGen code \citep{Jiang.etal.21, Green.etal.21}, a state-of-the-art, semi-analytical model (SAM) devised to generate statistical samples of satellite galaxy populations. Briefly, \SatGen constructs merger trees for host halos for a given cosmology using the extended Press-Schechter (EPS) method of \citet{Parkinson.etal.08}. These merger trees describe the complete MAHs of all progenitor halos that merge over time to produce a final host halo and are in excellent agreement, in a statistical sense, with merger trees extracted from $N$-body simulations \citep{Jiang.vdBosch.14, vdBosch.etal.14}. \SatGen assigns stellar masses and sizes to progenitor halos at accretion and integrates their orbits accounting for dynamical friction and tidal stripping using semi-analytic prescriptions that are calibrated against a large suite of DMO high-resolution idealized simulations. Throughout the process, we keep track of how much stellar mass each satellite loses due to tides, which ultimately contributes to the final ASH of the host halo system. 

\subsection{Merger Trees}
\label{sec:trees}

Throughout, we use three distinct samples of merger trees for MW-mass halos taken from \citet{Monzon.etal.24}. These have present-day masses that are drawn from a log-normal distribution centered on $\Mdm = 10^{12} \Msun$ and with a scatter of $\sigma_M$ dex. For the first sample, $S_0$, we set $\sigma_M=0$ so that all host halos have exactly the same halo mass of $10^{12} \Msun$. The subsequent samples, $S_{15}$ and $S_{30}$, were constructed with $\sigma_M = 0.15$ and $0.30$ dex, respectively. Each merger tree sample is made up of 10,000 unique accretion histories, yielding a total of 30,000 galaxies and an excellent handle on halo-to-halo variance. Each merger tree traces the evolution of each of its branches back to a ``leaf-mass'' of $10^{9} \Msun$\footnote{We have verified that pushing to a lower leaf-mass resolution of $10^8 \Msun$ does not significantly change the stellar mass budget of the ASHs shown in Section \ref{sec:results} or in the correlations discussed in Section \ref{sec:assembly_histories}.} which specifies our effective mass resolution. As we show, this is more than adequate for modeling the satellite galaxies that dominate the ASH mass budget of MW galaxies. In the following, we describe our model framework starting with the specific way we construct the ASH from the disrupted remains of satellites Throughout, masses that refer to the main progenitor halo or galaxy are indicated with an upper-case $M$, while those referring to subhalos or satellites are indicated with a lower-case $m$.

\subsection{The Accreted Stellar Halo}
\label{sec:ASH}

We refer to the final MW-sized halo at redshift $z=0$ as the ``host halo''. Halos that accrete directly onto the main progenitor of this host halo are called ``first-order'' subhalos, which themselves may contain subhalos that were accreted prior to merging with the main progenitor halo. We refer to such sub-subhalos as ``second-order'' subhalos, etc. Throughout, \SatGen keeps track of the order of each subhalo, which, as discussed below, can change with time. Note that the accretion redshift, $\zacc$, of a halo is defined as the redshift at which it first became a subhalo, which can be very different from the redshift at which it is accreted onto the main progenitor of the host halo.

At accretion, each subhalo is assigned
\begin{itemize}[leftmargin=0.27truecm, labelwidth=0.2truecm]

\item a stellar mass $m_*$, based on the a redshift dependent Stellar-Halo-Mass-Relation (SHMR; see Section \ref{sec:SHMR})

\item an effective size $\reff$, based on a redshift-independent empirical Size-Mass-Relation (SMR; see Section \ref{sec:SMR})

\item a halo concentration parameter $c$, based on its pre-accretion MAH (see Section \ref{sec:density_profiles})

\item an orbital energy and angular momentum drawn from an empirical redshift-independent probability distribution function (see Section \ref{sec:orbits}) 

\end{itemize}

The newly formed satellite system is then orbit-integrated in the potential of its direct parent halo, accounting for tidal stripping and dynamical friction, as outlined in Section \ref{sec:density_profiles}. This means that all subhalos of order $k$ are evolved in the potential of their parent halos of order $k-1$. Any stellar mass stripped from a satellite of order $k$ is instantaneously added to the ASH reservoir of its parent of order $k-1$. When that parent is accreted into a bigger halo, its ASH is added to that of its new parent. In this way, all stellar mass stripped, independent of the progenitor halo in which the stripping occurred, ends up as part of the ASH of the final host halo. Note that, at any redshift $z$, we refer to the ASH mass as the sum of the ASH masses of {\it all} progenitors at that redshift. During every time step in the orbit integration any of the following can happen:

\begin{itemize}[leftmargin=0.27truecm, labelwidth=0.2truecm]

\item {\bf Disruption:} When the mass of a subhalo drops below a critical value, $m_{\rm dis}$, we consider the satellite system disrupted. We define $m_{\rm dis} = f_{\rm dis} \times \macc$, where $f_{\rm dis}$ is a free parameter for which we adopt the fiducial value of $10^{-4}$ (see Section~\ref{sec:density_profiles}). If a satellite disrupts, all of its remaining stellar mass is added to the ASH of its direct parent.

\item {\bf Merging:} We assume that a satellite galaxy merges with the central galaxy of its direct parent if its orbital energy becomes too small. A fraction $\fcann$ of the remaining stellar mass of the satellite is then added to the central of its direct parent, while the remaining fraction ($1-\fcann$) is added to the ASH. If the satellite merges with the galaxy of the main progenitor, this fraction is also added to the $\Mex$ component (see Sections \ref{sec:central} and \ref{sec:mergers}). In this way we keep track of what fraction of the host galaxy's stellar mass has been accreted by cannibalizing satellites.

\item {\bf Order Dropping:} If a subhalo of order $k \geq 2$, orbits outside of the instantaneous tidal radius of its direct parent of order $k-1$, it has a probability of being released and becoming a subhalo of order $k-1$ (see Section \ref{sec:density_profiles}). In the event of release, the phase-space coordinates of the subhalo are updated to be the superposition of its original coordinates with respect to its old parent of order $k-1$ and its new parent of order $k-2$. The same release mechanism is triggered if a subhalo's direct parent is found to be disrupted or merged. 

\item {\bf Order Jumping:} If the direct parent of a subhalo is accreted into a bigger halo, the order of all its subhalos increase by one, i.e., $k \rightarrow k+1$. The subhalo in question continues to be integrated with respect to the potential of its direct parent.

\end{itemize}

Unlike numerical simulations that depend on subhalo-finding algorithms and are subject to artificial overmerging, this setup with \SatGen allows us to accurately track and account for the mass budgets of satellites and the ASH they produce. In the next sections we detail the different stellar mass components we consider in this work.
\begin{figure*}
    \centering
    \includegraphics{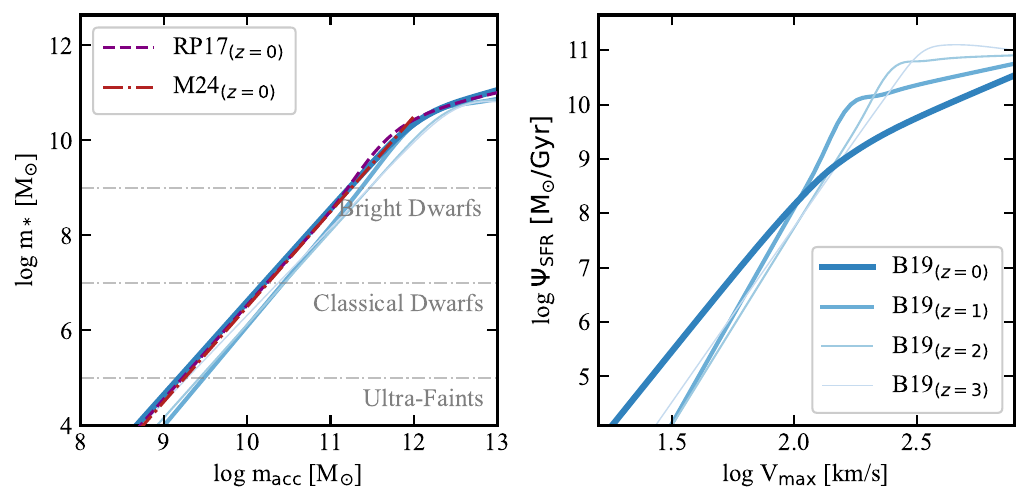}
    \caption{\textbf{Left panel:} The SHMR from \cite{Behroozi.etal.19} used to assign stellar masses to all the subhalos in our merger trees at accretion. We show the relation across several decades in dark matter mass despite only populating subhalos between $10^{9} \Msun \leq \macc \lesssim 10^{12} \Msun$. For clarity we don't show the constant 0.2 dex scatter that is assumed when drawing stellar masses. For reference we show the SHMR from \cite{RodriguezPuebla.etal.17} and the fiducial SHMR from \cite{Monzon.etal.24}. \textbf{Right panel:} The deterministic relation between star formation rate (SFR) and its maximum circular halo velocity $\Vmax$ from the UniverseMachine \citep{Behroozi.etal.19}. We use this relation to build up the ``in-situ'' component of central galaxies in our model by integrating SFRs across time snapshots.}
    \label{fig:SHMR}
\end{figure*}

\subsection{The Stellar Halo Mass Relation}
\label{sec:SHMR}

When a halo first becomes a subhalo, we assign it a stellar mass using the redshift-dependent SHMR i.e., $\mstar = \mstar(\macc,\zacc)$ given by equations (J1)-(J8) of \cite{Behroozi.etal.19}. Here, $\macc$ and $\zacc$ are the subhalo mass and redshift at accretion. To this mass, we add a random scatter of 0.2 dex to account for the intrinsic scatter in the SHMR. Figure~\ref{fig:SHMR} shows the SHMR of B19 over several decades in halo mass, and at four different redshifts, compared to similar empirical relations in the literature. For clarity we do not show the 0.2 dex scatter in stellar mass which we assume to be independent of halo mass or redshift\footnote{We thus ignore the fact that the results of B19 are consistent with a slight increase in the scatter towards lower halo masses and higher redshifts.}. Note that the empirical B19 model for the SHMR predicts very little redshift evolution in the slope of the SHMR at the low mass end. However, we emphasize that the results for $\macc \lta 10^{12} \Msun$ are mainly extrapolations from constraints at larger masses, especially for $z>0$, and that the SHMR at the low mass end remains poorly constrained \citep{Allen.etal.19, Nadler.etal.20, Munshi.etal.21, Santos.etal.22, Danieli.etal.23, OLeary.etal.23, Christensen.etal.24, Monzon.etal.24, Kado-Fong.etal.25}. Despite these uncertainties we adopt the SHMR of \citet[hereafter B19]{Behroozi.etal.19} as described above. In a fortcoming study we will test to what extent observations of stellar halos may be used to constrain the low-mass end of the SHMR \citep[eg.][]{Rey.and.Starkenburg.2022}.

\subsection{The Size Mass Relation}
\label{sec:SMR}

At accretion, we also assign the satellite galaxies an effective half-light radius, $\reff$, based on the stellar-mass radius relation (SMR) inferred from the Satellites Around Galactic Analogs survey (SAGA) \citep{Geha.etal.17,Mao.etal.21,Mao.etal.24}, which covers the stellar mass range $10^{6.75} \Msun < \mstar < 10^{10} \Msun$, and is consistent with results in the local volume \citep{Carlsten.etal.22}. Given a satellite's stellar mass at accretion, we draw its effective size from a log-normal distribution centered on the mean relation of \cite{Asali.etal.25};
\begin{equation}
     \log(\reff) =  0.27 - 2.11 \log(\mstar),
     \label{eq:SMR}
\end{equation}
assuming a scatter of 0.2 dex. The slope and intercept are the best-fit values from the SAGA DRIII ``Gold-Sample'' and we ignore the slight environmental dependence found by \cite{Asali.etal.25}. Since most of our satellites have an assigned stellar mass $\mstar < 10^9\Msun$, we ignore any potential redshift evolution in the SMR, which is poorly constrained for such low-mass galaxies. We also emphasize that in reality galaxy sizes are correlated with other galaxy properties, such as their star formation rate and morphology. Such correlations are not taken into account in our work but we do not expect this oversimplification to significantly impact the main results of this study.

\subsection{Stellar Mass of the Central Galaxy}
\label{sec:central}

The present-day stellar mass of the central is made up of two components; an in-situ component $M_{\rm in-situ}$, consisting of stars that formed in the main progenitor halo, and an ex-situ component $M_{\rm ex-situ}$, which consists of all the stars accreted by the central galaxy during satellite mergers. To build up the in-situ stellar component of the central host galaxy we follow the B19 prescription \citep[see also][for similar prescriptions]{Moster.etal.18, OLeary.etal.23} and assume that a galaxy's star formation rate $\psi_{\rm SFR}$, is a function of redshift and the host halo's instantaneous maximum circular velocity, $\Vmax$. The right-hand panel of Figure~\ref{fig:SHMR} shows the dependence of $\psi_{\rm SFR}$ on $\Vmax$ at four redshifts, as indicated. Using these star formation rates (SFRs), we calculate the final in-situ stellar mass of the central galaxy according to
\begin{equation}\label{Mstarinsitu}
\Min = \sum_{t_i < t} \psi_{\rm SFR}(t_i) \, \Delta t_i \, \left[ 1 - f_{\rm return}(t-t_i)\right]\,.
\end{equation} 
Here, the summation is over all discrete timesteps of the main progenitor's MAH, $f_{\rm return}$ is the return fraction that describes the mass that is returned to the interstellar medium due to stellar evolution, and $\psi_{\rm SFR}(t)$ is given by equations (4)-(11) of B19 assuming that $V_{\rm Mpeak} = V_{\rm max}(t)$. For the return fractions, we adopt
\begin{equation}
f_{\rm return}(t) = 0.05 \ln \left( 1 + \frac{t}{0.0014 \Gyr}\right)\,,
\end{equation}
which is taken from the Flexible Stellar Population Synthesis (FSPS) package \citep{Conroy.Gunn.10} assuming a \cite{Chabrier.03} initial mass function (IMF). 

\begin{figure*}
    \centering
    \includegraphics{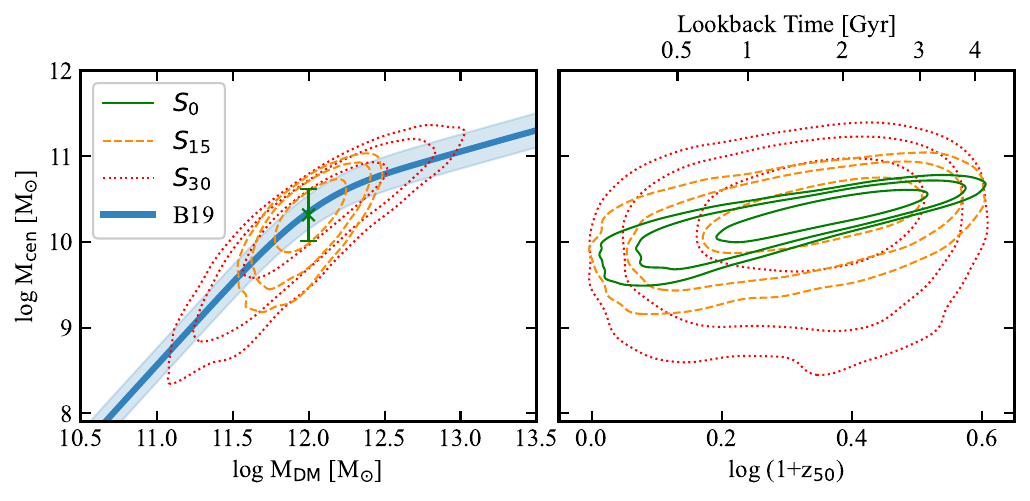}
    \caption{\textbf{Left panel}: The distribution of our three merger tree samples ($S_0, S_{15}, S_{30}$) in the SHMR plane assuming central galaxies grow according to the $\psi_{\rm SF} - \Vmax$ relation defined in \cite{Behroozi.etal.19}. For reference we show the present-day SHMR (thick blue line) along with its 0.2 dex scatter (blue shaded region). \textbf{Right panel}: The distribution of halo formation times (z$_{50}$) against present-day central stellar mass. Notice how the $S_0$ sample spans a similar range to the other two samples despite being restricted to present-day DM mass of $10^{12} \Msun$. This illustrates what we refer to as \htoh variance.}
    \label{fig:SHMR_samples}
\end{figure*}

Rather than simply drawing a stellar mass from a log-normal distribution centered on the SHMR, which is the method we use to assign stellar masses to our satellites at infall, this prescription automatically introduces a correlation between the stellar mass of the central of the main progenitor and its MAH. This can be seen in the right-hand panel of Figure~\ref{fig:SHMR_samples} which shows a weak correlation between $\Mcen$ and $\zfive$, defined as the redshift at which the host halo reaches 50\% of its present-day dark matter mass. Note that this correlation becomes weaker in the $S_{15}$ and $S_{30}$ samples due to the fact that the final halo mass ($\Mdm$) is a better indicator of $\Mcen$ than $z_{50}$. As shown in the left-hand panel of Figure~\ref{fig:SHMR_samples}, the stellar masses of our centrals are in perfect agreement with the SHMR of B19 used to assign stellar masses to the satellites, both in terms of the normalization as well as the scatter. Hence, our method of assigning stellar masses to satellites and centrals is self-consistent.

\subsection{Dark Matter Density Profiles}
\label{sec:density_profiles}

The fiducial \SatGen model initializes all subhalos with a Navarro-Frenk-White \citep{Navarro.etal.97} dark matter density profile;
\begin{align}
    \rho_{\rm NFW}(r) &= \rho_0 \, \left(\frac{r}{r_s}\right)^{-1} \, \left[1+ \left( \frac{r}{r_s}\right) \right]^{-2}.
\end{align}
Here, $\rho_0$ is a characteristic density and $\rs = \rvir/c$ is the scale radius, with $\rvir$ the virial radius and $c$ the halo concentration parameter. Throughout, halo concentrations are computed using the model of \cite{Zhao.etal.09} relating $c$ to the average density of the Universe at the redshift at which the halo's main progenitor reaches a mass that is 4\% of its mass at accretion.

As a subhalo orbits its parent, it is exposed to tides that strip away the outer layers of its mass distribution. In \SatGen this tidal stripping is modeled using the method originally pioneered by \citet{Taylor.Babul.01} and \citet{Zentner.Bullock.03}, according to which 
\begin{equation}
    \frac{\Delta m}{\Delta t} =-\alpha_{\rm s} \frac{m\left(>r_{\mathrm{tid}}\right)}{t_{\mathrm{dyn}}}\,.\\
\end{equation}
Here, $\alpha_{\rm s}$ is the stripping efficiency modeled as a function of the ratio between the concentration parameters of the host and subhalo at accretion \citep{Jiang.etal.21, Green.etal.21}, $t_{\rm dyn} = \sqrt{3 \pi/16 G \bar{\rho}(r)}$ is the dynamical time with $\bar{\rho}(r)$ the average density of the direct parent halo inside the instantaneous orbital radius $r$, and $r_{\rm tid}$ is the instantaneous tidal radius, which is given by the root of
\begin{equation}
    r_{\mathrm{tid}} = \left[\frac{G m\left(<r_{\mathrm{tid}}\right)}{\Omega^2-\left.\frac{\mathrm{d}^2 \Phi}{\mathrm{~d} r^2}\right|_r}\right]^{1 / 3}.
    \label{eq:tidal_radius}
\end{equation}
\citep{King.62, Tollet.etal.17, vdBosch.etal.18a}, with $\Omega$ the subhalo's instantaneous angular velocity and $\Phi$ the gravitational potential of the subhalo's direct parent halo. 

In order to model the density profiles of subhalo remnants that have been stripped of some of their mass, we assume that the structural properties of tidally stripped subhalos depend solely on their initial density profile and on the total amount of matter lost since accretion \citep{Penarrubia.etal.08, Penarrubia.etal.10, Errani.etal.18, Chiang.etal.24}. This means that the structural properties of subhalos evolve along ``tidal tracks'' and are independent of their orbit. Hence, the evolved density profile can be modeled using the initial density profile at accretion and a ``transfer function'' $H(r| f_b)$ that depends only on the subhalo's instantaneous bound mass fraction $f_b \equiv m(t)/\macc$. In particular, we have that
\begin{equation}\label{eq:transfer_function}
  \rho(r,t) = H(r | f_b(t)) \, \rho_{\rm NFW}(r,t_{\rm acc}).
\end{equation}
Throughout, we use the transfer function given by equations (5)-(8) in \citet{Green.vdBosch.19} that has been carefully calibrated and validated using a large suite of high-resolution idealized simulations \citep{Ogiya.etal.19, Green.vdBosch.19, Green.etal.21}.

Following \cite{Monzon.etal.24}, we assume that a subhalo (and its associated satellite galaxy) is disrupted whenever the instantaneous subhalo mass drops below a critical mass $m_{\rm dis} \equiv f_{\rm dis} \macc$. As mentioned above, we adopt $f_{\rm dis} = 10^{-4}$ as our fiducial value, which implies that subhalos disrupt once they have lost more than 99.99\% of their initial accretion mass. We have verified that setting $f_{\rm dis} = 10^{-3}$ or $f_{\rm dis}=10^{-5}$ has no significant impact on any of our results.

\subsection{Evolution of Satellite Stellar Masses}
\label{sec:satevo}

In order to model the effects of tidal stripping on the satellite stellar masses, we proceed as follows. Tidal tracks analogous to those describing the tidal evolution of the dark matter properties can also be used to evolve the stellar mass ($m_{*}$) and size ($\reff$) of an embedded satellite. Here we follow \cite{Errani.etal.18} and assume that the ratios $m_{*}(t) / m_{*}$ and $r_{\rm eff}(t) / r_{\rm eff}$ follow tidal tracks characterized by
\begin{equation}
    g(x) = \left(\frac{x_s + 1}{x_s + f_b}\right)^{\mu} f_b^{\eta}\,.
    \label{eq:errani}
\end{equation}
Here, $g(x)$ refers to either $m_{*}(t) / m_{*}$ or $r_{\rm eff}(t) / r_{\rm eff}$, and $f_b$ is the instantaneous bound mass fraction of the corresponding dark matter subhalo. The additional parameters $\mu$, $\eta$ and $x_s$ depend on the ``cuspiness'' of the subhalo at accretion, parameterized by the inner logarithmic slope of the subhalo's density profile, and on the ratio $r_{\rm eff}/r_{\rm max}$. Here, $r_{\rm max}$ is the radius at which the subhalo reaches its maximum circular velocity,$v_{\rm max}$, which for an unevolved NFW density profile is $\sim 2.16 \, r_{\rm s}$. \SatGen uses a simple interpolation scheme to determine the values of $\mu$, $\eta$ and $x_s$ based on the values inferred by \cite{Errani.etal.18} using a discrete set of simulation results that explored the relevant parameter space\footnote{Since those simulations modeled the stellar bodies as Plummer spheres, we are implicitly assuming the same.} 

\begin{figure}
    \centering
    \includegraphics{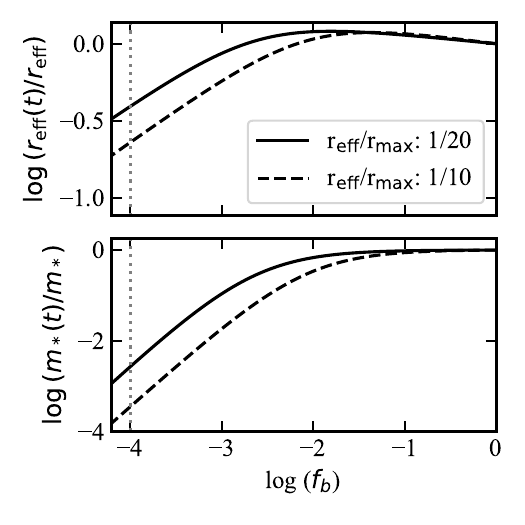}
    \caption{The stellar tidal tracks defined in \cite{Errani.etal.18} that we use to evolve the satellites. \textbf{Top panel}: The effective radius of the satellite as a function of the bound fraction of dark matter mass ($f_b$). The blue lines correspond to an initially cuspy (NFW) density profile for the dark matter. The solid vs. dashed lines indicate the initial sizes of the satellite galaxies relative to their dark matter subhalos as indicated. \textbf{Bottom panel}: The stellar mass of the satellites as a function of the $f_b$. They grey dotted line in both panels indicates our disruption criteria. Notice that, according to these models, a significant fraction of dark matter mass must be lost before any stars can be stripped.}
    \label{fig:errani}
\end{figure}

Figure~\ref{fig:errani} shows the evolution of the size (top panel) and stellar mass (bottom panel) of satellites as a function of the subhalo's bound mass fraction, based on the model of \cite{Errani.etal.18}. Results are shown for two different values of the $r_{\rm eff}/r_{\rm max}$ ratio, as indicated, to illustrate how the size of a satellite affects its tidal evolution. Note that both the stellar mass and size of satellites remain largely unaffected until the subhalo has lost roughly 99\% of its accretion mass. This simply reflects that the stellar body of a satellite galaxy is located near the center of the halo, where it is shielded from the tidal effects until the halo has been stripped down to the radius where the stars reside. This also explains why satellites with a larger effective radius experience more mass loss and a larger reduction in size for the same bound mass fraction of the subhalo remnant.

Any stellar mass that is stripped by tides is instantaneously deposited into the ASH. Note that we do not explicitly model the subsequent evolution of stripped stars and are therefore agnostic to where density profile of the ASH. Throughout, we make the simplifying assumption that any stellar mass lost to tides remains within the virial radius of the host galaxy and thus a part of the ASH of the final host halo. 

\subsection{Subhalo Orbits}
\label{sec:orbits}

Under the assumption that halos are spherical potentials, the orbital energy and angular momentum of a subhalo can be specified by two independent parameters; the infall velocity $v_{\rm inf}$ and the angle $\theta$ between the position and velocity vectors. For each subhalo, we initialize its orbit at accretion by randomly drawing $v_{\rm inf}$ and $\theta$ from their corresponding PDFs defined in Equations~(1) and~(2) of \citet{Li.etal.20}. 

We follow the standard \SatGen procedure and integrate subhalo orbits by treating them as point masses \citep[see][for details]{Jiang.etal.21}. Each subhalo is integrated in the potential of its direct parent, accounting for a dynamical friction force
\begin{equation}
F_{\rm DF}=-4 \pi \frac{G^2 m^2 }{v^2} \, \ln{\Lambda} \, \rho(<v) \, \frac{v}{| v | } 
\end{equation}
\citep{Chandrasekhar.43}. Here, $m$ is the instantaneous subhalo mass, $\ln \Lambda$ is the Coulomb logarithm, $v$ is the relative velocity of the subhalo with respect to its parent halo, and $\rho(<v)$ is the local density of dark matter particles in the parent halo with a speed less than $|v|$, which is computed under the standard assumption that the particles follow an isotropic Maxwellian velocity distribution \citep[see][]{Binney.Tremaine.08}. The Coulomb logarithm is modeled using
\begin{equation}\label{coulomb}
\ln \Lambda= \beta_{\mathrm{DF}} \, \ln \left(M / m\right)\,,
\end{equation}
where $\beta_{\rm DF}$ is a free parameter that regulates the overall strength of dynamical friction. As fiducial value, we set $\beta_{\rm DF} = 1$. In Appendix~\ref{App:freeparam} we demonstrate that increasing or decreasing $\beta_{\rm DF}$ by  a factor of two only has a small impact on the final masses of the ASH and the surviving satellite population.

\subsection{Mergers}
\label{sec:mergers}

\begin{table*}[]
    \centering
    \begin{tabular}{l | c | l }
         \hline
         \hline
         symbol  & equation & component description \\
         \hline
         $\Min$  &  (2) & stellar mass of main progenitor galaxy at $z=0$ that formed in-situ due to star formation\\
         $\Mex$  & (15) & stellar mass of main progenitor galaxy at $z=0$ accreted from satellites \\
         $\Mcen$ & (14) & total stellar mass of the main progenitor galaxy at $z=0$\\
         $\Mash$ & (16) & stellar mass of the accreted stellar halo at $z=0$\\
         $\Msat$ & (17) & total stellar mass of surviving satellites at $z=0$\\
         $\Mtot$ & (18) & total stellar mass of all satellites at their respective $\zacc$\\
         \hline
    \end{tabular}
    \caption{Various mass components of the final halo at $z=0$ relevant to the results presented in Section \ref{sec:results}. The first column lists the symbol used throughout the text, the second column lists the number of the defining equation, and the third column gives a brief description.}
    \label{tab:mass_comp}
\end{table*}

Although \SatGen does not allow satellites to merge with one another, they are allowed to merge with the central galaxy of their direct parent halo. Specifically, we merge a satellite galaxy with its parent if, at any point in its orbit integration, {\it both} its relative position ($r$) and speed ($v$) obey the following criterion
\begin{align}
\log \left(\frac{r_k}{r_{\rm{max}, k-1}}\right) < X_{\rm merge} \,\,\,\,{\rm AND} \,\,\,\,\log \left(\frac{v_k}{v_{\rm{max}, k-1}}\right) < X_{\rm merge}.    
\label{eq:merger}
\end{align} 
Here, $r_{\rm max}$ is the radius at which the parent halo reaches its maximum circular velocity, $v_{\rm max}$, and $X_{\rm merge}$ is a free parameter that allows us to control the frequency of central-satellite mergers.

If and when a satellite galaxy merges with its parent galaxy, we assume that a fraction $\fcann$ of its instantaneous stellar mass is cannibalized by the parent. The remaining fraction is deposited into the ASH. In general, $\fcann$ and $X_{\rm merge}$ depend on the mass ratio of the central and satellite and on the detailed orbit of the latter. Throughout, we ignore such details and instead treat both as free parameters that control the frequency of central-satellite mergers and their impact on the final masses of the ASH and the central. However, as we demonstrate in Appendix \ref{App:freeparam}, our qualitative results are robust to reasonable changes to both $\fcann$ and $X_{\rm merge}$.

Because of the possibility of mergers, the final stellar mass of the main progenitor galaxy can be written as
\begin{equation}
    M_{\rm cen} = M_{\rm in-situ} + M_{\rm ex-situ}.
\end{equation}
where $M_{\rm in-situ}$ is given by equation~(\ref{Mstarinsitu}) and represents the stellar mass that formed in-situ due to star formation in the central galaxy of the main progenitor, while 
\begin{equation}\label{eq:exsitu}
    M_{\rm ex-situ} = \sum_i^{N_{\rm merged}^{k=1}}  \fcann \, m_{*, i}
\end{equation}
is the ex-situ component due to the accretion of satellite galaxies. Note that the summation is only over first-order ($k=1$) satellite systems, as these are the only ones that can directly merge onto the main central. As discussed in Section~\ref{sec:central}, the final stellar masses of our central galaxies, which include both the in-situ and ex-situ components, are consistent with the SHMR presented in B19 in terms of both the normalization and the scatter. Importantly, the ex-situ component makes a negligible contribution (typically $\Mex < 10^{-3}\Min$) to the stellar mass of the main central.

Similarly, we can write the final mass of the ASH as a sum over three contributions
\begin{equation}
    \Mash = \sum_{i=1}^{N_{\rm surviving}} m_{*,i}^{\rm lost} +
            \sum_{i=1}^{N_{\rm disrupt}} m_{*,i} + \sum_{i=1}^{N_{\rm merged}} \left( m_{*,i}^{\rm lost} +  m_{*,i}^{\rm deposit} \right).
    \label{eq:ash}
\end{equation}
The first sum is over all surviving satellite galaxies, where $m_{*,i}^{\rm lost}$ is the total stellar mass that the satellite has lost to tidal stripping since its accretion. The second sum is over all satellites that have been completely disrupted and therefore contribute their entire stellar mass budget to the ASH. The third and final sum is over all satellites that have merged with their parents. Each of these contribute the sum of $m_{*,i}^{\rm lost}$ plus $m_{*,i}^{\rm deposit}$ to the ASH. The former is the stellar mass lost to tidal stripping prior to merging with the parent, while 
\begin{equation}
    m_{*,i}^{\rm deposit} = (1-\fcann) \, (m_{*,i} - m_{*,i}^{\rm lost})\,
\end{equation}
is the stellar mass of the satellite that is deposited into the ASH as a consequence of the merger.

At $z=0$, the sum of all stellar masses assigned to satellite galaxies at accretion is made up of  three components
\begin{equation}
    M_{\rm tot} = \Mex + \Mash + \Msat
    \label{eq:mtot}
\end{equation}
Here, $\Msat$ is the total stellar mass of the surviving satellites, which is given by
\begin{equation}
    M_{\rm sat} = \sum_{i=1}^{N_{\rm surviving}}  \left( m_{*, i} - m_{*,i}^{\rm lost} \right).
\end{equation}
Note that, depending on its orbit and final fate, a satellite can at most contribute to two of these three mass components. For convenience, Table \ref{tab:mass_comp} lists the various masses that are relevant to the results in Section \ref{sec:results} below. Finally, we caution against assigning too much value to the distinction between merged and disrupted systems. Whether a satellite merges with its parent or experiences complete disruption is sensitive to the specific merger and disruption criteria adopted, which are somewhat ad hoc. In both cases, the satellite ends up depositing most of its stellar mass to the ASH. Hence, whether the satellite disrupts or merges has little to no impact on any of our main results.

\section{Results}
\label{sec:results}

We now use the model described above to examine how the ASH of a MW-mass galaxy builds up over time. After presenting predictions for our fiducial model, we demonstrate the important role played by a handful of the most massive progenitors and discuss the impact of various sources of stochasticity.

\subsection{Fiducial Model}
\label{sec:fiducial}

\begin{figure*}
  \centering
  \includegraphics{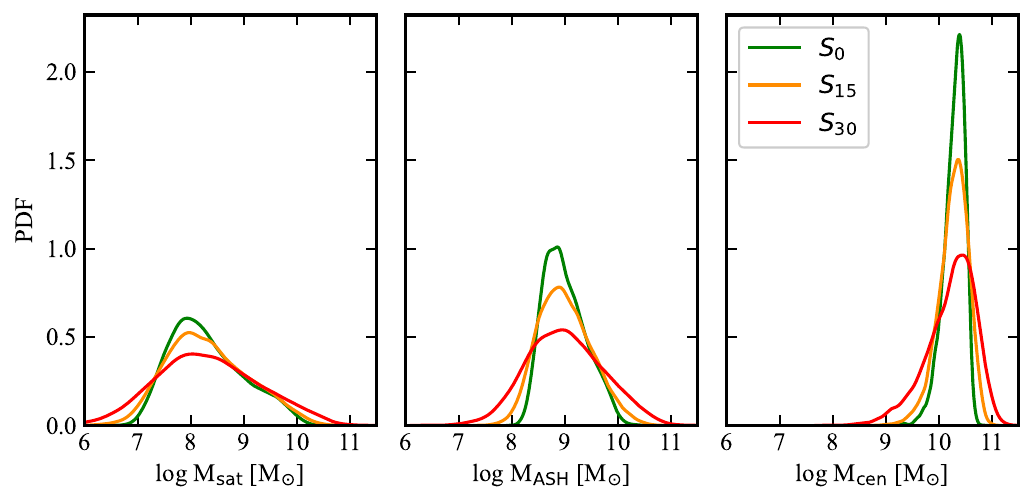}
  \caption{The fiducial distributions of the main mass components in our model for all three merger tree samples: $S_0, S_{15}, S_{30}$. \textbf{Left panel:} shows the PDFs of the $\Msat$ component which are all centered on $\sim 10^8 \Msun$. \textbf{Middle panel:} the PDFs of the $\Mash$ component which are all centered on $\sim 10^9 \Msun$. \textbf{Right panel:} shows the PDFs of the $\Mcen$ component which are all centered on $\sim 10^{10} \Msun$. Notice that all distributions widen as the amount of host halo mass-mixing increases.}
  \label{fig:fiducial}
\end{figure*}

The merger trees and the treatment of the tidal evolution of dark matter subhalos in \SatGen have been carefully calibrated and tested in a number of previous studies \citep{vdBosch.Jiang.16, Jiang.vdBosch.17, Green.vdBosch.19, Jiang.etal.21, Green.etal.21, Monzon.etal.24}. The specific implementation used here, and described above, has a number of additional ingredients that are not standard for the \SatGen model. These include the assignment of stellar masses and sizes at accretion, discussed in Sections~\ref{sec:SHMR} and~\ref{sec:SMR}, respectively, and three free parameters related to the treatment of mergers. These are the parameter $\beta_{\rm DF}$ that controls the strength of dynamical friction (see equation~[\ref{coulomb}]), the parameter $X_{\rm merge}$ that specifies the merger criterion (equation~[\ref{eq:merger}]), and the parameter $f_{\rm cann}$ that characterizes the fraction of stellar mass of the satellite that is cannibalized by the parent galaxy during a merger (equation~[\ref{eq:exsitu}]). As shown in Appendix~\ref{App:freeparam}, the three main mass components of interest, $\Mcen$, $\Mash$ and $\Msat$, are relatively insensitive to our choice of free parameters. In what follows, we therefore adopt the following fiducial values; $\beta_{\rm DF} = 1$, $X_{\rm merge} = -2.0$ and $f_{\rm cann}=0.8$. 

Figure~\ref{fig:fiducial} shows the resulting stellar mass distributions of the central galaxy (left panel), the ASH (middle panel), and the combined stellar mass of all surviving satellites (right panel) for this fiducial model. Each color shows the PDFs obtained from the $10,000$ unique merger trees in each of the three merger tree samples, as indicated. Typically, the stellar mass of the central galaxy is at least an order of magnitude larger than that of the ASH, which in turn is about an order of magnitude larger than that of the surviving satellite population. Note the large variance in the $\Mash$ and $\Msat$ distributions which, as expected, increases with the inclusion of host halo mass-mixing (i.e., going from sample $S_0$ to $S_{30}$). 
\subsection{Satellite Survival}
\label{sec:survival}

\begin{figure}
    \centering
    \includegraphics{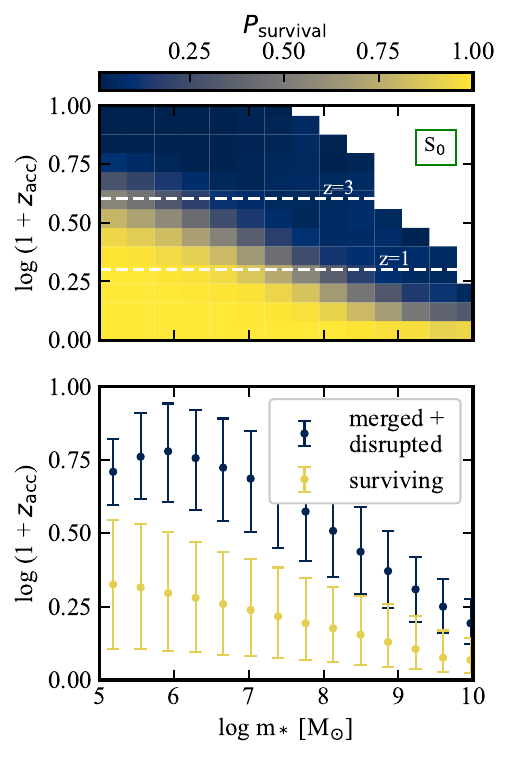}
    \caption{\textbf{Top Panel}: The stellar masses ($\mstar$) at accretion vs accretion redshift ($\zacc$) for all 1st order satellites in the $S_0$ sample. The grid indicates a simple 15 x 15 binning scheme in which survival probabilities were measured. Here, the lighter yellow cells denote higher probabilities of survival as indicated. Notice that, for a fixed $\zacc$, increasing stellar mass coincides with lower survival probability. \textbf{Bottom Panel}: The distribution of accretion redshifts binned by stellar mass. Here each point marks the median and 16-84 percentile, and the color indicates the satellite population. This clearly shows that for a fixed stellar mass, systems that survive are those which were accreted more recently.}
    \label{fig:survival}
\end{figure}

On average, $\sim 66\%$ of all satellites in our fiducial model are classified as merged or disrupted and are responsible for $\sim 95\%$ of the final mass of the ASH.\footnote{Most of this comes from the stellar mass lost to tides prior to the merger, rather than mass that is deposited into the ASH during the merger event itself.} This is broadly consistent with the AURIGA simulations \citep{Shipp.etal.24}, which show that the majority of all accreted satellites around MW-mass hosts have been disrupted. The remaining $\sim 34\%$ of all satellites that survive to the present-day only end up contributing $\sim 5\%$ to the present-day ASH. This is a direct consequence of the fact that the stellar components of satellites are shielded by their dark matter halos, such that a satellite only starts to experience significant stellar mass loss after its subhalo has lost $\sim$ 99\% of its initial mass (see Figure \ref{fig:errani}). Hence, most of the surviving satellites are still largely intact, with stellar masses comparable to those at accretion. In fact, for the $S_0$ sample, we find that 93\% of all surviving satellites have lost less than 10\% of their stellar mass at accretion.

$N$-body simulations and semi-analytic models predict that satellite survival depends strongly on satellite mass, accretion epoch, and host galaxy structure \citep{Bullock.etal.01, Taffoni.etal.03, Bullock.Johnston.05}. Several studies have examined the statistics of satellite (or subhalo) survival in MW-mass host halos  \citep[e.g.][]{Diemand.etal.07a, Geen.etal.13, Joshi.etal.24, Grimozzi.etal.24, Joshi.etal.25b, Pathak.etal.25} Unfortunately, these studies are all based on numerical simulations of at most a handful of MW-mass systems, and thus subject to relatively poor statistics. With the model presented here, we are in a unique position to study satellite survival in MW-like hosts with unprecedented statistical precision.

Using the 10,000 host galaxies in the $S_0$ merger tree sample, we select every subhalo that was accreted onto the main progenitor as a first-order\footnote{To facilitate a comparison with simulation results, we restrict our analysis to first-order ($k=1$) subhalos of the main progenitor.} subhalo and with an associated stellar mass above $10^5 \Msun$ (roughly our completeness limit). For each, we register the stellar mass at accretion, $\mstar$, the redshift of accretion onto the main progenitor, $z_{\rm acc}$, and whether the satellite survives to the present-day or not. The top panel of Figure~\ref{fig:survival} indicates the survival probability ($P_{\rm survival}$) of satellites, as a function of $\mstar$ and $\zacc$, measured using a simple 15 x 15 binning scheme. Here, the lighter yellow cells denote higher probabilities of survival as indicated. Figure~\ref{fig:survival} clearly shows that satellites accreted at redshifts $z>3$ only have a small probability of surviving to the present \citep[see also][]{Geen.etal.13}, and only if they are low mass ($\mstar \lesssim 10^7 \Msun$). More specifically, the accretion redshift above which the survival probability is less than 50\% decreases from $\zacc = 2.89$ for satellites with $\mstar \sim 10^{5} \Msun$, to $\zacc = 0.70$ for $\mstar \sim 10^{7}$, to $\zacc=0.32$ for $\mstar \sim 10^9 \Msun$. The fact that $P_{\rm survival}$ decreases with increasing stellar mass is due to the fact that more massive satellites experience stronger dynamical friction, but also to the fact that more massive satellites are accreted at lower redshift which is evident from the absence of data in the upper-right corner of the panel. 

The bottom panel of Figure~\ref{fig:survival} shows the median accretion redshift of satellites as a function of their stellar mass at accretion and the error bars indicate the corresponding 16-84 percentile ranges. Yellow and blue symbols denote surviving and disrupted/merged satellites, respectively. Two trends are noteworthy; more massive satellites are accreted later (independent of whether they survive to the present or not) and at fixed stellar mass, a satellite's accretion redshift is the primary determinant of its survivability. In fact, the probability that {\it any} surviving satellite was accreted prior to $z=1.0$ ($z=3.0$) is $0.45$ ($0.06$). Among the surviving systems, more massive satellites have been accreted more recently. We measure the median accretion redshift for the most-massive surviving satellite across the $S_0$ sample as $\zacc = 0.36$ with a 16-84 percentile range of $0.11 < \zacc < 0.89$.

Finally, for every system that does not survive, we also compute the time interval between disruption and infall ($\Delta t_{\rm dis}$). When selecting satellites that lie above the ``classical" dwarf galaxy regime ($\mstar > 10^7 \Msun$), we measure a median disruption timescale of 4.20 Gyr with a wide 16-84 percentile range of 1.08 - 7.56 Gyr. This is remarkably consistent with the range of timescales reported in simulation-based analyses \citep{Grimozzi.etal.24, Pathak.etal.25, Joshi.etal.25b}. 

\subsection{Dominant Progenitors}
\label{sec:fates}

\begin{figure}
    \centering
    \includegraphics{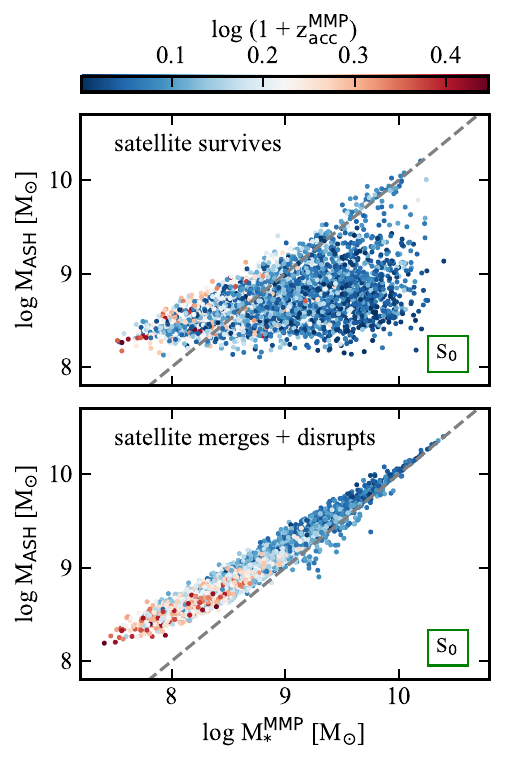}
    \caption{A scatter plot of the stellar mass at accretion of the massive progenitor satellite vs. the accreted stellar halo mass $\Mash$ in $S_0$ sample. The points are colored by the accretion redshift of the satellite $\zacc$ and the grey dashed line shows the one-to-one relation. \textbf{Top panel}: The host halos whose most massive progenitor survives to present-day. \textbf{Bottom panel}: The host halos whose most massive progenitor merges or disrupts. The merged+disrupted populations sit clustered near the grey dashed line because if the most massive progenitor does not survive as a satellite, it dominates the $\Mash$ component. The surviving population covers the same parameter space but extends to the bottom right of the panel where galaxies have low $\Mash$ but high $\MMP$. This figure illustrates that a single satellite can significantly alter the stellar mass budget in $\Mash$ component. }
    \label{fig:fates}
\end{figure}

Previous studies based on simulations have shown that the ASHs of MW-mass galaxies are predominantly built up from only a handful of the most massive accreted satellites \citep{Bullock.Johnston.05, Cooper.etal.10, Deason.etal.16, Fattahi.etal.19, Monachesi.etal.19, Orkney.etal.22, Khoperskov.etal.23, Pu.etal.25}. In this section we examine which progenitors contribute the most to the ASH in our model using two summary statistics; the stellar mass at accretion of the most massive\footnote{Here ``most massive'' refers to the stellar rather than the dark matter mass.} progenitor ($\MMP$), and the number of mass-ranked progenitors responsible for 90\% of the present-day ASH mass ($\Nnine$). 

Figure~\ref{fig:fates} shows scatter plots of $\Mash$ as a function of $\MMP$. Here every point represents a host galaxy from the $S_0$ sample color-coded based on the accretion redshift of its most massive progenitor. For clarity, we forgo showing results for the $S_{15}$ and $S_{30}$ samples, but they are qualitatively the same. Surviving satellites are shown in the top panel, and merged or disrupted satellites in the bottom panel. 

The bottom panel of Figure~\ref{fig:fates} shows that if the most massive progenitor disrupts or merges, it typically contributes a very substantial fraction of the total mass of the ASH.  The slight deviation away from the one-to-one relation (grey-dashed line) at lower masses is simply due to the fact that more than just the most massive progenitor can deposit stars into the ASH. Since massive progenitors typically experience significant dynamical friction, a large fraction of the most massive progenitors sink to the center of their host where they either merge with their parent galaxy or experience complete disruption due to the strong tides. Indeed, only $\sim 31\%$ of the most massive progenitors in the $S_{0}$ sample survive to the present-day. Interestingly, this is comparable to the survival fraction of the entire satellite population (34\%), which seems at odds with the fact that more massive satellites experience stronger dynamical friction. The reason is that the most massive progenitor is typically accreted relatively late compared to the average satellite as seen by the lack of a clear color gradient in the surviving population.

If the most massive progenitor survives to the present-day, either because it was accreted recently or because it ended up on a more circular orbit along which the tides are weaker, its contribution to the ASH is typically much smaller. In fact, the lack of points well above the one-to-one line at the massive end indicates that massive ASHs in MW-mass galaxies are primarily built through the accretion and eventual disruption of their most massive progenitor satellites \citep{Deason.etal.16}. Altogether, these results emphasize how the final fate of a single massive satellite can have a significant effect on $\Mash$. 

\begin{figure}
    \centering
    \includegraphics[width=0.5\textwidth]{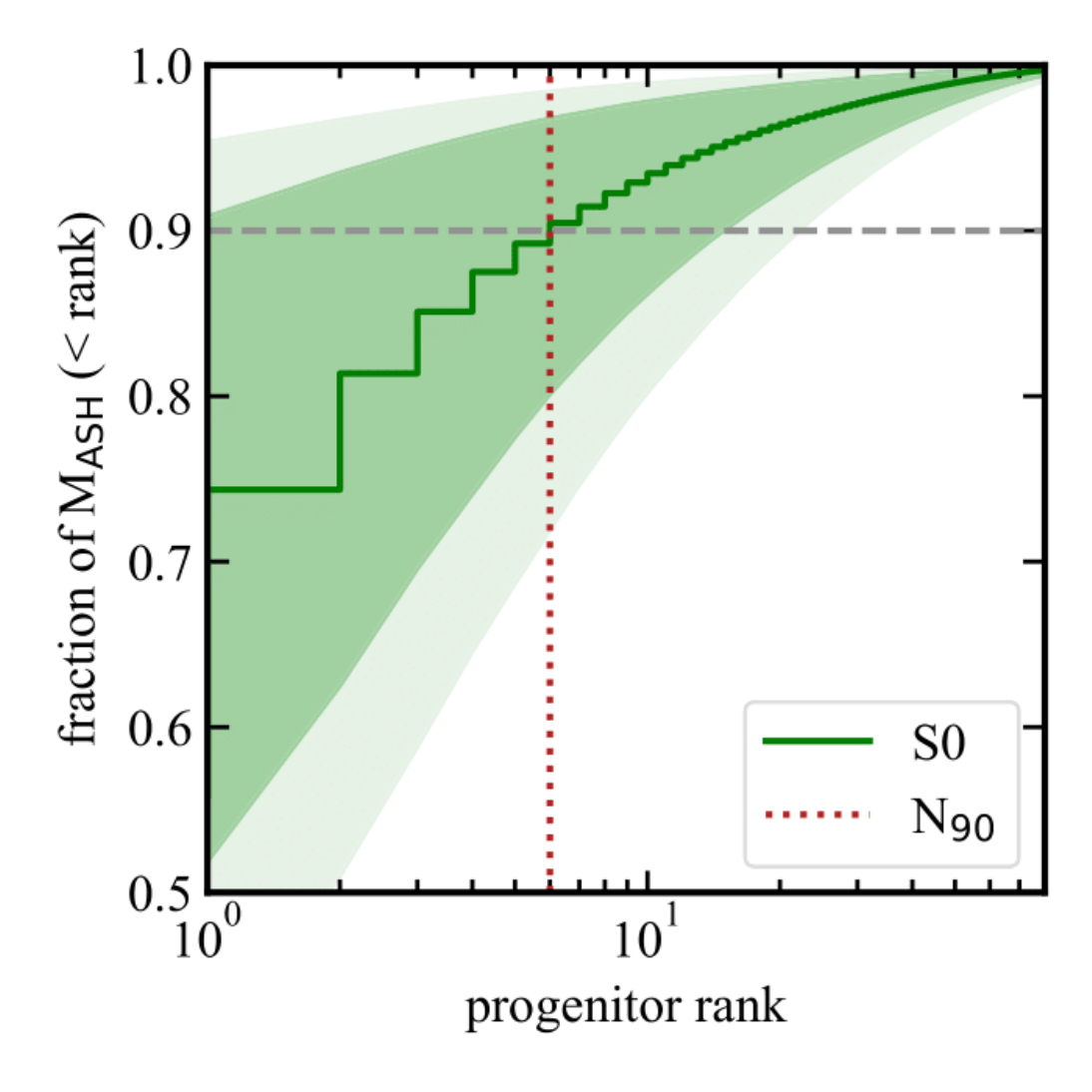}
    \caption{The cumulative distribution of the present-day ASHs as function of progenitor rank from the $S_0$ sample. The solid green line shows the median value across each progenitor rank and the shaded regions show the 16-84 (darker shade) and 5-95 (lighter shade) percentiles across the sample. The grey dashed line indicates 90\% of the present-day $\Mash$ which we take define the number of significant progenitors $\Nnine$. The maroon dotted line indicates the median value of $\Nnine = 6$.}
    \label{fig:progenitors}
\end{figure}

Next we rank-order the progenitors by the amount of stars they contribute to the ASH. Note that the highest ranked progenitors are not necessarily the most massive. Instead, highly-ranked progenitors are preferentially those that were assigned more eccentric orbits which results in smaller peri-centric distances and therefore in more tidal mass loss \citep{Penarrubia.etal.06, Proctor.etal.24}. Figure \ref{fig:progenitors} shows the cumulative distribution of the $\Mash$ component as a function of these mass-ranked progenitors. Since the results are virtually identical across all three merger tree samples, we only show results for the $S_0$ sample. The solid green line shows the median value for the entire sample, while the shaded regions show the 16-84 and 5-95 percentiles. Using the same formalism as in \cite{Monachesi.etal.19}, we calculate the number $\Nnine$ of mass-ranked progenitors that, when combined, are responsible for at least 90\% of the final ASH mass. The dotted red line in Figure \ref{fig:progenitors} shows the median value of $\Nnine$ which, for our fiducial $S_0$ model, is 6 progenitors. This is in excellent agreement with  \cite{Monachesi.etal.19} who reported a median value of $\Nnine \approx 6.5$ based on the Auriga suite of 30 MW-mass galaxies. Importantly, we measure the 16-84 percentile range to be 1 to 11, which is also consistent with \cite{Monachesi.etal.19} who report $\Nnine$ ranging from 1 to 14.

\subsection{Sources of Stochasticity}
\label{sec:stochasticity}

The semi-analytical framework of \SatGen allows for a straightforward investigation of the various sources of stochasticity that ultimately give rise to the large variance in $\Mash$. The main sources of stochasticity are (i) the variance in mass accretion histories of the host halo, (ii) the random orbit that is assigned to each satellite galaxy at accretion, (iii) the assumed scatter in the empirical relations used to assign satellite sizes and stellar masses at accretion, and (iv) the variance in host halo masses (in samples $S_{15}$ and $S_{30}$). Except for a weak correlation between host halo mass and assembly history \citep[more massive halos assemble later, see e.g.,][]{vdBosch.02}, these four sources of stochasticity are independent of each other.

\begin{figure}
    \centering
    \includegraphics{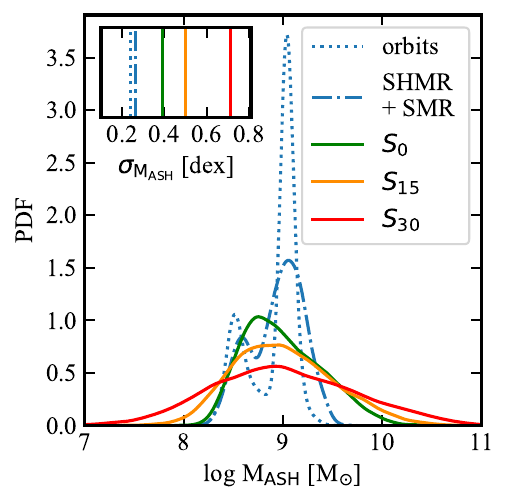}
    \caption{The PDFs of $\Mash$ across the different stochasticity models discussed in Section \ref{sec:stochasticity}. Here the blue lines correspond to a single merger tree realization while the green, orange and red lines correspond to the $S_0$, $S_{15}$ and $S_{30}$ merger tree samples, respectively. The small inset panel on the top left shows the standard deviation across the samples. Following the legend from top to bottom, the blue dotted curve is a model that only includes stochasticity from orbit initialization, the blue dash-dotted curve adds in stochasticity from assumed scatter in the empirical assignment of stellar mass and size, the green curve adds in the stochasticity from halo-to-halo variance, and finally the orange and red lines add in stochasticity from host halo mass-mixing.}
    \label{fig:scatter}
\end{figure}

In order to quantify how each of these sources of stochasticity impact the ASH, we select a single merger tree whose $\Mash$ lies near the median of the $S_0$ sample. We use this merger tree to construct $1,000$ realizations of the stellar halo, each time assigning the satellites different orbits at accretion, but keeping the same stellar masses and sizes. Hence, each realization yields a different $\Mash$, whose variance is due solely to the stochasticity associated with the assignment of satellite orbits. The blue dotted curve in Figure \ref{fig:scatter} shows the resulting PDF for $\Mash$. The pronounced double peak is yet another manifestation of the prominent role played by the most massive progenitor; the left peak corresponds to orbital draws for which the most massive progenitor survives to the present-day, thus contributing little to the ASH, while the right peak corresponds to orbital draws that result in either complete disruption or merging of the most massive progenitor. The vertical blue dotted line in the inset shows that orbital stochasticity alone contributes a little more than 0.2 dex to the scatter in the final $\Mash$.

Next, we repeat the same exercise except that this time, in addition to resampling the orbits, we also resample the stellar masses and sizes at accretion. Hence, we add the impact of stochasticity resulting from the scatter in the SHMR. The corresponding results are shown as the blue dash-dotted curve. Although the two peaks of the PDF have both broadened somewhat, the PDF is still clearly bimodal. The inset shows that the overall variance in $\Mash$ has barely increased. Hence, the impact of orbit stochasticity dominates that of scatter in the assumed empirical relations.

The solid green line shows the PDF for $\Mash$ for the entire $S_0$ sample (identical to the middle panel in Figure~\ref{fig:fiducial}), which displays the combined impact of stochasticity due to orbit initialization, scatter in the SHMR, and halo-to-halo variance in MAHs. As is apparent from the inset, the added stochasticity due to the latter significantly boosts the scatter in $\Mash$ to $\sim 0.4$ dex. The main reason is that, as demonstrated in Section~\ref{sec:fates} above, the ASH is dominated by the contribution of only a handful of satellites. Hence, $\Mash$ is susceptible to discreteness noise originating from variance in the masses and accretion redshifts of the most massive progenitor halos. Note that this relative impact of the massive progenitors is amplified by the steep slope of the SHMR at the low mass end, which ensures that massive progenitors contribute a stellar mass that is disproportionally high. Hence, it is expected that the level of stochasticity contributed by the halo-to-halo variance of MAHs is larger (smaller) when the low-mass slope of the SHMR is steeper (shallower). See Section \ref{sec:correlations} below for more detailed discussion.

Finally, the orange and red curves in Figure~\ref{fig:scatter} show the effects of including host halo mass-mixing. As expected, this significantly increases the variance in $\Mash$ and ends up being the dominant source of stochasticity simply because more (less) massive host halos accrete more (less) massive substructure. Comparing the PDFs for the $S_{15}$ and $S_{30}$ samples, it is evident that the additional $0.15$ dex increase in the scatter of the host halo mass translates to $\sim 0.2$ dex increase in the scatter of $\Mash$. 

\subsection{Example Mass Accretion Histories}
\label{sec:examples}

\begin{figure}
    \centering
    \includegraphics{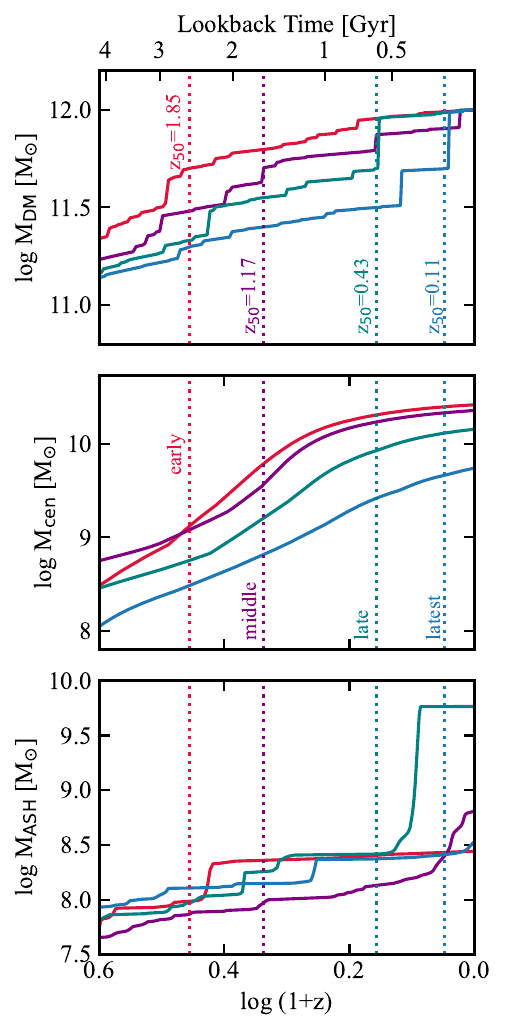}
    \caption{The MAHs of the four illustrative example galaxies described in Section \ref{sec:stochasticity}. \textbf{Top panel}: The dark matter MAHs and corresponding $\zfive$ values. There is significant variance across the four examples despite them all having the same present-day mass of $\Mdm = 10^{12} \Msun$. \textbf{Second panel}: The central galaxy MAHs whose shape is dominated by the $\Min$ component. Notice that the earliest forming galaxy has the largest $\Mcen$ by $z=0$. \textbf{Third panel}: The stellar halo MAHs whose shapes are a convolution of the dark matter MAHs and the subsequent satellite orbit evolution. }
    \label{fig:examples_MAH}
\end{figure}

The above analysis shows that variance in MAHs translates into variance in $\Mash$. One of the main goals of this work is to investigate to what extent these properties are correlated. To build some insight, we select only 4 merger trees from the $S_0$ sample with different formation times that roughly sample the full range of $\zfive$. The corresponding MAHs are shown in the top panel of Figure \ref{fig:examples_MAH}, where each small step in an individual MAH represents a subhalo that was accreted onto the main progenitor at that time. For ease of reference, in what follows we refer to these MAHs as \textit{earliest} ($\zfive = 1.85$),  \textit{early} ($\zfive = 1.17$), \textit{late} ($\zfive = 0.43$) and \textit{latest} ($\zfive = 0.11$).  

The middle panel shows the growth histories of the central galaxies. The fact that these four halos have unique MAHs results in four distinct central stellar masses despite having identical host halo masses (see Section \ref{sec:central}). Notice the relative smoothness in the growth of the stellar mass of the central component compared to the step-like increase of the mass of the dark matter halo. This is a consequence of the fact that the build-up of stellar mass involves an integration of the star formation rate across cosmic time, buffered by a time-dependent return fraction (see Section~\ref{sec:central}). At $z=0$, there is a clear positive correlation between $\Mcen$ and $\zfive$ (see also Figure \ref{fig:SHMR}), with the halos that assemble earlier hosting a more massive central galaxy \citep{Correa.etal.20}. This correlation is a consequence of the enhanced SFRs at early times in the in-situ model that dominates the $\Mcen$ component.

Finally, the bottom panel shows the evolution of the ASH mass. Notice the step like cadence similar to that in the top-most panel. This simply reflects that the ASH builds up it mass from the accretion and subsequent stripping of individual satellite galaxies (see also Figure 8 of \cite{Cooper.etal.10}). Despite this correspondence between the assembly history of the host halo and that of its ASH, it is clear that the final ASH mass is poorly correlated with host halo formation time. In particular, the most massive ASH has formed in the \textit{late} host, which has an intermediate formation time, whereas the \textit{early} and \textit{latest} hosts have ASH masses that are extremely similar, despite having halo formation times that are very different. The main reason for this weak correlation is the impact of the other sources of stochasticity, in particular that arising from the random orbits assigned to the satellites. For example, the \textit{latest} tree accretes its most massive progenitor ($\macc = 10^{11.67} \Msun$) at $z=0.11$ which is initialized with an orbit that neither merges with the central galaxy nor disrupts by $z=0$. In contrast, when the \textit{late} tree accretes its most massive ($\macc = 10^{11.60} \Msun$) progenitor at $z = 0.43$, it is initialized with an orbit that sees it merge with the central in less than $\sim 2$ Gyrs. Dynamical friction, tidal stripping and the eventual merger event transfer almost the entirety of its $10^{9.74} \Msun$ stellar budget to the $\Mash$ component. This large deposit of stars is evident as the sharp jump in $\Mash$ for the \textit{late} system at $z \sim 0.09$, roughly 2 Gyrs after the satellite system was accreted. 

These four examples demonstrate that the stellar mass of both the central galaxy and the ASH depend strongly on the MAH of the host halo, even at fixed present-day halo mass. However, other stochastic elements, in particular the orbital parameters of the accreted satellites, also strongly impact the buildup of the stellar halo, thereby reducing the correlation between $\Mash$ and the system's assembly history.  Whether or not the correlation remains strong enough to put meaningful constraints on a galaxy's DM MAH from measurements of its ASH is the topic of the next section.

\subsection{Median Stellar Halo Assembly Histories}
\label{sec:ash_mah}

\begin{figure*}
    \centering
    \includegraphics{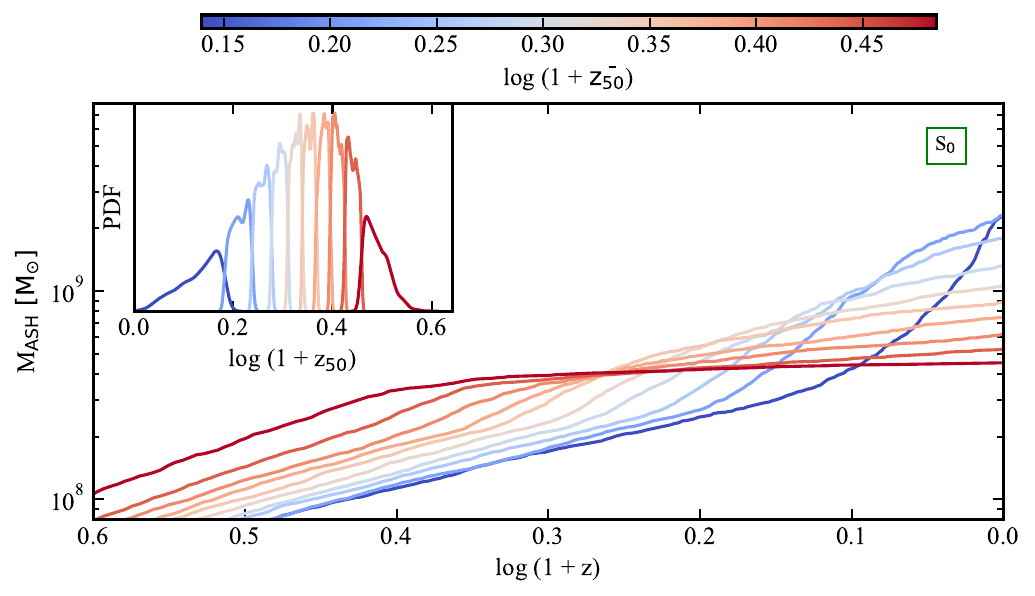}
    \caption{Accreted stellar halo assembly histories for the $S_0$ sample, binned by host halo formation redshift according to the scheme shown in the sub-panel. Each line shows the median assembly history across 1000 trees in a given $\zfive$ bin, colored by the median formation redshift. Clearly, earlier (later) forming host galaxies assemble less massive (more massive) stellar halos. We caution that the strength of this trend with $\zfive$ is a consequence of our stacking procedure (see Sections \ref{sec:stochasticity} and \ref{sec:examples}).}
    \label{fig:ashmah}
\end{figure*}

In Section 3.5, we focused on a small number of representative accretion histories to build intuition for how our model operates. Here, instead, we consider ensemble trends in stellar halo assembly histories by stacking merger trees according to host halo formation redshift. We sort the $S_0$ sample (for clarity we exclude the $S_{15}$ and $S_{30}$ samples as they are qualitatively similar) by $\zfive$ and bin it such that each subsample contains 1000 unique merger trees. The inset panel in the top-left corner of Figure~\ref{fig:ashmah} shows the resulting binned $\zfive$ distribution. The color-bar indicates the median $\log(1+\zfive)$ in each bin, over which we compute the median assembly histories, with earlier (later) forming host galaxies plotted in red (blue).

In contrast to the individual examples in Figure~\ref{fig:examples_MAH}, these ensemble histories are significantly smoother and do not exhibit the discrete jumps associated with individual accretion events. They do, however, clearly demonstrate that earlier (later) forming host galaxies assemble less massive (more massive) accreted stellar halos. The relatively sharp late-time growth seen in the bluer curves highlights that the most massive stellar halos are assembled recently through the disruption of a small number of massive progenitors. In contrast, the redder curves reach their present-day values earlier and subsequently plateau. We discuss these trends in detail in Sections~\ref{sec:assembly_histories} and \ref{sec:discussion}, and shown analogous visualizations of the data in Figures~\ref{fig:zcorrelation} and \ref{fig:stack}.

\section{Inferring Galaxy Assembly Histories}
\label{sec:assembly_histories}

In this section, we quantify how a galaxy’s mass accretion history (MAH) is encoded in its present-day halo observables. To do so, we assume that the following mass components are, in principle, observable for every galaxy: the stellar mass of the central galaxy ($\Mcen$), the accreted stellar mass in the halo ($\Mash$), and the cumulative stellar mass of all surviving satellites ($\Msat$). We refer to this set of observables, [$\Mcen, \Mash, \Msat$], as the ``input features''. From these inputs, we aim to infer three distinct ``target variables'' that characterize the galaxy’s MAH: (i) the redshift at which the host halo assembled 50\% of its present-day mass ($\zfive$); (ii) the number of progenitor galaxies contributing 90\% of the present-day stellar halo ($\Nnine$); and (iii) the dark matter mass of the most massive progenitor halo ($\MMPdm$) at accretion.

\begin{table}[ht]
    \vskip0.1in
    \small\centering
    \begin{tabular}{c|c|c|c}
    \hline
    \hline
    sample & \textcolor{ForestGreen}{S$_0$} & \textcolor{YellowOrange}{S$_{15}$} & \textcolor{red}{S$_{30}$}\\
    \hline
    & \multicolumn{3}{c}{$\rho_S$ ($\Mcen$)}\\
    \hline
    $\zfive$ & 0.852 & 0.528 & 0.225\\
    $\Nnine$ & 0.452 & 0.262 & 0.139\\
    $\MMPdm$ & -0.648 & 0.006 & 0.525\\
    \hline
    & \multicolumn{3}{c}{$\rho_S$ ($\Mash$)}\\
    \hline
    $\zfive$ & -0.538 & -0.458 & -0.422 \\
    $\Nnine$ & -0.965 & -0.785 & -0.567 \\
    $\MMPdm$ & 0.669 & 0.771 & 0.884 \\
    \hline
    & \multicolumn{3}{c}{$\rho_S$ ($\Msat$)}\\
    \hline
    $\zfive$ & -0.342 & -0.352 & -0.365 \\
    $\Nnine$ & 0.257 & 0.219 & 0.134 \\
    $\MMPdm$ & 0.090 & 0.327 & 0.611\\
    \end{tabular}
\caption{The Spearman rank-order correlation coefficients between the input features and target variables. Each measurement is made across the 10,000 galaxies in each merger tree sample (separated by columns). See Figure \ref{fig:stack} for the accompanying 2D contour plots.}
\label{tab:correlations}
\end{table}

We first measure the Spearman rank-order correlation coefficients ($\rho_S$) in each merger tree sample for each pair of input feature and target variable. The resulting measurements are listed in Table \ref{tab:correlations}. Notice how, for almost each target variable, the inclusion of host halo mass-mixing decreases the correlation strength as measured by $\vert \rho_S \vert$ (See Appendix~\ref{App:stack} for a more detailed discussion of these correlation coefficients and their physical interpretations). Below we quantify how each input feature correlates with different definitions of halo formation time and we train a Random Forest Regression algorithm (RFRa) on subsets of our merger tree samples to see how well they recover truth values in our target variables. Details regarding how the RFRa works can be found in Appendix \ref{App:RFR}, along with  the definition of the $R^2$ metric discussed in Sections \ref{sec:z50}, \ref{sec:N90} and \ref{sec:MMP}.

\subsection{Correlation Strengths}
\label{sec:correlations}

It is common to quantify the halo formation (or assembly) time by $\zfive$, the redshift at which the main progenitor has assembled 50\% of its final mass. However, this is fairly arbitrary and it might well be that other epochs in a halo's formation history are more strongly correlated with the final properties of the halo's stellar content. Our \SatGen model is ideally suited to examine this in detail. For each of our three input features ($\Mcen$, $\Mash$, and $\Msat$) and for each of our three samples ($S_0$, $S_{15}$ and $S_{30}$) we compute the Spearman rank-order correlation coefficient between the input feature and the halo formation redshift $z_f$, defined as the redshift at which the main progenitor has assembled a fraction $f$ of the final $z=0$ halo mass. We do this for a range of different values of $f$, ranging from 0.01 to 0.99 in steps of 0.01.

\begin{figure}
    \centering
    \includegraphics{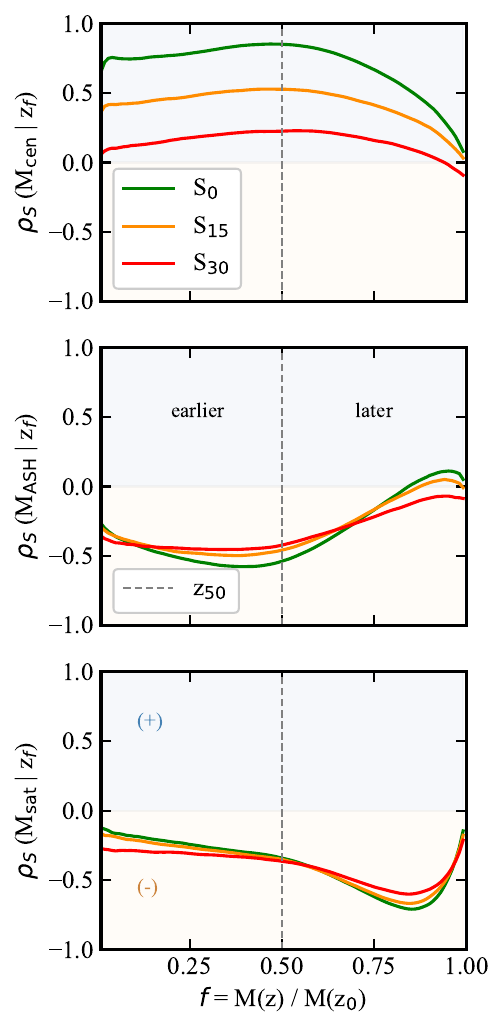}
    \caption{The correlations between each observable stellar mass component and halo formation times. Here we compute $z_f$ on a discrete grid of mass fractions $f = M(z) / M(z_0)$ (where $z_{37}$ would correspond to the redshift at which a halo had formed 37\% of its present-day mass). The colored lines show the rank-order correlation coefficient as measured across the 10,000 galaxies in each of the three merger tree samples ($S_0, S_{15}$ and $S_{30}$). The vertical dashed line indicates $\zfive$ and the shaded colored regions distinguish between positive and negative correlations. \textbf{Top Panel:} the correlations between $z_f$ and $\Mcen$, \textbf{Middle Panel:} the correlations between $z_f$ and $\Mash$ and finally, \textbf{Bottom Panel:} the correlations between $z_f$ and $\Msat$.}
    \label{fig:zcorrelation}
\end{figure}

The results are shown in Figure \ref{fig:zcorrelation}, which shows $\rho_S$ as a function of $f$. Different panels correspond to different input features, while different colors correspond to different halo samples, as indicated. Note that whereas $M_{\rm cen}$ is positively correlated with $z_\rmf$, both $\Mash$ and $\Msat$ are anti-correlated with halo formation time. The positive correlation between $M_{\rm cen}$ and halo formation time is an outcome of the fact that star formation is more efficient at higher redshifts, such that halos that assemble earlier, build up a larger stellar mass (see Section \ref{sec:central}). The anti-correlations between $\Mash$, $\Msat$, and $z_f$ arise from the assumed SHMR. Hierarchical structure formation dictates that numerous low-mass progenitors dominate accretion onto host galaxies at early times while more massive progenitors are typically accreted at later times. If the slope at the low-mass end of the SHMR is steeper than unity (i.e. $\rmd\log \mstar/\rmd\log \Mdm > 1$), more massive progenitors bring in a stellar mass that is disproportionally high. As a consequence, host galaxies that assemble a large fraction of their mass at late times, through the accretion and (potential) disruption of a handful of massive satellites, will naturally exhibit larger $\Mash$ and $\Msat$ \citep{Cooper.etal.13, Amorisco.etal.17, Elias.etal.18, Rey.and.Starkenburg.2022}. Hence, the anti-correlations with $\zfive$ seen in the lower two panels of Figure~\ref{fig:zcorrelation}, arise from the fact that our assumed SHMR has a slope of $\rmd\log \mstar/\rmd\log \Mdm \simeq 2$ below $\Mdm \sim 10^{12}\Msun$.

This has an interesting corollary. The slope of the SHMR is known to change from greater than unity to less than unity around a halo mass of $\sim 10^{12} \Msun$ \citep[e.g.,][]{Moster.etal.10, Yang.etal.12, Behroozi.etal.19}. Hence, halos with a mass less than (or comparable to) that of the MW will typically accrete subhalos that fall in the range where $\rmd\log \mstar/\rmd\log \Mdm > 1$, thus giving rise to an anti-correlation. However, halos with $M \gta 10^{13} \Msun$ are expected to accrete most of their mass from progenitors that fall in the range where $\rmd\log \mstar/\rmd\log \Mdm < 1$. Consequently, we expect their $\Mash$ and $\Msat$ to be positively correlated with the halo formation time. Using a particle tagging technique, \cite{Dacunha.etal.25} measured the correlation strengths between the present day stellar mass \textit{gaps} (difference in mass between the central galaxy and the brightest satellite) and host halo mass assembly histories in a manner analogous to that shown in Figure \ref{fig:zcorrelation}. They find that, at the cluster scale, the present day stellar mass gap is most strongly correlated ($\rho_S \sim 0.4$) with early time host formation. We investigate this in more detail in a forthcoming study (Monzon et al. in prep).

As is evident from Figure~\ref{fig:zcorrelation}, each of the three input features reveals a different dependence on $f$. In particular, while $\Mcen$ is most strongly correlated with halo assembly time for $f=0.5$, $\Msat$ and $\Mash$ reveal their strongest correlation for $f \sim 0.85$ and $f \sim 0.45$, respectively. This indicates that the total mass in surviving satellites is a good indicator of the halo's more recent assembly history, while the stellar mass of the central and its ASH are more indicative of the host halo's earlier assembly history. These trends are easy to understand. The stellar mass of the central and its ASH built up over the entire assembly history. In particular, the star formation rate of the central galaxy closely follows the assembly history of the main progenitor (see Section \ref{sec:central}) while the final ASH is the aggregate of stars stripped from satellites in all progenitors (see Section \ref{sec:mergers}). In the case of satellite galaxies, the situation is different since the vast majority either completely disrupt, merge with the central, or lose a substantial fraction of their mass to the ASH. Because of this, only satellites that were accreted relatively recently make a significant contribution to the final stellar mass of surviving satellites, which explains why $\Msat$ is a good gauge of the halo's recent accretion history (see Section \ref{sec:survival} for further discussion). 

Comparing the results for our three different halo samples, it is clear that host halo mass-mixing drastically suppresses the correlation between halo formation time and central stellar mass. This is basically a consequence of the fact that more massive halos typically form more massive centrals, which implies a positive correlation between halo mass and $\Mcen$. Since more massive halos assemble later, this implies a negative correlation between $z_\rmf$ and $\Mcen$. However, at fixed halo mass, the results for the $S_0$ sample clearly reveal a positive correlation between $z_\rmf$ and $\Mcen$. These two opposing trends explain the suppression of the correlation strength with increased mass-mixing. Interestingly, the correlation strength for the other two input features appear to be largely impervious to mass-mixing.

Despite the disparity in which formation time maximizes the correlation strength, in what follows, we proceed with using $\zfive$ as our principal MAH summary statistic. This is motivated by the fact that $f=0.5$ is a good compromise for which all three observables reveal a significant correlation strength and by the fact that $\zfive$ is widely used in the literature, thereby facilitating comparisons with other studies.

\subsection{Training the Algorithms}
\label{sec:training}

\begin{figure*}
    \centering
    \includegraphics{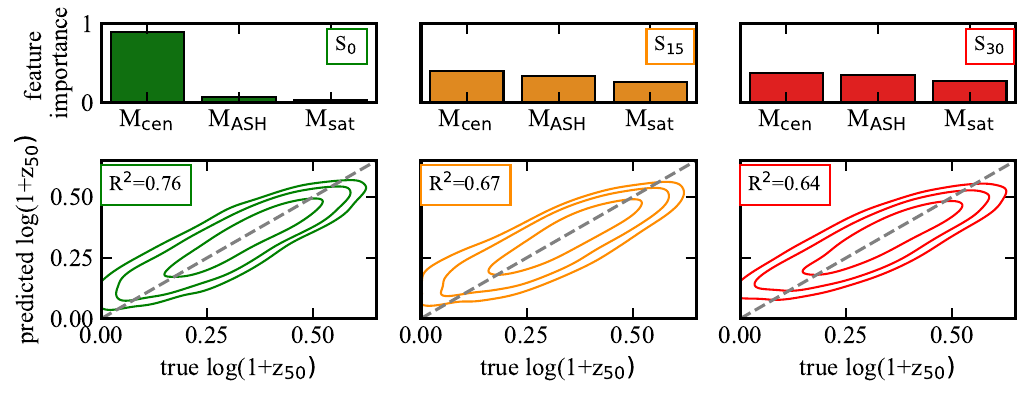}
    \caption{The results of the RFRa for the $\zfive$ target variable. \textbf{Top three panels}: The relative importance of the three input features [$\Mcen, \Mash$ and $\Msat$] for all three merger tree samples. \textbf{Bottom three panels}: The True vs. Predicted contours for the 2000 galaxies used in the validation step. The contours show the 1, 2, and 3$\sigma$ confidence intervals from a kernel density estimator. Legends show the algorithm's $R^2$ score which can be interpreted as proportion of variance in $\zfive$ that can be explained by the independent features in the model. Notice that the $S_0$ inference test is almost entirely reliant on the $\Mcen$ input feature but, when host halo mass-mixing is included, $\Mash$ and $\Msat$ become more relevant. Furthermore, because the $R^2$ score worsens from $S_{0}$ to $S_{30}$, its clear that $\Mdm$ should be constrained before inferring $\zfive$.}
    \label{fig:RF_z50}
\end{figure*}

\begin{figure*}
    \centering
    \includegraphics{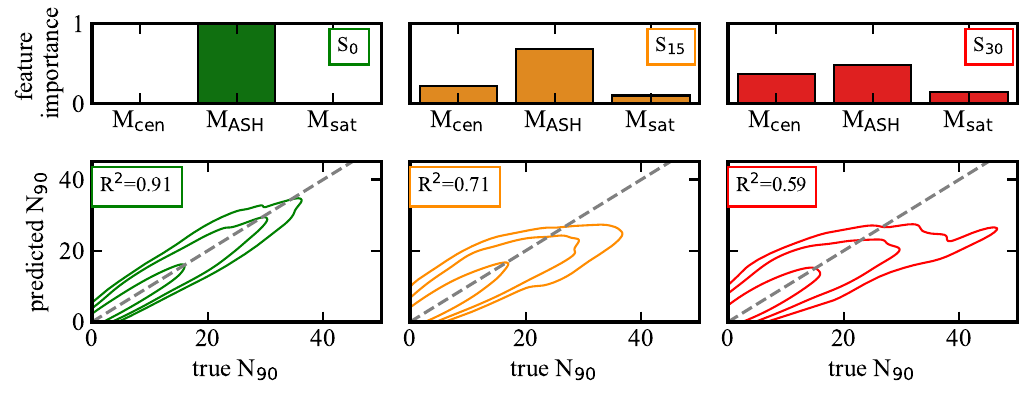}
    \caption{The same as Figure \ref{fig:RF_z50} but for the $\Nnine$ target variable. Notice that the $S_0$ inference test is entirely reliant on the $\Mash$ input feature but, when host halo mass-mixing is included, $\Mcen$ and $\Msat$ become more relevant. Once again, because the $R^2$ score worsens from $S_{0}$ to $S_{30}$, its important to get a handle on the underlying host halo masses before trying to infer $\Nnine$.}
    \label{fig:RF_N90}
\end{figure*}

\begin{figure*}
    \centering
    \includegraphics{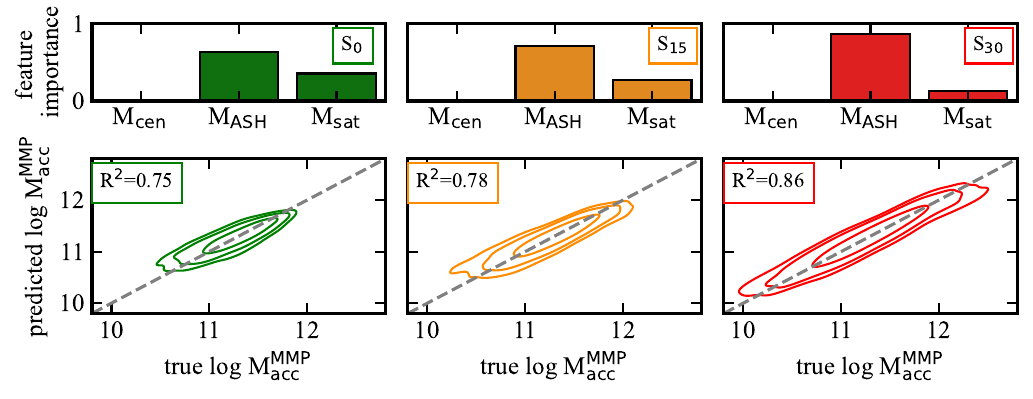}
    \caption{The same as Figure \ref{fig:RF_z50} but for the $\MMPdm$ target variable. Here the $\Mcen$ feature holds almost effectively no information relevant to $\MMPdm$ and $\Mash$ is dominant across all three samples. Unlike the other two target variables, the $R^2$ scores actually increase from $S_0$ to $S_{30}$ which means that host halo mass-mixing does not muddle inference on the $\MMPdm$ target variable.}
    \label{fig:RF_MMP}
\end{figure*}

Throughout the analysis, we use an 80\% / 20\% split for the training and validation sets of the RFRa, respectively. Each inference test is performed using data from one of our three merger tree samples ($S_0, S_{15}, S_{30}$). Recall that each merger tree sample has 10,000 unique galaxy realizations. Therefore, we train each algorithm on 8,000 trees selected at random during initialization and measure $R^2$ (see Appendix \ref{App:RFR} for details) with respect to the remaining 2,000 trees. In this way, the algorithms are never evaluated on data that were used for training.

As mentioned previously, the algorithms are designed to learn the mapping between a set of input features and a single target variable. This means that for every merger tree sample, we train three RFRs, one for each of the target variables. We adopt the following shorthand: an RFR trained on the $S_{15}$ sample to recover the $\Nnine$ target variable is denoted $f_{15}^{\Nnine}$. Similarly, an RFR trained on the $S_{0}$ sample to recover the $\zfive$ target variable would be denoted $f_{0}^{\zfive}$, etc. Using only the fiducial input features [$\Mcen, \Mash$ and $\Msat$], we train a total of $3 \times 3 = 9$ algorithms. We show the results of our RFR-based inference tests in Figures \ref{fig:RF_z50}, \ref{fig:RF_N90}, \ref{fig:RF_MMP} for the $\zfive$, $\Nnine$, and $\MMPdm$ target variables, respectively. The top rows of these figures show the relative feature importance, and the bottom rows show the True vs. Predicted scatter plots.

\subsubsection{The $\zfive$ Target Variable} 
\label{sec:z50}

The inference tests on the $\zfive$ target variable differ significantly across the three merger tree samples. This is evident from the top three panels of Figure \ref{fig:RF_z50} that show the relative feature importance of each test. Starting with the $f_{0}^{\zfive}$ algorithm, it finds almost no relevant information in the $\Mash$ and $\Msat$ input features, but performs well with an $R^2 = 0.76$. This is a direct consequence of how we model $\Min$, which results in more massive centrals in halos that assemble earlier (see Section~\ref{sec:central}) and the fact that the $S_0$ sample is free of host halo mass-mixing. In other words, the $f_{0}^{\zfive}$ algorithm illustrates ``a best case scenario'' where the halo mass of the host galaxy is known perfectly.

The $f_{15}^{\zfive}$ and $f_{30}^{\zfive}$ algorithms rely less on the direct link between $\Mcen$ and $\zfive$ and more on the information content of the other two features. Unfortunately, the host-halo mass-mixing causes a downgrade in $R^2$ scores compared to the $f_{0}^{\zfive}$ algorithm. This makes sense given that a significant fraction of the variance in the target variable comes from variance in host halo masses, rather than from variance in the input features. Interestingly, comparing the $f_{15}^{\zfive}$ and $f_{30}^{\zfive}$ algorithms, the additional 0.15 dex scatter in host halo masses in the $S_{30}$ sample does not significantly impact the inference of $z_{\rm 50}$. This is likely due to the fact that their underlying $\zfive$ distributions are very similar (see Figures \ref{fig:SHMR_samples} \& \ref{fig:stack}).

\subsubsection{The $\Nnine$ Target Variable} 
\label{sec:N90}

Figure~\ref{fig:RF_N90} shows the results for the $\Nnine$ target variable. The $f_{0}^{\Nnine}$ algorithm is entirely informed by the $\Mash$ input feature, but performs remarkably well with $R^2 = 0.91$. This highlights the strong anti-correlation discussed in Section \ref{sec:correlations}, showing that halos with more massive ASHs are formed by fewer significant progenitors. 

As discussed in Section~\ref{sec:correlations}, this anti-correlation is a consequence of the relatively steep slope of the SHMR at the low mass end, which implies that more massive subhalos contribute a stellar mass that is disproportionally high. Obviously, the success of the $f_{0}^{\Nnine}$ algorithm is also aided by the fact that host halo mass-mixing is neglected.

In the case of the $f_{15}^{\Nnine}$ and $f_{30}^{\Nnine}$ algorithms, the inclusion of host halo mass-mixing makes the algorithms less reliant on $\Mash$. As with the $\zfive$ target variable, this reduces the $R^2$ performance relative to the $f_{0}^{\Nnine}$ algorithm. Clearly, by increasing host halo mass-mixing, the inference of $\Nnine$ from the input features becomes less reliable. In fact, both the $f_{15}^{\Nnine}$ and $f_{30}^{\Nnine}$ algorithms have a tendency to under predict $\Nnine$ for galaxies with the largest number of significant progenitors.

\subsubsection{The $\MMPdm$ Target Variable} 
\label{sec:MMP}

Unlike $\zfive$ and $\Nnine$, the inference tests on the $\MMPdm$ target do not vary significantly across the three merger tree samples. Figure \ref{fig:RF_MMP} shows that both the $\Mash$ and $\Msat$ features are important for all three samples. This can be understood as follows. If $\MMPdm$ is large, and remains largely intact, it makes a large contribution to $\Msat$. On the other hand, if it does not survive, it makes a large contribution to $\Mash$.

Interestingly, $\MMPdm$ is the only target variable whose prediction accuracy increases with the inclusion of host halo mass-mixing. This owes to the fact that $\MMPdm$, $\Mash$, and $\Msat$ are all strongly correlated with host halo mass (more massive hosts have more massive progenitors, more massive stellar halos and a more massive satellite population). Consequently, the range of $\MMPdm$ values increases with increased mass-mixing, as is clearly evident in the lower panels of Figure~\ref{fig:RF_MMP}. Given that realistic samples will always include a non-negligible amount of mass-mixing, we can therefore conclude that observations of $\Mash$ and $\Msat$ can give a surprisingly good handle on $\MMPdm$.

\section{Discussion}
\label{sec:discussion}

\subsection{Galaxy Assembly Bias}
\label{sec:assembly_bias}

\begin{figure*}
    \centering
    \includegraphics{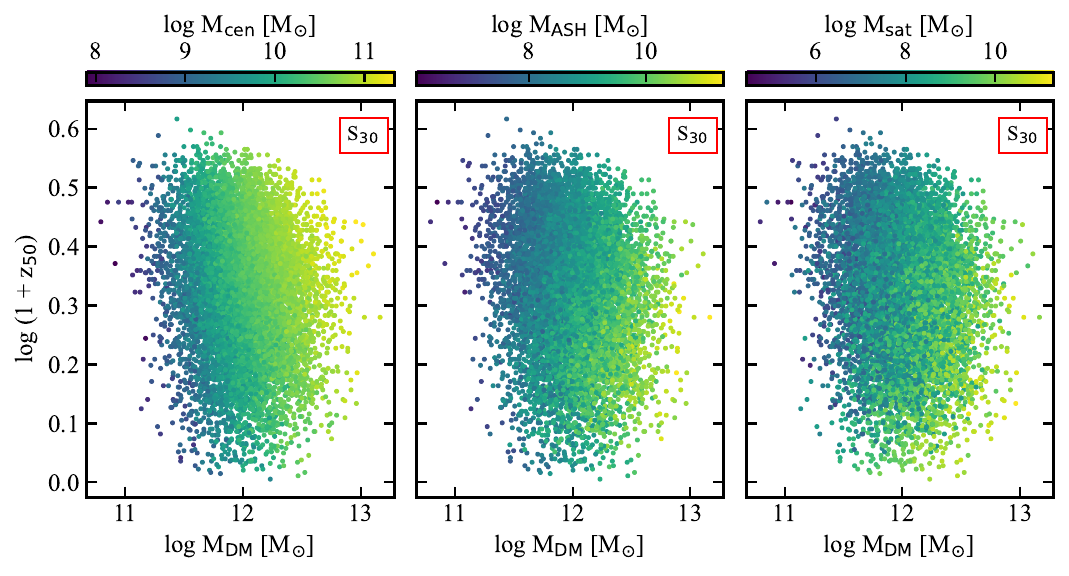}
    \caption{Scatter plots of $\Mdm$ and $\zfive$ in the $S_{30}$ merger tree sample. \textbf{Left Panel:} Here the distribution is colored by the $\Mcen$ input feature and shows a clear color gradient in the x-direction indicative of the assumed SHMR. \textbf{Middle Panel:} Here the distribution is colored by the $\Mash$ input feature and shows a clear color gradient in the x-direction alongside a noisy color gradient in the y-direction indicative of the fact that later forming host halos have more massive accreted stellar halos. \textbf{Right Panel:} Here the distribution is colored by the $\Msat$ input feature and qualitatively demonstrates the same behavior as the $\Mash$ component.}
    \label{fig:assemblybias}
\end{figure*}

An important outstanding question in galaxy formation theory is whether there is evidence for galaxy assembly bias; the correlation between galaxy properties and halo formation time at fixed halo mass \citep[see e.g.,][]{MBW10, Zentner.etal.14, Hearin.etal.15, Zu.Mandelbaum.18, Wechsler.Tinker.18, Wang.etal.22}. The main problem lies in that it is difficult to obtain reliable estimates for the formation times of individual halos. As shown in Section~\ref{sec:assembly_histories}, our model predicts that $\Mcen$, $\Mash$, and $\Msat$ are all correlated with halo formation time, albeit only weakly (see also Figure~\ref{fig:stack} in Appendix~\ref{App:stack}). This suggests that observations of $\Mcen$, $\Mash$ and $\Msat$, can in principle be used to constrain $\zfive$. Indeed, we have demonstrated the ability to infer an estimate of $\log[1+\zfive]$ with an error that is roughly 50 to 60 percent of its variance (as estimated from $\sqrt{1-R^2}$). Although far from perfect, this should be sufficient to split any sample of galaxies in early- and late-forming systems and test for galaxy assembly bias.

However, this comes with an important caveat: tests of assembly bias must control for halo mass and are therefore ideally performed using halos of identical mass. Unfortunately, selecting galaxies directly by halo mass is extremely difficult. Typically, samples are constructed by imposing cuts on stellar mass, luminosity, or surface brightness in an attempt to select a narrow range of host halo mass \citep{Carlsten.etal.22, Mao.etal.24, Guzman24}. Given the non-zero scatter in the SHMR, realistic data samples will unavoidably be subject to significant host halo mass-mixing. Given the shape and scatter of the SHMR, the extent of this mass-mixing is expected to be larger for brighter galaxies \citep[see e.g.,][]{More.etal.09a}. Unfortunately, as shown in Figure~\ref{fig:RF_z50}, host halo mass-mixing significantly weakens the precision of the inferred halo formation time.  Hence, unless halo mass estimates can be made on individual galaxies, for example from gravitational lensing or satellite kinematics, the unknown distribution of $\Mdm$ will significantly weaken the inference of assembly time from observational measurements of $\Mcen$, $\Mash$ and $\Msat$.

To illustrate the extent of this challenge, Figure~\ref{fig:assemblybias} plots the host halo masses vs. formation times of galaxies in the $S_{30}$ sample, color coded based on $\Mcen$ (left panel), $\Mash$ (middle panel) and $\Msat$ (right panel). In all three cases, the color gradient is much more pronounced in the $\Mdm$-direction than in the direction of $\log(1+\zfive)$, indicating that all three `observables' depend more strongly on halo mass than on halo formation time. This highlights the challenges one faces when trying to infer properties such as formation time from a sample of galaxies for which the host halo masses are unknown or uncertain. Nevertheless, our tests based on samples $S_{15}$ and $S_{30}$ suggest that reliable measurement of $\Mcen$, $\Mash$, and $\Msat$ still contains valuable information useful for inferring MAHs. In forthcoming work, we extent this analysis by examining whether such observables can be used to simultaneously infer both halo mass and halo formation time.

\subsection{Caveats}
\label{sec:caveats}

The model that we have used here to study the stellar halos of Milky Way-like galaxies is subject to a number of caveats:

\subsubsection{In-Situ Stellar Halos}
\label{sec:insitu}

Stellar halos are expected to contain a mixture of ex-situ and in-situ populations, with the relative contribution depending on galactocentric radius, the star formation history of the central galaxy, and the assembly history of the host halo \citep{Bullock.Johnston.05, Cooper.etal.10, Deason.etal.16}.  However, throughout this work, we only model the \textit{ex-situ} component. As a result, our predictions are best interpreted as lower limits on the total stellar halo mass of MW-mass galaxies.

In-situ halo stars are generally confined to the inner halo \citep{Cooper.etal.13, Font.etal.20}, where they arise from processes such as disk heating \citep{Tissera.etal.13} and bulge-driven evolution, including feedback and secular mechanisms \citep{Zolotov.etal.09}. Beyond galactocentric radii of $\sim 20-30$ kpc, the stellar halos of MW-mass galaxies are expected to be dominated by accreted (ex-situ) stars \citep{Cooper.etal.15, Font.etal.20, Wright.etal.24}. Simulation-based studies find that the in-situ contribution to the stellar halo mass of MW-mass systems is typically subdominant, though with substantial scatter across different suites. Reported in-situ fractions span a wide range from $\sim$ 5\% to 40\%, depending on numerical resolution and subgrid prescriptions \citep{RodriguezGomez.etal.16, Davison.etal.20, Santistevan.etal.20, Cooper.etal.25, Celiz.etal.25}. It is unclear whether this diversity is due to physical halo-to-halo variance or to uncertain baryonic physics and/or ambiguity in defining the stellar halo. 

From an observational perspective, extragalactic measurements of stellar halos rely heavily on projected radial cuts \citep[e.g.,][]{Merritt.etal.16}, which tend to mix in-situ and ex-situ populations \citep{Sanderson.etal.18}. This makes it unfeasible to isolate and quantify the in-situ contribution in a model-independent manner \citep{RodriguezGomez.etal.16}. Nevertheless, recent observational studies of individual systems are broadly consistent with the picture that ex-situ stars dominate the density profiles and substructure signatures of MW-mass stellar halos, particularly at large radii \citep{Bell.etal.08, Williams.etal.25}.

By restricting our analysis to the ex-situ component, we focus on the stellar material that directly traces satellite accretion and hierarchical assembly. This modeling choice enables robust comparisons with simulation-based studies that can trivially measure ex-situ fractions by selecting star particles that were not formed inside the main progenitor. Finally, because our $\Mash$ component does not constitute the entire stellar halo mass budget, direct comparisons with observational measurements should be interpreted with appropriate caution.

\subsubsection{Satellite Quenching}

By assigning satellite galaxies stellar masses based on the SHMR at the redshift of accretion, we have effectively assumed that galaxies quench their star formation upon accretion. Although it is well known that satellite galaxies experience enhanced quenching compared to centrals \citep[e.g.,][]{Weinmann.etal.06, vdBosch.etal.08a},  the assumption of instantaneous quenching upon accretion is clearly oversimplified. Indeed, a significant fraction of satellite galaxies are star forming \citep{Weinmann.etal.06, Geha.etal.24} and several studies have shown that there is a typical delay time of the order of 1-2 Gyr between satellite accretion and quenching \citep{Wetzel.etal.13, Maier.etal.19, Akins.etal.21, Pathak.etal.25}. These delays are believed to roughly reflect the time it takes a satellite to reach its first peri-centric passage, where ram-pressure-stripping is most effective at removing gas and quenching the system \citep{Gunn.Gott.72, Tollet.etal.17}. However, as long as satellite galaxies do not significantly increase their stellar mass due to ongoing star formation after accretion—as found by several recent studies \citep[e.g.,][]{Munshi.etal.21, Engler.etal.21a, Shipp.etal.24, RodriguezCardoso.etal.25}—this oversimplification should not have a large effect on the results presented here. 

\subsubsection{Dark Matter Density Profiles}

Throughout, we have assumed that all dark matter halos have cuspy NFW density profiles. In reality, baryonic processes can significantly modify the (central) density profiles of halos \citep[][]{Blumenthal.etal.86, Pontzen.Governato.12, Zolotov.etal.12}. which in turn can have a significant impact on their tidal evolution \citep[][]{Penarrubia.etal.10, Errani.Penarrubia.20}. Similarly, if dark matter is self-interacting rather than collisionless, the halos may be cored or have undergone core-collapse \citep[][]{Spergel.Steinhardt.00,Balberg.etal.02}, both of which will significantly impact their tidal evolution. We leave it for future work to examine how modifications of the subhalo/satellite density profiles impact the statistics of $\Mash$ and $\Msat$.

\section{Conclusions}
\label{sec:conclusions}

We have presented a new model based on \SatGen for the construction of stellar halos from the disrupted remains of accreted satellite galaxies. Unlike computationally expensive hydrodynamical simulations that are hampered by limited sample sizes and numerical artifacts, our semi-analytic approach efficiently samples the full diversity of merger histories essential for probing the stochastic processes that shape galaxy formation. Using three merger tree samples, $S_0$, $S_{15}$, and $S_{30}$, which span a log-normal distribution in present-day halo mass centered on $\Mdm = 10^{12},\Msun$ with scatter values of $\sigma_M = 0$, $0.15$, and $0.30$ dex, respectively, we investigate correlations between several observables and characteristics of the host halo's assembly history. Each merger tree sample contains $10,000$ unique accretion histories enabling a robust characterization of halo-to-halo variance.

We have specifically focused on three main present-day observables: $\Mcen$, the stellar mass of the main host galaxy, $\Mash$ the accreted stellar halo mass, and $\Msat$ the cumulative stellar mass in surviving satellites. We studied how these quantities are impacted by various sources of stochasticity in the assembly of MW-mass galaxies, and how they are correlated with characteristics of the host halo's MAH.  Our main results are as follows:

\begin{itemize}[leftmargin=0.27truecm, labelwidth=0.2truecm]

    \item On average, $\sim$ 95\% of the present-day ASH comes from satellite systems that have been completely disrupted by tidal forces or merged directly with the central. The remaining $\sim$ 5\% originates from surviving satellite systems that are in the process of being tidally stripped. 

    \item Because of dynamical friction, more massive satellites are less likely to survive to the present. Specifically for the $S_0$ sample, we measure the accretion redshift above which the survival probability is less than 50\% decreases from $\zacc = 2.89$ for satellites with $\mstar \sim 10^{5} \Msun$, to $\zacc = 0.70$ for $\mstar \sim 10^{7}$, to $\zacc=0.32$ for $\mstar \sim 10^9 \Msun$. 
        
    \item Among the surviving systems, more massive satellites have been accreted more recently. The most massive surviving satellite in the $S_0$ sample was accreted between $z=0.11$ and $z=0.89$ (16-84 percentile range). Furthermore, we measure the probability that {\it any} first-order surviving satellite (with a stellar mass at accretion $\mstar > 10^5$) was accreted prior to $z=1.0$ ($z=3.0$) is $0.45$ ($0.06$).

    \item Despite the prevalence of satellites being cannibalized by the central galaxy, the ex-situ fraction of the final stellar mass of the central galaxy is negligible ($\lesssim 1\%$) even when our free parameters are set to maximize the contribution (see Appendix~\ref{App:freeparam}). This is in agreement with several previous studies \citep[e.g.,][]{Yang.etal.13, Lu.etal.15}, that find the vast majority stars in the central galaxy of MW-mass hosts formed in-situ.
    
    \item Given the steep slope of the low-mass end of the SHMR assumed, the most massive progenitor satellite typically plays a decisive role in setting $\Mash$ \citep{Cooper.etal.13, Amorisco.etal.17, Elias.etal.18, Rey.and.Starkenburg.2022}. This is particularly true for late-forming galaxies (i.e., those with a small $\zfive$) where the survival or destruction/merger of the most massive progenitor leads to order-of-magnitude differences in the accreted stellar halo mass (see Figure~\ref{fig:fates}).

    \item The majority of the present-day ASH originates from a small number of progenitors. The median number of mass-ranked progenitors that, when combined, contribute 90\% to the accreted stellar halo, is 6 with a broad 16-84 percentile range of 1 to 11 (see Figure \ref{fig:progenitors}). This indicates that accreted stellar halos of MW-mass galaxies are subject to large halo-to-halo variance sourced by the stochasticity of the accretion masses, accretion redshifts, and orbits of a handful of the most massive accretion events \citep[see also][]{Deason.etal.16, Monachesi.etal.19}.
    
    \item At fixed halo mass, the main sources of variance in $\Mash$ and $\Msat$ are stochasticity in mass accretion histories and initial orbital satellite properties. Scatter in the SHMR of accreted satellites makes a negligible contribution and will therefore be difficult to constrain given observations of stellar halos and/or the surviving satellite populations
    \citep[see also][]{Monzon.etal.24}.

    \item Correlating our three observables with the redshift $z_f$ at which the main progenitor has assembled a fraction $f$ of its final mass, for different values of $f$, we find that different stellar components trace different formation epochs of dark matter assembly (see Figure \ref{fig:zcorrelation}). For $\Mcen$ and $\Mash$ the correlations are maximized for $f \sim 0.5$, indicating that they carry information regarding the early formation history of the host halo. On the other hand, $\Msat$ is more indicative of the host halo's more recent assembly, with a maximum (anti)correlation for $f \sim 0.85$.

    \item In realistic samples of stellar halos, host halo mass-mixing is unavoidable and is likely to be the dominant source of scatter in $\Mash$ and $\Msat$ (see Figure \ref{fig:scatter}). We find that adding $0.15$ ($0.3$) dex of scatter in host halo mass boosts the scatter in $\Mash$ from $\sim 0.4$ dex to $\sim 0.5$ ($0.7$) dex. In general, host halo mass-mixing significantly weakens correlations between observable input features and target variables. In particular, host halo mass-mixing leads to a degradation in the recovery of $\zfive$ and $\Nnine$ (see Figures~\ref{fig:RF_z50} and~\ref{fig:RF_N90}). In contrast, because $\MMPdm$ traces $\Mdm$, recovery performance actually improves when host halo masses are mixed (see Figure~\ref{fig:RF_MMP}). 

    \item  Using random forest regression, we have demonstrated that the combination of $\Mcen$, $\Mash$, and $\Msat$ observables can be used to infer properties of the host halo's assembly history despite the large variance due to stochasticity in the orbital properties of satellite galaxies.
\end{itemize}

In future work, we will use the model developed here to examine stellar halos and satellite populations across a much wider range in host halo mass and investigate how their properties may inform characteristics of the SHMR, especially at the low-mass end.


\section{Acknowledgments}

The authors thank Andrew Cooper, Rebekka Coles-Bieri, Oscar Agertz, Yao-Yuan Mao, Marla Geha, Yasmeen Asali, Ethan Nadler, Raphael Errani and Peter Behroozi for valuable discussions. The authors are grateful to the anonymous referees for their insightful comments that have significantly improved the manuscript. FvdB is supported by the National Science Foundation (NSF) through grant AST-2307280. This work was performed in part at the Kavli Institute for Theoretical Physics (KITP) in Santa Barbara, which is supported in part by the National Science Foundation under Grant No. NSF PHY-174895.


\bibliographystyle{mnras}
\bibliography{references.bib}

@ARTICLE{Abraham.etal.14,
       author = {{Abraham}, Roberto G. and {van Dokkum}, Pieter G.},
        title = "{Ultra-Low Surface Brightness Imaging with the Dragonfly Telephoto Array}",
      journal = {\pasp},
         year = 2014,
        month = jan,
       volume = {126},
       number = {935},
        pages = {55},
          doi = {10.1086/674875},
archivePrefix = {arXiv},
       eprint = {1401.5473},
 primaryClass = {astro-ph.IM},
       adsurl = {https://ui.adsabs.harvard.edu/abs/2014PASP..126...55A}
}

@ARTICLE{Amorisco.etal.17,
       author = {{Amorisco}, Nicola C.},
        title = "{The accreted stellar halo as a window on halo assembly in L$^{*}$ galaxies}",
      journal = {\mnras},
         year = 2017,
        month = jul,
       volume = {469},
       number = {1},
        pages = {L48-L52},
          doi = {10.1093/mnrasl/slx044},
archivePrefix = {arXiv},
       eprint = {1701.02741},
 primaryClass = {astro-ph.GA},
       adsurl = {https://ui.adsabs.harvard.edu/abs/2017MNRAS.469L..48A}
}

@article{Aihara.etal.18,
  author       = {H. Aihara and the HSC Collaboration},
  title        = {The Hyper Suprime-Cam SSP Survey: Overview and Survey Design},
  journal      = {Publications of the Astronomical Society of Japan},
  year         = {2018},
  volume       = {70},
  pages        = {S4},
  doi          = {10.1093/pasj/psx066},
  url          = {https://arxiv.org/abs/1704.05858}
}

@ARTICLE{Abadi.etal.03,
       author = {{Abadi}, Mario G. and {Navarro}, Julio F. and {Steinmetz}, Matthias and {Eke}, Vincent R.},
        title = "{Simulations of Galaxy Formation in a {\ensuremath{\Lambda}} Cold Dark Matter Universe. I. Dynamical and Photometric Properties of a Simulated Disk Galaxy}",
      journal = {\apj},
         year = 2003,
        month = jul,
       volume = {591},
       number = {2},
        pages = {499-514},
          doi = {10.1086/375512},
archivePrefix = {arXiv},
       eprint = {astro-ph/0211331},
 primaryClass = {astro-ph},
       adsurl = {https://ui.adsabs.harvard.edu/abs/2003ApJ...591..499A}
}

@ARTICLE{Akins.etal.21,
       author = {{Akins}, Hollis B. and {Christensen}, Charlotte R. and {Brooks}, Alyson M. and {Munshi}, Ferah and {Applebaum}, Elaad and {Engelhardt}, Anna and {Chamberland}, Lucas},
        title = "{Quenching Timescales of Dwarf Satellites around Milky Way-mass Hosts}",
      journal = {\apj},
         year = 2021,
        month = mar,
       volume = {909},
       number = {2},
          eid = {139},
        pages = {139},
          doi = {10.3847/1538-4357/abe2ab},
archivePrefix = {arXiv},
       eprint = {2008.02805},
 primaryClass = {astro-ph.GA},
       adsurl = {https://ui.adsabs.harvard.edu/abs/2021ApJ...909..139A}
}

@ARTICLE{Allen.etal.19,
       author = {{Allen}, Magdelena and {Behroozi}, Peter and {Ma}, Chung-Pei},
        title = "{Constraining scatter in the stellar mass-halo mass relation for haloes less massive than the Milky Way}",
      journal = {\mnras},
         year = 2019,
        month = oct,
       volume = {488},
       number = {4},
        pages = {4916-4925},
          doi = {10.1093/mnras/stz2067},
archivePrefix = {arXiv},
       eprint = {1812.05733},
 primaryClass = {astro-ph.GA},
       adsurl = {https://ui.adsabs.harvard.edu/abs/2019MNRAS.488.4916A}
}

@ARTICLE{Asali.etal.25,
       author = {{Asali}, Yasmeen and {Geha}, Marla and {Kado-Fong}, Erin and {Mao}, Yao-Yuan and {Wechsler}, Risa H. and {de los Reyes}, Mithi A.~C. and {Pasha}, Imad and {Kallivayalil}, Nitya and {Nadler}, Ethan O. and {Tollerud}, Erik J. and {Wang}, Yunchong and {Weiner}, Benjamin and {Wu}, John F.},
        title = "{The SAGA Survey. VI. The Size-Mass Relation for Low-Mass Galaxies Across Environments}",
      journal = {arXiv e-prints},
         year = 2025,
        month = sep,
          eid = {arXiv:2509.25335},
        pages = {arXiv:2509.25335},
          doi = {10.48550/arXiv.2509.25335},
archivePrefix = {arXiv},
       eprint = {2509.25335},
 primaryClass = {astro-ph.GA},
       adsurl = {https://ui.adsabs.harvard.edu/abs/2025arXiv250925335A}
}

@ARTICLE{Balberg.etal.02,
       author = {{Balberg}, Shmuel and {Shapiro}, Stuart L. and {Inagaki}, Shogo},
        title = "{Self-Interacting Dark Matter Halos and the Gravothermal Catastrophe}",
      journal = {\apj},
         year = 2002,
        month = apr,
       volume = {568},
       number = {2},
        pages = {475-487},
          doi = {10.1086/339038},
archivePrefix = {arXiv},
       eprint = {astro-ph/0110561},
 primaryClass = {astro-ph},
       adsurl = {https://ui.adsabs.harvard.edu/abs/2002ApJ...568..475B}
}

@ARTICLE{Behroozi.etal.19,
       author = {{Behroozi}, Peter and {Wechsler}, Risa H. and {Hearin}, Andrew P. and {Conroy}, Charlie},
        title = "{UNIVERSEMACHINE: The correlation between galaxy growth and dark matter halo assembly from z = 0-10}",
      journal = {\mnras},
         year = 2019,
        month = sep,
       volume = {488},
       number = {3},
        pages = {3143-3194},
          doi = {10.1093/mnras/stz1182},
archivePrefix = {arXiv},
       eprint = {1806.07893},
 primaryClass = {astro-ph.GA},
       adsurl = {https://ui.adsabs.harvard.edu/abs/2019MNRAS.488.3143B}
}

@ARTICLE{Belokurov.etal.18,
       author = {{Belokurov}, V. and {Erkal}, D. and {Evans}, N.~W. and {Koposov}, S.~E. and {Deason}, A.~J.},
        title = "{Co-formation of the disc and the stellar halo}",
      journal = {\mnras},
         year = 2018,
        month = jul,
       volume = {478},
       number = {1},
        pages = {611-619},
          doi = {10.1093/mnras/sty982},
archivePrefix = {arXiv},
       eprint = {1802.03414},
 primaryClass = {astro-ph.GA},
       adsurl = {https://ui.adsabs.harvard.edu/abs/2018MNRAS.478..611B}
}

@ARTICLE{Bell.etal.08,
       author = {{Bell}, Eric F. and {Zucker}, Daniel B. and {Belokurov}, Vasily and {Sharma}, Sanjib and {Johnston}, Kathryn V. and {Bullock}, James S. and {Hogg}, David W. and {Jahnke}, Knud and {de Jong}, Jelte T.~A. and {Beers}, Timothy C. and {Evans}, N.~W. and {Grebel}, Eva K. and {Ivezi{\'c}}, {\v{Z}}eljko and {Koposov}, Sergey E. and {Rix}, Hans-Walter and {Schneider}, Donald P. and {Steinmetz}, Matthias and {Zolotov}, Adi},
        title = "{The Accretion Origin of the Milky Way's Stellar Halo}",
      journal = {\apj},
         year = 2008,
        month = jun,
       volume = {680},
       number = {1},
        pages = {295-311},
          doi = {10.1086/588032},
archivePrefix = {arXiv},
       eprint = {0706.0004},
 primaryClass = {astro-ph},
       adsurl = {https://ui.adsabs.harvard.edu/abs/2008ApJ...680..295B}
}

@BOOK{Binney.Tremaine.08,
   author = {{Binney}, J. and {Tremaine}, S.},
    title = "{Galactic Dynamics: Second Edition}",
booktitle = {Galactic Dynamics: Second Edition, by James Binney and Scott Tremaine.~ISBN 978-0-691-13026-2 (HB).~Published by Princeton University Press, Princeton, NJ USA, 2008.},
     year = 2008,
publisher = {Princeton University Press},
   adsurl = {http://adsabs.harvard.edu/abs/2008gady.book.....B}
}

@ARTICLE{Blumenthal.etal.86,
       author = {{Blumenthal}, G.~R. and {Faber}, S.~M. and {Flores}, R. and {Primack}, J.~R.},
        title = "{Contraction of Dark Matter Galactic Halos Due to Baryonic Infall}",
      journal = {\apj},
         year = 1986,
        month = feb,
       volume = {301},
        pages = {27},
          doi = {10.1086/163867},
       adsurl = {https://ui.adsabs.harvard.edu/abs/1986ApJ...301...27B}
}

@ARTICLE{Bonaca.etal.25,
       author = {{Bonaca}, Ana and {Price-Whelan}, Adrian M.},
        title = "{Stellar streams in the Gaia era}",
      journal = {\nar},
         year = 2025,
        month = jun,
       volume = {100},
          eid = {101713},
        pages = {101713},
          doi = {10.1016/j.newar.2024.101713},
archivePrefix = {arXiv},
       eprint = {2405.19410},
 primaryClass = {astro-ph.GA},
       adsurl = {https://ui.adsabs.harvard.edu/abs/2025NewAR.10001713B}
}

@ARTICLE{Breiman.01,
       author = {{Breiman}, Leo},
        title = "{Random Forests.}",
      journal = {Machine Learning},
         year = 2001,
        month = jan,
       volume = {45},
        pages = {5-32},
          doi = {10.1023/A:1010933404324},
       adsurl = {https://ui.adsabs.harvard.edu/abs/2001MachL..45....5B}
}

@book{Breiman.84,
  author    = {Breiman, Leo and Friedman, Jerome and Olshen, Richard A. and Stone, Charles J.},
  title     = {Classification and Regression Trees},
  year      = {1984},
  edition   = {1st},
  publisher = {Chapman and Hall/CRC},
  doi       = {10.1201/9781315139470}
}

@ARTICLE{Bullock.etal.01,
   author = {{Bullock}, J.~S. and {Kolatt}, T.~S. and {Sigad}, Y. and {Somerville}, R.~S. and 
	{Kravtsov}, A.~V. and {Klypin}, A.~A. and {Primack}, J.~R. and 
	{Dekel}, A.},
    title = "{Profiles of dark haloes: evolution, scatter and environment}",
  journal = {\mnras},
   eprint = {astro-ph/9908159},
     year = 2001,
    month = mar,
   volume = 321,
    pages = {559-575},
      doi = {10.1046/j.1365-8711.2001.04068.x},
   adsurl = {http://adsabs.harvard.edu/abs/2001MNRAS.321..559B}
}

@ARTICLE{Bullock.Johnston.05,
       author = {{Bullock}, James S. and {Johnston}, Kathryn V.},
        title = "{Tracing Galaxy Formation with Stellar Halos. I. Methods}",
      journal = {\apj},
         year = 2005,
        month = dec,
       volume = {635},
       number = {2},
        pages = {931-949},
          doi = {10.1086/497422},
archivePrefix = {arXiv},
       eprint = {astro-ph/0506467},
 primaryClass = {astro-ph},
       adsurl = {https://ui.adsabs.harvard.edu/abs/2005ApJ...635..931B}
}

@ARTICLE{Carlsten.etal.22,
       author = {{Carlsten}, Scott G. and {Greene}, Jenny E. and {Beaton}, Rachael L. and {Danieli}, Shany and {Greco}, Johnny P.},
        title = "{The Exploration of Local VolumE Satellites (ELVES) Survey: A Nearly Volume-limited Sample of Nearby Dwarf Satellite Systems}",
      journal = {\apj},
         year = 2022,
        month = jul,
       volume = {933},
       number = {1},
          eid = {47},
        pages = {47},
          doi = {10.3847/1538-4357/ac6fd7},
archivePrefix = {arXiv},
       eprint = {2203.00014},
 primaryClass = {astro-ph.GA},
       adsurl = {https://ui.adsabs.harvard.edu/abs/2022ApJ...933...47C}
}

@ARTICLE{Celiz.etal.25,
       author = {{Celiz}, Bruno M. and {Navarro}, Julio F. and {Abadi}, Mario G.},
        title = "{Accreted stars and stellar haloes of simulated galaxies in TNG50}",
      journal = {arXiv e-prints},
         year = 2025,
        month = oct,
          eid = {arXiv:2510.18971},
        pages = {arXiv:2510.18971},
          doi = {10.48550/arXiv.2510.18971},
archivePrefix = {arXiv},
       eprint = {2510.18971},
 primaryClass = {astro-ph.GA},
       adsurl = {https://ui.adsabs.harvard.edu/abs/2025arXiv251018971C}
}

@ARTICLE{Chabrier.03,
       author = {{Chabrier}, Gilles},
        title = "{Galactic Stellar and Substellar Initial Mass Function}",
      journal = {\pasp},
         year = 2003,
        month = jul,
       volume = {115},
       number = {809},
        pages = {763-795},
          doi = {10.1086/376392},
archivePrefix = {arXiv},
       eprint = {astro-ph/0304382},
 primaryClass = {astro-ph},
       adsurl = {https://ui.adsabs.harvard.edu/abs/2003PASP..115..763C}
}

@ARTICLE{Chandrasekhar.43,
   author = {{Chandrasekhar}, S.},
    title = "{Dynamical Friction. I. General Considerations: the Coefficient of Dynamical Friction.}",
  journal = {\apj},
     year = 1943,
    month = mar,
   volume = 97,
    pages = {255},
      doi = {10.1086/144517},
   adsurl = {http://adsabs.harvard.edu/abs/1943ApJ....97..255C}
}

@ARTICLE{Chiang.etal.24,
       author = {{Chiang}, Barry T. and {van den Bosch}, Frank C. and {Schive}, Hsi-Yu},
        title = "{The tidal evolution of anisotropic subhaloes: A new pathway to creating isotropic and cored satellites}",
      journal = {arXiv e-prints},
         year = 2024,
        month = nov,
          eid = {arXiv:2411.03192},
        pages = {arXiv:2411.03192},
          doi = {10.48550/arXiv.2411.03192},
archivePrefix = {arXiv},
       eprint = {2411.03192},
 primaryClass = {astro-ph.GA},
       adsurl = {https://ui.adsabs.harvard.edu/abs/2024arXiv241103192C}
}

@ARTICLE{Christensen.etal.24,
       author = {{Christensen}, Charlotte R. and {Brooks}, Alyson M. and {Munshi}, Ferah and {Riggs}, Claire and {Van Nest}, Jordan and {Akins}, Hollis and {Quinn}, Thomas R. and {Chamberland}, Lucas},
        title = "{Environment Matters: Predicted Differences in the Stellar Mass{\textendash}Halo Mass Relation and History of Star Formation for Dwarf Galaxies}",
      journal = {\apj},
         year = 2024,
        month = feb,
       volume = {961},
       number = {2},
          eid = {236},
        pages = {236},
          doi = {10.3847/1538-4357/ad0c5a},
archivePrefix = {arXiv},
       eprint = {2311.04975},
 primaryClass = {astro-ph.GA},
       adsurl = {https://ui.adsabs.harvard.edu/abs/2024ApJ...961..236C}
}

@ARTICLE{Conroy.Gunn.10,
       author = {{Conroy}, Charlie and {Gunn}, James E.},
        title = "{The Propagation of Uncertainties in Stellar Population Synthesis Modeling. III. Model Calibration, Comparison, and Evaluation}",
      journal = {\apj},
         year = 2010,
        month = apr,
       volume = {712},
       number = {2},
        pages = {833-857},
          doi = {10.1088/0004-637X/712/2/833},
archivePrefix = {arXiv},
       eprint = {0911.3151},
 primaryClass = {astro-ph.CO},
       adsurl = {https://ui.adsabs.harvard.edu/abs/2010ApJ...712..833C}
}

@ARTICLE{Cooper.etal.10,
       author = {{Cooper}, A.~P. and {Cole}, S. and {Frenk}, C.~S. and {White}, S.~D.~M. and {Helly}, J. and {Benson}, A.~J. and {De Lucia}, G. and {Helmi}, A. and {Jenkins}, A. and {Navarro}, J.~F. and {Springel}, V. and {Wang}, J.},
        title = "{Galactic stellar haloes in the CDM model}",
      journal = {\mnras},
         year = 2010,
        month = aug,
       volume = {406},
       number = {2},
        pages = {744-766},
          doi = {10.1111/j.1365-2966.2010.16740.x},
archivePrefix = {arXiv},
       eprint = {0910.3211},
 primaryClass = {astro-ph.GA},
       adsurl = {https://ui.adsabs.harvard.edu/abs/2010MNRAS.406..744C}
}

@ARTICLE{Cooper.etal.13,
       author = {{Cooper}, Andrew P. and {D'Souza}, Richard and {Kauffmann}, Guinevere and {Wang}, Jing and {Boylan-Kolchin}, Michael and {Guo}, Qi and {Frenk}, Carlos S. and {White}, Simon D.~M.},
        title = "{Galactic accretion and the outer structure of galaxies in the CDM model}",
      journal = {\mnras},
         year = 2013,
        month = oct,
       volume = {434},
       number = {4},
        pages = {3348-3367},
          doi = {10.1093/mnras/stt1245},
archivePrefix = {arXiv},
       eprint = {1303.6283},
 primaryClass = {astro-ph.CO},
       adsurl = {https://ui.adsabs.harvard.edu/abs/2013MNRAS.434.3348C}
}

@ARTICLE{Cooper.etal.15,
       author = {{Cooper}, Andrew P. and {Parry}, Owen H. and {Lowing}, Ben and {Cole}, Shaun and {Frenk}, Carlos},
        title = "{Formation of in situ stellar haloes in Milky Way-mass galaxies}",
      journal = {\mnras},
         year = 2015,
        month = dec,
       volume = {454},
       number = {3},
        pages = {3185-3199},
          doi = {10.1093/mnras/stv2057},
archivePrefix = {arXiv},
       eprint = {1501.04630},
 primaryClass = {astro-ph.GA},
       adsurl = {https://ui.adsabs.harvard.edu/abs/2015MNRAS.454.3185C}
}

@ARTICLE{Cooper.etal.25,
       author = {{Cooper}, Andrew P. and {Frenk}, Carlos S. and {Hellwing}, Wojciech A. and {Bose}, Sownak},
        title = "{Simulations of the accreted stellar halos of low-mass field galaxies}",
      journal = {arXiv e-prints},
         year = 2025,
        month = jan,
          eid = {arXiv:2501.13317},
        pages = {arXiv:2501.13317},
          doi = {10.48550/arXiv.2501.13317},
archivePrefix = {arXiv},
       eprint = {2501.13317},
 primaryClass = {astro-ph.GA},
       adsurl = {https://ui.adsabs.harvard.edu/abs/2025arXiv250113317C}
}

@ARTICLE{Correa.etal.20,
       author = {{Correa}, Camila A. and {Schaye}, Joop},
        title = "{The dependence of the galaxy stellar-to-halo mass relation on galaxy morphology}",
      journal = {\mnras},
         year = 2020,
        month = dec,
       volume = {499},
       number = {3},
        pages = {3578-3593},
          doi = {10.1093/mnras/staa3053},
archivePrefix = {arXiv},
       eprint = {2010.01186},
 primaryClass = {astro-ph.GA},
       adsurl = {https://ui.adsabs.harvard.edu/abs/2020MNRAS.499.3578C}
}

@ARTICLE{Danieli.etal.23,
       author = {{Danieli}, Shany and {Greene}, Jenny E. and {Carlsten}, Scott and {Jiang}, Fangzhou and {Beaton}, Rachael and {Goulding}, Andy D.},
        title = "{ELVES. IV. The Satellite Stellar-to-halo Mass Relation Beyond the Milky Way}",
      journal = {\apj},
         year = 2023,
        month = oct,
       volume = {956},
       number = {1},
          eid = {6},
        pages = {6},
          doi = {10.3847/1538-4357/acefbd},
archivePrefix = {arXiv},
       eprint = {2210.14233},
 primaryClass = {astro-ph.GA},
       adsurl = {https://ui.adsabs.harvard.edu/abs/2023ApJ...956....6D}
}

@ARTICLE{Davison.etal.20,
       author = {{Davison}, Thomas A. and {Norris}, Mark A. and {Pfeffer}, Joel L. and {Davies}, Jonathan J. and {Crain}, Robert A.},
        title = "{An EAGLE's view of ex situ galaxy growth}",
      journal = {\mnras},
         year = 2020,
        month = sep,
       volume = {497},
       number = {1},
        pages = {81-93},
          doi = {10.1093/mnras/staa1816},
archivePrefix = {arXiv},
       eprint = {2006.08590},
 primaryClass = {astro-ph.GA},
       adsurl = {https://ui.adsabs.harvard.edu/abs/2020MNRAS.497...81D}
}

@ARTICLE{Deason.etal.15,
       author = {{Deason}, A.~J. and {Belokurov}, V. and {Weisz}, D.~R.},
        title = "{The progenitors of the Milky Way stellar halo: big bricks favoured over little bricks.}",
      journal = {\mnras},
         year = 2015,
        month = mar,
       volume = {448},
        pages = {L77-L81},
          doi = {10.1093/mnrasl/slv001},
archivePrefix = {arXiv},
       eprint = {1501.02806},
 primaryClass = {astro-ph.GA},
       adsurl = {https://ui.adsabs.harvard.edu/abs/2015MNRAS.448L..77D}
}

@ARTICLE{Deason.etal.16,
       author = {{Deason}, Alis J. and {Mao}, Yao-Yuan and {Wechsler}, Risa H.},
        title = "{The Eating Habits of Milky Way-mass Halos: Destroyed Dwarf Satellites and the Metallicity Distribution of Accreted Stars}",
      journal = {\apj},
         year = 2016,
        month = apr,
       volume = {821},
       number = {1},
          eid = {5},
        pages = {5},
          doi = {10.3847/0004-637X/821/1/5},
archivePrefix = {arXiv},
       eprint = {1601.07905},
 primaryClass = {astro-ph.GA},
       adsurl = {https://ui.adsabs.harvard.edu/abs/2016ApJ...821....5D}
}

@ARTICLE{Deason.etal.19,
       author = {{Deason}, Alis J. and {Belokurov}, Vasily and {Sanders}, Jason L.},
        title = "{The total stellar halo mass of the Milky Way}",
      journal = {\mnras},
         year = 2019,
        month = dec,
       volume = {490},
       number = {3},
        pages = {3426-3439},
          doi = {10.1093/mnras/stz2793},
archivePrefix = {arXiv},
       eprint = {1908.02763},
 primaryClass = {astro-ph.GA},
       adsurl = {https://ui.adsabs.harvard.edu/abs/2019MNRAS.490.3426D}
}

@ARTICLE{Deason.etal.24,
       author = {{Deason}, Alis J. and {Belokurov}, Vasily},
        title = "{Galactic Archaeology with Gaia}",
      journal = {\nar},
         year = 2024,
        month = dec,
       volume = {99},
          eid = {101706},
        pages = {101706},
          doi = {10.1016/j.newar.2024.101706},
archivePrefix = {arXiv},
       eprint = {2402.12443},
 primaryClass = {astro-ph.GA},
       adsurl = {https://ui.adsabs.harvard.edu/abs/2024NewAR..9901706D}
}

@ARTICLE{Diemand.etal.07a,
   author = {{Diemand}, J. and {Kuhlen}, M. and {Madau}, P.},
    title = "{Formation and Evolution of Galaxy Dark Matter Halos and Their Substructure}",
  journal = {\apj},
   eprint = {astro-ph/0703337},
     year = 2007,
    month = oct,
   volume = 667,
    pages = {859-877},
      doi = {10.1086/520573},
   adsurl = {http://adsabs.harvard.edu/abs/2007ApJ...667..859D}
}

@ARTICLE{Diemer.etal.24,
       author = {{Diemer}, Benedikt and {Behroozi}, Peter and {Mansfield}, Philip},
        title = "{Haunted haloes: tracking the ghosts of subhaloes lost by halo finders}",
      journal = {\mnras},
         year = 2024,
        month = oct,
       volume = {533},
       number = {4},
        pages = {3811-3827},
          doi = {10.1093/mnras/stae2007},
archivePrefix = {arXiv},
       eprint = {2305.00993},
 primaryClass = {astro-ph.CO},
       adsurl = {https://ui.adsabs.harvard.edu/abs/2024MNRAS.533.3811D}
}

@ARTICLE{Dropulic.etal.25,
       author = {{Dropulic}, Adriana and {Shipp}, Nora and {Kim}, Stacy and {Mezghanni}, Zeineb and {Necib}, Lina and {Lisanti}, Mariangela},
        title = "{StreamGen: Connecting Populations of Streams and Shells to Their Host Galaxies}",
      journal = {\apj},
         year = 2025,
        month = sep,
       volume = {990},
       number = {2},
          eid = {162},
        pages = {162},
          doi = {10.3847/1538-4357/adf1a7},
archivePrefix = {arXiv},
       eprint = {2409.13810},
 primaryClass = {astro-ph.GA},
       adsurl = {https://ui.adsabs.harvard.edu/abs/2025ApJ...990..162D}
}

@ARTICLE{Dacunha.etal.25,
       author = {{Dacunha}, Tara and {Mansfield}, Phil and {Wechsler}, Risa H.},
        title = "{Memoirs of Mass Accretion: Probing the Edges of Intracluster Light in Simulated Galaxy Clusters}",
      journal = {\apj},
         year = 2025,
        month = dec,
       volume = {994},
       number = {2},
          eid = {274},
        pages = {274},
          doi = {10.3847/1538-4357/ae1031},
archivePrefix = {arXiv},
       eprint = {2508.02837},
 primaryClass = {astro-ph.GA},
       adsurl = {https://ui.adsabs.harvard.edu/abs/2025ApJ...994..274D}
}

@ARTICLE{Elias.etal.18,
       author = {{Elias}, Lydia M. and {Sales}, Laura V. and {Creasey}, Peter and {Cooper}, Michael C. and {Bullock}, James S. and {Rich}, R. Michael and {Hernquist}, Lars},
        title = "{Stellar halos in Illustris: probing the histories of Milky Way-mass galaxies}",
      journal = {\mnras},
         year = 2018,
        month = sep,
       volume = {479},
       number = {3},
        pages = {4004-4016},
          doi = {10.1093/mnras/sty1718},
archivePrefix = {arXiv},
       eprint = {1801.07273},
 primaryClass = {astro-ph.GA},
       adsurl = {https://ui.adsabs.harvard.edu/abs/2018MNRAS.479.4004E}
}

@ARTICLE{Engler.etal.21a,
       author = {{Engler}, Christoph and {Pillepich}, Annalisa and {Joshi}, Gandhali D. and {Nelson}, Dylan and {Pasquali}, Anna and {Grebel}, Eva K. and {Lisker}, Thorsten and {Zinger}, Elad and {Donnari}, Martina and {Marinacci}, Federico and {Vogelsberger}, Mark and {Hernquist}, Lars},
        title = "{The distinct stellar-to-halo mass relations of satellite and central galaxies: insights from the IllustrisTNG simulations}",
      journal = {\mnras},
         year = 2021,
        month = jan,
       volume = {500},
       number = {3},
        pages = {3957-3975},
          doi = {10.1093/mnras/staa3505},
archivePrefix = {arXiv},
       eprint = {2002.11119},
 primaryClass = {astro-ph.GA},
       adsurl = {https://ui.adsabs.harvard.edu/abs/2021MNRAS.500.3957E}
}

@ARTICLE{Errani.etal.18,
       author = {{Errani}, Rapha{\"e}l and {Pe{\~n}arrubia}, Jorge and {Walker}, Matthew G.},
        title = "{Systematics in virial mass estimators for pressure-supported systems}",
      journal = {\mnras},
         year = 2018,
        month = dec,
       volume = {481},
       number = {4},
        pages = {5073-5090},
          doi = {10.1093/mnras/sty2505},
archivePrefix = {arXiv},
       eprint = {1805.00484},
 primaryClass = {astro-ph.GA},
       adsurl = {https://ui.adsabs.harvard.edu/abs/2018MNRAS.481.5073E}
}

@ARTICLE{Errani.Penarrubia.20,
       author = {{Errani}, Rapha{\"e}l and {Pe{\~n}arrubia}, Jorge},
        title = "{Can tides disrupt cold dark matter subhaloes?}",
      journal = {\mnras},
         year = 2020,
        month = feb,
       volume = {491},
       number = {4},
        pages = {4591-4601},
          doi = {10.1093/mnras/stz3349},
archivePrefix = {arXiv},
       eprint = {1906.01642},
 primaryClass = {astro-ph.GA},
       adsurl = {https://ui.adsabs.harvard.edu/abs/2020MNRAS.491.4591E}
}

@ARTICLE{Fattahi.etal.19,
       author = {{Fattahi}, Azadeh and {Belokurov}, Vasily and {Deason}, Alis J. and {Frenk}, Carlos S. and {G{\'o}mez}, Facundo A. and {Grand}, Robert J.~J. and {Marinacci}, Federico and {Pakmor}, R{\"u}diger and {Springel}, Volker},
        title = "{The origin of galactic metal-rich stellar halo components with highly eccentric orbits}",
      journal = {\mnras},
         year = 2019,
        month = apr,
       volume = {484},
       number = {4},
        pages = {4471-4483},
          doi = {10.1093/mnras/stz159},
archivePrefix = {arXiv},
       eprint = {1810.07779},
 primaryClass = {astro-ph.GA},
       adsurl = {https://ui.adsabs.harvard.edu/abs/2019MNRAS.484.4471F}
}

@ARTICLE{Forourhar.etal.25,
       author = {{Forourhar Moreno}, Victor J. and {Fattahi}, Azadeh and {Deason}, Alis J. and {Henstridge}, Fergus and {Ben{\'\i}tez-Llambay}, Alejandro},
        title = "{The accreted stellar haloes of Milky Way-mass galaxies as a probe of the nature of the dark matter}",
      journal = {\mnras},
         year = 2025,
        month = aug,
          doi = {10.1093/mnras/staf1344},
archivePrefix = {arXiv},
       eprint = {2407.05899},
 primaryClass = {astro-ph.GA},
       adsurl = {https://ui.adsabs.harvard.edu/abs/2025MNRAS.tmp.1296F}
}

@ARTICLE{Font.etal.20,
       author = {{Font}, Andreea S. and {McCarthy}, Ian G. and {Poole-Mckenzie}, Robert and {Stafford}, Sam G. and {Brown}, Shaun T. and {Schaye}, Joop and {Crain}, Robert A. and {Theuns}, Tom and {Schaller}, Matthieu},
        title = "{The ARTEMIS simulations: stellar haloes of Milky Way-mass galaxies}",
      journal = {\mnras},
         year = 2020,
        month = oct,
       volume = {498},
       number = {2},
        pages = {1765-1785},
          doi = {10.1093/mnras/staa2463},
archivePrefix = {arXiv},
       eprint = {2004.01914},
 primaryClass = {astro-ph.GA},
       adsurl = {https://ui.adsabs.harvard.edu/abs/2020MNRAS.498.1765F}
}

@ARTICLE{Kado-Fong.etal.25,
       author = {{Kado-Fong}, Erin and {Mao}, Yao-Yuan and {Asali}, Yasmeen and {Geha}, Marla and {Wechsler}, Risa H. and {de los Reyes}, Mithi A.~C. and {Wang}, Yunchong and {Nadler}, Ethan O. and {Kallivayalil}, Nitya and {Tollerud}, Erik J. and {Weiner}, Benjamin},
        title = "{SAGAbg. III. Environmental Stellar Mass Functions, Self-quenching, and the Stellar-to-halo Mass Relation in the Dwarf Galaxy Regime}",
      journal = {\apj},
         year = 2025,
        month = dec,
       volume = {994},
       number = {2},
          eid = {231},
        pages = {231},
          doi = {10.3847/1538-4357/ae102d},
archivePrefix = {arXiv},
       eprint = {2509.20444},
 primaryClass = {astro-ph.GA},
       adsurl = {https://ui.adsabs.harvard.edu/abs/2025ApJ...994..231K}
}

@ARTICLE{Geen.etal.13,
       author = {{Geen}, Sam and {Slyz}, Adrianne and {Devriendt}, Julien},
        title = "{Satellite survival in highly resolved Milky Way class haloes}",
      journal = {\mnras},
         year = 2013,
        month = feb,
       volume = {429},
       number = {1},
        pages = {633-651},
          doi = {10.1093/mnras/sts364},
archivePrefix = {arXiv},
       eprint = {1204.3327},
 primaryClass = {astro-ph.CO},
       adsurl = {https://ui.adsabs.harvard.edu/abs/2013MNRAS.429..633G}
}

@ARTICLE{Geha.etal.17,
       author = {{Geha}, Marla and {Wechsler}, Risa H. and {Mao}, Yao-Yuan and {Tollerud}, Erik J. and {Weiner}, Benjamin and {Bernstein}, Rebecca and {Hoyle}, Ben and {Marchi}, Sebastian and {Marshall}, Phil J. and {Mu{\~n}oz}, Ricardo and {Lu}, Yu},
        title = "{The SAGA Survey. I. Satellite Galaxy Populations around Eight Milky Way Analogs}",
      journal = {\apj},
         year = 2017,
        month = sep,
       volume = {847},
       number = {1},
          eid = {4},
        pages = {4},
          doi = {10.3847/1538-4357/aa8626},
archivePrefix = {arXiv},
       eprint = {1705.06743},
 primaryClass = {astro-ph.GA},
       adsurl = {https://ui.adsabs.harvard.edu/abs/2017ApJ...847....4G}
}

@ARTICLE{Geha.etal.24,
       author = {{Geha}, Marla and {Mao}, Yao-Yuan and {Wechsler}, Risa H. and {Asali}, Yasmeen and {Kado-Fong}, Erin and {Kallivayalil}, Nitya and {Nadler}, Ethan O. and {Tollerud}, Erik J. and {Weiner}, Benjamin and {de los Reyes}, Mithi A.~C. and {Wang}, Yunchong and {Wu}, John F.},
        title = "{The SAGA Survey. IV. The Star Formation Properties of 101 Satellite Systems around Milky Way-mass Galaxies}",
      journal = {arXiv e-prints},
         year = 2024,
        month = apr,
          eid = {arXiv:2404.14499},
        pages = {arXiv:2404.14499},
archivePrefix = {arXiv},
       eprint = {2404.14499},
 primaryClass = {astro-ph.GA},
       adsurl = {https://ui.adsabs.harvard.edu/abs/2024arXiv240414499G}
}

@ARTICLE{Gilhuly.etal.22,
       author = {{Gilhuly}, Colleen and {Merritt}, Allison and {Abraham}, Roberto and {Danieli}, Shany and {Lokhorst}, Deborah and {Liu}, Qing and {van Dokkum}, Pieter and {Conroy}, Charlie and {Greco}, Johnny},
        title = "{Stellar Halos from the The Dragonfly Edge-on Galaxies Survey}",
      journal = {\apj},
         year = 2022,
        month = jun,
       volume = {932},
       number = {1},
          eid = {44},
        pages = {44},
          doi = {10.3847/1538-4357/ac6750},
archivePrefix = {arXiv},
       eprint = {2204.06596},
 primaryClass = {astro-ph.GA},
       adsurl = {https://ui.adsabs.harvard.edu/abs/2022ApJ...932...44G}
}

@ARTICLE{Green.vdBosch.19,
       author = {{Green}, Sheridan B. and {van den Bosch}, Frank C.},
        title = "{The tidal evolution of dark matter substructure - I. subhalo density profiles}",
      journal = {\mnras},
         year = 2019,
        month = dec,
       volume = {490},
       number = {2},
        pages = {2091-2101},
          doi = {10.1093/mnras/stz2767},
archivePrefix = {arXiv},
       eprint = {1908.08537},
 primaryClass = {astro-ph.GA},
       adsurl = {https://ui.adsabs.harvard.edu/abs/2019MNRAS.490.2091G}
}

@ARTICLE{Green.etal.21,
       author = {{Green}, Sheridan B. and {van den Bosch}, Frank C. and {Jiang}, Fangzhou},
        title = "{The tidal evolution of dark matter substructure - II. The impact of artificial disruption on subhalo mass functions and radial profiles}",
      journal = {\mnras},
         year = 2021,
        month = may,
       volume = {503},
       number = {3},
        pages = {4075-4091},
          doi = {10.1093/mnras/stab696},
archivePrefix = {arXiv},
       eprint = {2103.01227},
 primaryClass = {astro-ph.GA},
       adsurl = {https://ui.adsabs.harvard.edu/abs/2021MNRAS.503.4075G}
}

@ARTICLE{Grimozzi.etal.24,
       author = {{Grimozzi}, Salvador E. and {Font}, Andreea S. and {De Rossi}, Mar{\'\i}a Emilia},
        title = "{Differences in the properties of disrupted and surviving satellites of Milky-Way-mass galaxies in relation to their host accretion histories.}",
      journal = {\mnras},
         year = 2024,
        month = may,
       volume = {530},
       number = {1},
        pages = {95-116},
          doi = {10.1093/mnras/stae878},
archivePrefix = {arXiv},
       eprint = {2401.04182},
 primaryClass = {astro-ph.GA},
       adsurl = {https://ui.adsabs.harvard.edu/abs/2024MNRAS.530...95G}
}

@ARTICLE{Gunn.Gott.72,
       author = {{Gunn}, James E. and {Gott}, J. Richard, III},
        title = "{On the Infall of Matter Into Clusters of Galaxies and Some Effects on Their Evolution}",
      journal = {\apj},
         year = 1972,
        month = aug,
       volume = {176},
        pages = {1},
          doi = {10.1086/151605},
       adsurl = {https://ui.adsabs.harvard.edu/abs/1972ApJ...176....1G}
}

@INPROCEEDINGS{Guzman24,
       author = {{Guzm{\'a}n}, Rafael},
        title = "{ARRAKIHS: The New ESA F-Class Mission to Investigate the Nature of Dark Matter}",
    booktitle = {EAS2024, European Astronomical Society Annual Meeting},
         year = 2024,
        month = jul,
          eid = {1990},
        pages = {1990},
       adsurl = {https://ui.adsabs.harvard.edu/abs/2024eas..conf.1990G}
}

@ARTICLE{Harmsen.etal.17,
       author = {{Harmsen}, Benjamin and {Monachesi}, Antonela and {Bell}, Eric F. and {de Jong}, Roelof S. and {Bailin}, Jeremy and {Radburn-Smith}, David J. and {Holwerda}, Benne W.},
        title = "{Diverse stellar haloes in nearby Milky Way mass disc galaxies}",
      journal = {\mnras},
         year = 2017,
        month = apr,
       volume = {466},
       number = {2},
        pages = {1491-1512},
          doi = {10.1093/mnras/stw2992},
archivePrefix = {arXiv},
       eprint = {1611.05448},
 primaryClass = {astro-ph.GA},
       adsurl = {https://ui.adsabs.harvard.edu/abs/2017MNRAS.466.1491H}
}

@ARTICLE{Hearin.etal.15,
   author = {{Hearin}, A.~P. and {Watson}, D.~F. and {van den Bosch}, F.~C.},
    title = "{Beyond halo mass: galactic conformity as a smoking gun of central galaxy assembly bias}",
  journal = {\mnras},
archivePrefix = "arXiv",
   eprint = {1404.6524},
     year = 2015,
    month = sep,
   volume = 452,
    pages = {1958-1969},
      doi = {10.1093/mnras/stv1358},
   adsurl = {http://adsabs.harvard.edu/abs/2015MNRAS.452.1958H}
}

@ARTICLE{Helmi.etal.18,
       author = {{Helmi}, Amina and {Babusiaux}, Carine and {Koppelman}, Helmer H. and {Massari}, Davide and {Veljanoski}, Jovan and {Brown}, Anthony G.~A.},
        title = "{The merger that led to the formation of the Milky Way's inner stellar halo and thick disk}",
      journal = {\nat},
         year = 2018,
        month = oct,
       volume = {563},
       number = {7729},
        pages = {85-88},
          doi = {10.1038/s41586-018-0625-x},
archivePrefix = {arXiv},
       eprint = {1806.06038},
 primaryClass = {astro-ph.GA},
       adsurl = {https://ui.adsabs.harvard.edu/abs/2018Natur.563...85H}
}

@ARTICLE{Helmi.etal.20,
       author = {{Helmi}, Amina},
        title = "{Streams, Substructures, and the Early History of the Milky Way}",
      journal = {\araa},
         year = 2020,
        month = aug,
       volume = {58},
        pages = {205-256},
          doi = {10.1146/annurev-astro-032620-021917},
archivePrefix = {arXiv},
       eprint = {2002.04340},
 primaryClass = {astro-ph.GA},
       adsurl = {https://ui.adsabs.harvard.edu/abs/2020ARA&A..58..205H}
}

@ARTICLE{Jiang.vdBosch.14,
       author = {{Jiang}, Fangzhou and {van den Bosch}, Frank C.},
        title = "{Generating merger trees for dark matter haloes: a comparison of methods}",
      journal = {\mnras},
         year = 2014,
        month = may,
       volume = {440},
       number = {1},
        pages = {193-207},
          doi = {10.1093/mnras/stu280},
archivePrefix = {arXiv},
       eprint = {1311.5225},
 primaryClass = {astro-ph.CO},
       adsurl = {https://ui.adsabs.harvard.edu/abs/2014MNRAS.440..193J}
}

@ARTICLE{Jiang.vdBosch.17,
   author = {{Jiang}, F. and {van den Bosch}, F.~C.},
    title = "{Statistics of dark matter substructure - III. Halo-to-halo variance}",
  journal = {\mnras},
archivePrefix = "arXiv",
   eprint = {1610.02399},
     year = 2017,
    month = nov,
   volume = 472,
    pages = {657-674},
      doi = {10.1093/mnras/stx1979},
   adsurl = {http://adsabs.harvard.edu/abs/2017MNRAS.472..657J}
}

@ARTICLE{Jiang.etal.21,
       author = {{Jiang}, Fangzhou and {Dekel}, Avishai and {Freundlich}, Jonathan and {van den Bosch}, Frank C. and {Green}, Sheridan B. and {Hopkins}, Philip F. and {Benson}, Andrew and {Du}, Xiaolong},
        title = "{SatGen: a semi-analytical satellite galaxy generator - I. The model and its application to Local-Group satellite statistics}",
      journal = {\mnras},
         year = 2021,
        month = mar,
       volume = {502},
       number = {1},
        pages = {621-641},
          doi = {10.1093/mnras/staa4034},
archivePrefix = {arXiv},
       eprint = {2005.05974},
 primaryClass = {astro-ph.GA},
       adsurl = {https://ui.adsabs.harvard.edu/abs/2021MNRAS.502..621J}
}

@ARTICLE{Joshi.etal.24,
       author = {{Joshi}, Gandhali D. and {Pontzen}, Andrew and {Agertz}, Oscar and {Rey}, Martin P. and {Read}, Justin and {Renaud}, Florent},
        title = "{VINTERGATAN-GM: How do mergers affect the satellite populations of MW-like galaxies?}",
      journal = {\mnras},
         year = 2024,
        month = feb,
       volume = {528},
       number = {2},
        pages = {2346-2357},
          doi = {10.1093/mnras/stae129},
archivePrefix = {arXiv},
       eprint = {2307.02206},
 primaryClass = {astro-ph.GA},
       adsurl = {https://ui.adsabs.harvard.edu/abs/2024MNRAS.528.2346J}
}

@ARTICLE{Joshi.etal.25b,
       author = {{Joshi}, Gandhali D. and {Pontzen}, Andrew and {Agertz}, Oscar and {Read}, Justin and {Rey}, Martin P.},
        title = "{The PARADIGM project II: the lifetimes and quenching of satellites in Milky Way-mass haloes}",
      journal = {\mnras},
         year = 2025,
        month = dec,
       volume = {544},
       number = {3},
        pages = {2811-2834},
          doi = {10.1093/mnras/staf1871},
archivePrefix = {arXiv},
       eprint = {2507.05401},
 primaryClass = {astro-ph.GA},
       adsurl = {https://ui.adsabs.harvard.edu/abs/2025MNRAS.544.2811J}
}

@ARTICLE{King.62,
   author = {{King}, I.},
    title = "{The structure of star clusters. I. an empirical density law}",
  journal = {\aj},
     year = 1962,
    month = oct,
   volume = 67,
    pages = {471},
      doi = {10.1086/108756},
   adsurl = {http://adsabs.harvard.edu/abs/1962AJ.....67..471K}
}

@ARTICLE{Kong.etal.25,
       author = {{Kong}, Hyunsu and {Boylan-Kolchin}, Michael and {Bullock}, James S.},
        title = "{Bloodhound Unleashed: Particle-based Substructure Tracking for Cosmological Simulations}",
      journal = {arXiv e-prints},
         year = 2025,
        month = mar,
          eid = {arXiv:2503.10766},
        pages = {arXiv:2503.10766},
          doi = {10.48550/arXiv.2503.10766},
archivePrefix = {arXiv},
       eprint = {2503.10766},
 primaryClass = {astro-ph.GA},
       adsurl = {https://ui.adsabs.harvard.edu/abs/2025arXiv250310766K}
}

@ARTICLE{Khoperskov.etal.23,
       author = {{Khoperskov}, Sergey and {Minchev}, Ivan and {Libeskind}, Noam and {Haywood}, Misha and {Di Matteo}, Paola and {Belokurov}, Vasily and {Steinmetz}, Matthias and {Gomez}, Facundo A. and {Grand}, Robert J.~J. and {Hoffman}, Yehuda and {Knebe}, Alexander and {Sorce}, Jenny G. and {Spaare}, Martin and {Tempel}, Elmo and {Vogelsberger}, Mark},
        title = "{The stellar halo in Local Group Hestia simulations. II. The accreted component}",
      journal = {\aap},
         year = 2023,
        month = sep,
       volume = {677},
          eid = {A90},
        pages = {A90},
          doi = {10.1051/0004-6361/202244233},
archivePrefix = {arXiv},
       eprint = {2206.04522},
 primaryClass = {astro-ph.GA},
       adsurl = {https://ui.adsabs.harvard.edu/abs/2023A&A...677A..90K}
}

@ARTICLE{Koppelman.etal.18,
       author = {{Koppelman}, Helmer and {Helmi}, Amina and {Veljanoski}, Jovan},
        title = "{One Large Blob and Many Streams Frosting the nearby Stellar Halo in Gaia DR2}",
      journal = {\apjl},
         year = 2018,
        month = jun,
       volume = {860},
       number = {1},
          eid = {L11},
        pages = {L11},
          doi = {10.3847/2041-8213/aac882},
archivePrefix = {arXiv},
       eprint = {1804.11347},
 primaryClass = {astro-ph.GA},
       adsurl = {https://ui.adsabs.harvard.edu/abs/2018ApJ...860L..11K}
}

@ARTICLE{Li.etal.20,
       author = {{Li}, Zhao-Zhou and {Zhao}, Dong-Hai and {Jing}, Y.~P. and {Han}, Jiaxin and {Dong}, Fu-Yu},
        title = "{Orbital Distribution of Infalling Satellite Halos across Cosmic Time}",
      journal = {\apj},
         year = 2020,
        month = dec,
       volume = {905},
       number = {2},
          eid = {177},
        pages = {177},
          doi = {10.3847/1538-4357/abc481},
archivePrefix = {arXiv},
       eprint = {2008.05710},
 primaryClass = {astro-ph.CO},
       adsurl = {https://ui.adsabs.harvard.edu/abs/2020ApJ...905..177L}
}

@ARTICLE{Lu.etal.15,
       author = {{Lu}, Zhankui and {Mo}, H.~J. and {Lu}, Yu and {Katz}, Neal and {Weinberg}, Martin D. and {van den Bosch}, Frank C. and {Yang}, Xiaohu},
        title = "{Star formation and stellar mass assembly in dark matter haloes: from giants to dwarfs}",
      journal = {\mnras},
         year = 2015,
        month = jun,
       volume = {450},
       number = {2},
        pages = {1604-1617},
          doi = {10.1093/mnras/stv667},
archivePrefix = {arXiv},
       eprint = {1406.5068},
 primaryClass = {astro-ph.GA},
       adsurl = {https://ui.adsabs.harvard.edu/abs/2015MNRAS.450.1604L}
}

@ARTICLE{Maier.etal.19,
       author = {{Maier}, C. and {Ziegler}, B.~L. and {Haines}, C.~P. and {Smith}, G.~P.},
        title = "{Slow-then-rapid quenching as traced by tentative evidence for enhanced metallicities of cluster galaxies at z {\ensuremath{\sim}} 0.2 in the slow quenching phase}",
      journal = {\aap},
         year = 2019,
        month = jan,
       volume = {621},
          eid = {A131},
        pages = {A131},
          doi = {10.1051/0004-6361/201834290},
archivePrefix = {arXiv},
       eprint = {1809.07675},
 primaryClass = {astro-ph.GA},
       adsurl = {https://ui.adsabs.harvard.edu/abs/2019A&A...621A.131M}
}

@ARTICLE{Malhan.etal.22,
       author = {{Malhan}, Khyati and {Ibata}, Rodrigo A. and {Sharma}, Sanjib and {Famaey}, Benoit and {Bellazzini}, Michele and {Carlberg}, Raymond G. and {D'Souza}, Richard and {Yuan}, Zhen and {Martin}, Nicolas F. and {Thomas}, Guillaume F.},
        title = "{The Global Dynamical Atlas of the Milky Way Mergers: Constraints from Gaia EDR3-based Orbits of Globular Clusters, Stellar Streams, and Satellite Galaxies}",
      journal = {\apj},
         year = 2022,
        month = feb,
       volume = {926},
       number = {2},
          eid = {107},
        pages = {107},
          doi = {10.3847/1538-4357/ac4d2a},
archivePrefix = {arXiv},
       eprint = {2202.07660},
 primaryClass = {astro-ph.GA},
       adsurl = {https://ui.adsabs.harvard.edu/abs/2022ApJ...926..107M}
}

@ARTICLE{Mao.etal.21,
       author = {{Mao}, Yao-Yuan and {Geha}, Marla and {Wechsler}, Risa H. and {Weiner}, Benjamin and {Tollerud}, Erik J. and {Nadler}, Ethan O. and {Kallivayalil}, Nitya},
        title = "{The SAGA Survey. II. Building a Statistical Sample of Satellite Systems around Milky Way-like Galaxies}",
      journal = {\apj},
         year = 2021,
        month = feb,
       volume = {907},
       number = {2},
          eid = {85},
        pages = {85},
          doi = {10.3847/1538-4357/abce58},
archivePrefix = {arXiv},
       eprint = {2008.12783},
 primaryClass = {astro-ph.GA},
       adsurl = {https://ui.adsabs.harvard.edu/abs/2021ApJ...907...85M}
}

@ARTICLE{Mao.etal.24,
       author = {{Mao}, Yao-Yuan and {Geha}, Marla and {Wechsler}, Risa H. and {Asali}, Yasmeen and {Wang}, Yunchong and {Kado-Fong}, Erin and {Kallivayalil}, Nitya and {Nadler}, Ethan O. and {Tollerud}, Erik J. and {Weiner}, Benjamin and {de los Reyes}, Mithi A.~C. and {Wu}, John F.},
        title = "{The SAGA Survey. III. A Census of 101 Satellite Systems around Milky Way-mass Galaxies}",
      journal = {arXiv e-prints},
         year = 2024,
        month = apr,
          eid = {arXiv:2404.14498},
        pages = {arXiv:2404.14498},
archivePrefix = {arXiv},
       eprint = {2404.14498},
 primaryClass = {astro-ph.GA},
       adsurl = {https://ui.adsabs.harvard.edu/abs/2024arXiv240414498M}
}

@ARTICLE{Mansfield.etal.24,
       author = {{Mansfield}, Philip and {Darragh-Ford}, Elise and {Wang}, Yunchong and {Nadler}, Ethan O. and {Diemer}, Benedikt and {Wechsler}, Risa H.},
        title = "{SYMFIND : Addressing the Fragility of Subhalo Finders and Revealing the Durability of Subhalos}",
      journal = {\apj},
         year = 2024,
        month = aug,
       volume = {970},
       number = {2},
          eid = {178},
        pages = {178},
          doi = {10.3847/1538-4357/ad4e33},
archivePrefix = {arXiv},
       eprint = {2308.10926},
 primaryClass = {astro-ph.CO},
       adsurl = {https://ui.adsabs.harvard.edu/abs/2024ApJ...970..178M}
}

@ARTICLE{Merritt.etal.16,
       author = {{Merritt}, Allison and {van Dokkum}, Pieter and {Abraham}, Roberto and {Zhang}, Jielai},
        title = "{The Dragonfly nearby Galaxies Survey. I. Substantial Variation in the Diffuse Stellar Halos around Spiral Galaxies}",
      journal = {\apj},
         year = 2016,
        month = oct,
       volume = {830},
       number = {2},
          eid = {62},
        pages = {62},
          doi = {10.3847/0004-637X/830/2/62},
archivePrefix = {arXiv},
       eprint = {1606.08847},
 primaryClass = {astro-ph.GA},
       adsurl = {https://ui.adsabs.harvard.edu/abs/2016ApJ...830...62M}
}

@ARTICLE{Merritt.etal.20,
       author = {{Merritt}, Allison and {Pillepich}, Annalisa and {van Dokkum}, Pieter and {Nelson}, Dylan and {Hernquist}, Lars and {Marinacci}, Federico and {Vogelsberger}, Mark},
        title = "{A missing outskirts problem? Comparisons between stellar haloes in the Dragonfly Nearby Galaxies Survey and the TNG100 simulation}",
      journal = {\mnras},
         year = 2020,
        month = jul,
       volume = {495},
       number = {4},
        pages = {4570-4604},
          doi = {10.1093/mnras/staa1164},
archivePrefix = {arXiv},
       eprint = {2004.11402},
 primaryClass = {astro-ph.GA},
       adsurl = {https://ui.adsabs.harvard.edu/abs/2020MNRAS.495.4570M}
}

@ARTICLE{Monachesi.etal.19,
       author = {{Monachesi}, Antonela and {G{\'o}mez}, Facundo A. and {Grand}, Robert J.~J. and {Simpson}, Christine M. and {Kauffmann}, Guinevere and {Bustamante}, Sebasti{\'a}n and {Marinacci}, Federico and {Pakmor}, R{\"u}diger and {Springel}, Volker and {Frenk}, Carlos S. and {White}, Simon D.~M. and {Tissera}, Patricia B.},
        title = "{The Auriga stellar haloes: connecting stellar population properties with accretion and merging history}",
      journal = {\mnras},
         year = 2019,
        month = may,
       volume = {485},
       number = {2},
        pages = {2589-2616},
          doi = {10.1093/mnras/stz538},
archivePrefix = {arXiv},
       eprint = {1804.07798},
 primaryClass = {astro-ph.GA},
       adsurl = {https://ui.adsabs.harvard.edu/abs/2019MNRAS.485.2589M}
}

@ARTICLE{Monzon.etal.24,
       author = {{Monzon}, J. Sebastian and {van den Bosch}, Frank C. and {Mitra}, Kaustav},
        title = "{Constraining the Low-mass End of the Stellar-to-halo Mass Relation with Surveys of Satellite Galaxies}",
      journal = {\apj},
         year = 2024,
        month = dec,
       volume = {976},
       number = {2},
          eid = {197},
        pages = {197},
          doi = {10.3847/1538-4357/ad834e},
archivePrefix = {arXiv},
       eprint = {2410.02873},
 primaryClass = {astro-ph.GA},
       adsurl = {https://ui.adsabs.harvard.edu/abs/2024ApJ...976..197M}
}

@BOOK{MBW10,
   author = {{Mo}, H. and {van den Bosch}, F.~C. and {White}, S.},
    title = "{Galaxy Formation and Evolution}",
booktitle = {Galaxy Formation and Evolution, by Houjun Mo , Frank van den Bosch , Simon White, Cambridge, UK: Cambridge University Press, 2010},
     year = 2010,
publisher = {Cambridge University Press},
    month = may,
   adsurl = {http://adsabs.harvard.edu/abs/2010gfe..book.....M}
}

@ARTICLE{More.etal.09a,
   author = {{More}, S. and {van den Bosch}, F.~C. and {Cacciato}, M.},
    title = "{Satellite kinematics - I. A new method to constrain the halo mass-luminosity relation of central galaxies}",
  journal = {\mnras},
archivePrefix = "arXiv",
   eprint = {0807.4529},
     year = 2009,
    month = jan,
   volume = 392,
    pages = {917-924},
      doi = {10.1111/j.1365-2966.2008.14114.x},
   adsurl = {http://adsabs.harvard.edu/abs/2009MNRAS.392..917M}
}

@ARTICLE{Moster.etal.10,
       author = {{Moster}, Benjamin P. and {Somerville}, Rachel S. and {Maulbetsch}, Christian and {van den Bosch}, Frank C. and {Macci{\`o}}, Andrea V. and {Naab}, Thorsten and {Oser}, Ludwig},
        title = "{Constraints on the Relationship between Stellar Mass and Halo Mass at Low and High Redshift}",
      journal = {\apj},
         year = 2010,
        month = feb,
       volume = {710},
       number = {2},
        pages = {903-923},
          doi = {10.1088/0004-637X/710/2/903},
archivePrefix = {arXiv},
       eprint = {0903.4682},
 primaryClass = {astro-ph.CO},
       adsurl = {https://ui.adsabs.harvard.edu/abs/2010ApJ...710..903M}
}

@ARTICLE{Moster.etal.18,
       author = {{Moster}, Benjamin P. and {Naab}, Thorsten and {White}, Simon D.~M.},
        title = "{EMERGE - an empirical model for the formation of galaxies since z {\ensuremath{\sim}} 10}",
      journal = {\mnras},
         year = 2018,
        month = jun,
       volume = {477},
       number = {2},
        pages = {1822-1852},
          doi = {10.1093/mnras/sty655},
archivePrefix = {arXiv},
       eprint = {1705.05373},
 primaryClass = {astro-ph.GA},
       adsurl = {https://ui.adsabs.harvard.edu/abs/2018MNRAS.477.1822M}
}

@ARTICLE{Munshi.etal.21,
       author = {{Munshi}, Ferah and {Brooks}, Alyson M. and {Applebaum}, Elaad and {Christensen}, Charlotte R. and {Quinn}, T. and {Sligh}, Serena},
        title = "{Quantifying Scatter in Galaxy Formation at the Lowest Masses}",
      journal = {\apj},
         year = 2021,
        month = dec,
       volume = {923},
       number = {1},
          eid = {35},
        pages = {35},
          doi = {10.3847/1538-4357/ac0db6},
archivePrefix = {arXiv},
       eprint = {2101.05822},
 primaryClass = {astro-ph.GA},
       adsurl = {https://ui.adsabs.harvard.edu/abs/2021ApJ...923...35M}
}

@ARTICLE{Nadler.etal.20,
       author = {{Nadler}, E.~O. and {Wechsler}, R.~H. and {Bechtol}, K. and {Mao}, Y. -Y. and {Green}, G. and {Drlica-Wagner}, A. and {McNanna}, M. and {Mau}, S. and {Pace}, A.~B. and {Simon}, J.~D. and {Kravtsov}, A. and {Dodelson}, S. and {Li}, T.~S. and {Riley}, A.~H. and {Wang}, M.~Y. and {Abbott}, T.~M.~C. and {Aguena}, M. and {Allam}, S. and {Annis}, J. and {Avila}, S. and {Bernstein}, G.~M. and {Bertin}, E. and {Brooks}, D. and {Burke}, D.~L. and {Rosell}, A. Carnero and {Kind}, M. Carrasco and {Carretero}, J. and {Costanzi}, M. and {da Costa}, L.~N. and {De Vicente}, J. and {Desai}, S. and {Evrard}, A.~E. and {Flaugher}, B. and {Fosalba}, P. and {Frieman}, J. and {Garc{\'\i}a-Bellido}, J. and {Gaztanaga}, E. and {Gerdes}, D.~W. and {Gruen}, D. and {Gschwend}, J. and {Gutierrez}, G. and {Hartley}, W.~G. and {Hinton}, S.~R. and {Honscheid}, K. and {Krause}, E. and {Kuehn}, K. and {Kuropatkin}, N. and {Lahav}, O. and {Maia}, M.~A.~G. and {Marshall}, J.~L. and {Menanteau}, F. and {Miquel}, R. and {Palmese}, A. and {Paz-Chinch{\'o}n}, F. and {Plazas}, A.~A. and {Romer}, A.~K. and {Sanchez}, E. and {Santiago}, B. and {Scarpine}, V. and {Serrano}, S. and {Smith}, M. and {Soares-Santos}, M. and {Suchyta}, E. and {Tarle}, G. and {Thomas}, D. and {Varga}, T.~N. and {Walker}, A.~R. and {DES Collaboration}},
        title = "{Milky Way Satellite Census. II. Galaxy-Halo Connection Constraints Including the Impact of the Large Magellanic Cloud}",
      journal = {\apj},
         year = 2020,
        month = apr,
       volume = {893},
       number = {1},
          eid = {48},
        pages = {48},
          doi = {10.3847/1538-4357/ab846a},
archivePrefix = {arXiv},
       eprint = {1912.03303},
 primaryClass = {astro-ph.GA},
       adsurl = {https://ui.adsabs.harvard.edu/abs/2020ApJ...893...48N}
}

@ARTICLE{Naidu.etal.21,
       author = {{Naidu}, Rohan P. and {Conroy}, Charlie and {Bonaca}, Ana and {Zaritsky}, Dennis and {Weinberger}, Rainer and {Ting}, Yuan-Sen and {Caldwell}, Nelson and {Tacchella}, Sandro and {Han}, Jiwon Jesse and {Speagle}, Joshua S. and {Cargile}, Phillip A.},
        title = "{Reconstructing the Last Major Merger of the Milky Way with the H3 Survey}",
      journal = {\apj},
         year = 2021,
        month = dec,
       volume = {923},
       number = {1},
          eid = {92},
        pages = {92},
          doi = {10.3847/1538-4357/ac2d2d},
archivePrefix = {arXiv},
       eprint = {2103.03251},
 primaryClass = {astro-ph.GA},
       adsurl = {https://ui.adsabs.harvard.edu/abs/2021ApJ...923...92N}
}

@ARTICLE{Navarro.etal.97,
   author = {{Navarro}, J.~F. and {Frenk}, C.~S. and {White}, S.~D.~M.},
    title = "{A Universal Density Profile from Hierarchical Clustering}",
  journal = {\apj},
   eprint = {astro-ph/9611107},
     year = 1997,
    month = dec,
   volume = 490,
    pages = {493-508},
      doi = {10.1086/304888},
   adsurl = {http://adsabs.harvard.edu/abs/1997ApJ...490..493N}
}

@ARTICLE{Ogiya.etal.19,
       author = {{Ogiya}, Go and {van den Bosch}, Frank C. and {Hahn}, Oliver and {Green}, Sheridan B. and {Miller}, Tim B. and {Burkert}, Andreas},
        title = "{DASH: a library of dynamical subhalo evolution}",
      journal = {\mnras},
         year = 2019,
        month = may,
       volume = {485},
       number = {1},
        pages = {189-202},
          doi = {10.1093/mnras/stz375},
archivePrefix = {arXiv},
       eprint = {1901.08601},
 primaryClass = {astro-ph.GA},
       adsurl = {https://ui.adsabs.harvard.edu/abs/2019MNRAS.485..189O}
}

@ARTICLE{OLeary.etal.23,
       author = {{O'Leary}, Joseph A. and {Steinwandel}, Ulrich P. and {Moster}, Benjamin P. and {Martin}, Nicolas and {Naab}, Thorsten},
        title = "{Predictions on the stellar-to-halo mass relation in the dwarf regime using the empirical model for galaxy formation EMERGE}",
      journal = {\mnras},
         year = 2023,
        month = mar,
       volume = {520},
       number = {1},
        pages = {897-916},
          doi = {10.1093/mnras/stad166},
archivePrefix = {arXiv},
       eprint = {2301.07122},
 primaryClass = {astro-ph.GA},
       adsurl = {https://ui.adsabs.harvard.edu/abs/2023MNRAS.520..897O}
}

@ARTICLE{Orkney.etal.22,
       author = {{Orkney}, Matthew D.~A. and {Laporte}, Chervin F.~P. and {Grand}, Robert J.~J. and {G{\'o}mez}, Facundo A. and {van de Voort}, Freeke and {Marinacci}, Federico and {Fragkoudi}, Francesca and {Pakmor}, Ruediger and {Springel}, Volker},
        title = "{The impact of two massive early accretion events in a Milky Way-like galaxy: repercussions for the buildup of the stellar disc and halo}",
      journal = {\mnras},
         year = 2022,
        month = nov,
       volume = {517},
       number = {1},
        pages = {L138-L142},
          doi = {10.1093/mnrasl/slac126},
archivePrefix = {arXiv},
       eprint = {2206.09246},
 primaryClass = {astro-ph.GA},
       adsurl = {https://ui.adsabs.harvard.edu/abs/2022MNRAS.517L.138O}
}

@ARTICLE{Parkinson.etal.08,
       author = {{Parkinson}, Hannah and {Cole}, Shaun and {Helly}, John},
        title = "{Generating dark matter halo merger trees}",
      journal = {\mnras},
         year = 2008,
        month = jan,
       volume = {383},
       number = {2},
        pages = {557-564},
          doi = {10.1111/j.1365-2966.2007.12517.x},
archivePrefix = {arXiv},
       eprint = {0708.1382},
 primaryClass = {astro-ph},
       adsurl = {https://ui.adsabs.harvard.edu/abs/2008MNRAS.383..557P}
}

@article{Pedregosa.etal.11,
  title={Scikit-learn: Machine Learning in {P}ython},
  author={Pedregosa, F. and Varoquaux, G. and Gramfort, A. and Michel, V.
          and Thirion, B. and Grisel, O. and Blondel, M. and Prettenhofer, P.
          and Weiss, R. and Dubourg, V. and Vanderplas, J. and Passos, A. and
          Cournapeau, D. and Brucher, M. and Perrot, M. and Duchesnay, E.},
  journal={Journal of Machine Learning Research},
  volume={12},
  pages={2825--2830},
  year={2011}
}

@ARTICLE{Penarrubia.etal.06,
       author = {{Pe{\~n}arrubia}, Jorge and {McConnachie}, Alan and {Babul}, Arif},
        title = "{On the Formation of Extended Galactic Disks by Tidally Disrupted Dwarf Galaxies}",
      journal = {\apjl},
         year = 2006,
        month = oct,
       volume = {650},
       number = {1},
        pages = {L33-L36},
          doi = {10.1086/508656},
archivePrefix = {arXiv},
       eprint = {astro-ph/0606101},
 primaryClass = {astro-ph},
       adsurl = {https://ui.adsabs.harvard.edu/abs/2006ApJ...650L..33P}
}

@ARTICLE{Penarrubia.etal.08,
       author = {{Pe{\~n}arrubia}, Jorge and {McConnachie}, Alan W. and {Navarro}, Julio F.},
        title = "{The Cold Dark Matter Halos of Local Group Dwarf Spheroidals}",
      journal = {\apj},
         year = 2008,
        month = jan,
       volume = {672},
       number = {2},
        pages = {904-913},
          doi = {10.1086/521543},
archivePrefix = {arXiv},
       eprint = {astro-ph/0701780},
 primaryClass = {astro-ph},
       adsurl = {https://ui.adsabs.harvard.edu/abs/2008ApJ...672..904P}
}

@ARTICLE{Penarrubia.etal.10,
   author = {{Pe{\~n}arrubia}, J. and {Benson}, A.~J. and {Walker}, M.~G. and 
	{Gilmore}, G. and {McConnachie}, A.~W. and {Mayer}, L.},
    title = "{The impact of dark matter cusps and cores on the satellite galaxy population around spiral galaxies}",
  journal = {\mnras},
archivePrefix = "arXiv",
   eprint = {1002.3376},
     year = 2010,
    month = aug,
   volume = 406,
    pages = {1290-1305},
      doi = {10.1111/j.1365-2966.2010.16762.x},
   adsurl = {http://adsabs.harvard.edu/abs/2010MNRAS.406.1290P}
}

@ARTICLE{Planck.18,
       author = {{Planck Collaboration} and {Aghanim}, N. and {Akrami}, Y. and {Ashdown}, M. and {Aumont}, J. and {Baccigalupi}, C. and {Ballardini}, M. and {Banday}, A.~J. and {Barreiro}, R.~B. and {Bartolo}, N. and {Basak}, S. and {Battye}, R. and {Benabed}, K. and {Bernard}, J. -P. and {Bersanelli}, M. and {Bielewicz}, P. and {Bock}, J.~J. and {Bond}, J.~R. and {Borrill}, J. and {Bouchet}, F.~R. and {Boulanger}, F. and {Bucher}, M. and {Burigana}, C. and {Butler}, R.~C. and {Calabrese}, E. and {Cardoso}, J. -F. and {Carron}, J. and {Challinor}, A. and {Chiang}, H.~C. and {Chluba}, J. and {Colombo}, L.~P.~L. and {Combet}, C. and {Contreras}, D. and {Crill}, B.~P. and {Cuttaia}, F. and {de Bernardis}, P. and {de Zotti}, G. and {Delabrouille}, J. and {Delouis}, J. -M. and {Di Valentino}, E. and {Diego}, J.~M. and {Dor{\'e}}, O. and {Douspis}, M. and {Ducout}, A. and {Dupac}, X. and {Dusini}, S. and {Efstathiou}, G. and {Elsner}, F. and {En{\ss}lin}, T.~A. and {Eriksen}, H.~K. and {Fantaye}, Y. and {Farhang}, M. and {Fergusson}, J. and {Fernandez-Cobos}, R. and {Finelli}, F. and {Forastieri}, F. and {Frailis}, M. and {Fraisse}, A.~A. and {Franceschi}, E. and {Frolov}, A. and {Galeotta}, S. and {Galli}, S. and {Ganga}, K. and {G{\'e}nova-Santos}, R.~T. and {Gerbino}, M. and {Ghosh}, T. and {Gonz{\'a}lez-Nuevo}, J. and {G{\'o}rski}, K.~M. and {Gratton}, S. and {Gruppuso}, A. and {Gudmundsson}, J.~E. and {Hamann}, J. and {Handley}, W. and {Hansen}, F.~K. and {Herranz}, D. and {Hildebrandt}, S.~R. and {Hivon}, E. and {Huang}, Z. and {Jaffe}, A.~H. and {Jones}, W.~C. and {Karakci}, A. and {Keih{\"a}nen}, E. and {Keskitalo}, R. and {Kiiveri}, K. and {Kim}, J. and {Kisner}, T.~S. and {Knox}, L. and {Krachmalnicoff}, N. and {Kunz}, M. and {Kurki-Suonio}, H. and {Lagache}, G. and {Lamarre}, J. -M. and {Lasenby}, A. and {Lattanzi}, M. and {Lawrence}, C.~R. and {Le Jeune}, M. and {Lemos}, P. and {Lesgourgues}, J. and {Levrier}, F. and {Lewis}, A. and {Liguori}, M. and {Lilje}, P.~B. and {Lilley}, M. and {Lindholm}, V. and {L{\'o}pez-Caniego}, M. and {Lubin}, P.~M. and {Ma}, Y. -Z. and {Mac{\'\i}as-P{\'e}rez}, J.~F. and {Maggio}, G. and {Maino}, D. and {Mandolesi}, N. and {Mangilli}, A. and {Marcos-Caballero}, A. and {Maris}, M. and {Martin}, P.~G. and {Martinelli}, M. and {Mart{\'\i}nez-Gonz{\'a}lez}, E. and {Matarrese}, S. and {Mauri}, N. and {McEwen}, J.~D. and {Meinhold}, P.~R. and {Melchiorri}, A. and {Mennella}, A. and {Migliaccio}, M. and {Millea}, M. and {Mitra}, S. and {Miville-Desch{\^e}nes}, M. -A. and {Molinari}, D. and {Montier}, L. and {Morgante}, G. and {Moss}, A. and {Natoli}, P. and {N{\o}rgaard-Nielsen}, H.~U. and {Pagano}, L. and {Paoletti}, D. and {Partridge}, B. and {Patanchon}, G. and {Peiris}, H.~V. and {Perrotta}, F. and {Pettorino}, V. and {Piacentini}, F. and {Polastri}, L. and {Polenta}, G. and {Puget}, J. -L. and {Rachen}, J.~P. and {Reinecke}, M. and {Remazeilles}, M. and {Renzi}, A. and {Rocha}, G. and {Rosset}, C. and {Roudier}, G. and {Rubi{\~n}o-Mart{\'\i}n}, J.~A. and {Ruiz-Granados}, B. and {Salvati}, L. and {Sandri}, M. and {Savelainen}, M. and {Scott}, D. and {Shellard}, E.~P.~S. and {Sirignano}, C. and {Sirri}, G. and {Spencer}, L.~D. and {Sunyaev}, R. and {Suur-Uski}, A. -S. and {Tauber}, J.~A. and {Tavagnacco}, D. and {Tenti}, M. and {Toffolatti}, L. and {Tomasi}, M. and {Trombetti}, T. and {Valenziano}, L. and {Valiviita}, J. and {Van Tent}, B. and {Vibert}, L. and {Vielva}, P. and {Villa}, F. and {Vittorio}, N. and {Wandelt}, B.~D. and {Wehus}, I.~K. and {White}, M. and {White}, S.~D.~M. and {Zacchei}, A. and {Zonca}, A.},
        title = "{Planck 2018 results. VI. Cosmological parameters}",
      journal = {\aap},
         year = 2020,
        month = sep,
       volume = {641},
          eid = {A6},
        pages = {A6},
          doi = {10.1051/0004-6361/201833910},
archivePrefix = {arXiv},
       eprint = {1807.06209},
 primaryClass = {astro-ph.CO},
       adsurl = {https://ui.adsabs.harvard.edu/abs/2020A&A...641A...6P}
}

@ARTICLE{Pontzen.Governato.12,
       author = {{Pontzen}, Andrew and {Governato}, Fabio},
        title = "{How supernova feedback turns dark matter cusps into cores}",
      journal = {\mnras},
         year = 2012,
        month = apr,
       volume = {421},
       number = {4},
        pages = {3464-3471},
          doi = {10.1111/j.1365-2966.2012.20571.x},
archivePrefix = {arXiv},
       eprint = {1106.0499},
 primaryClass = {astro-ph.CO},
       adsurl = {https://ui.adsabs.harvard.edu/abs/2012MNRAS.421.3464P}
}

@ARTICLE{Pathak.etal.25,
       author = {{Pathak}, Debosmita and {Christensen}, Charlotte R. and {Brooks}, Alyson M. and {Munshi}, Ferah and {Wright}, Anna C. and {Carter}, Courtney},
        title = "{Survivors and Zombies: The Quenching and Disruption of Satellites Around Milky Way Analogs}",
      journal = {\apj},
         year = 2025,
        month = aug,
       volume = {989},
       number = {2},
          eid = {178},
        pages = {178},
          doi = {10.3847/1538-4357/adec94},
archivePrefix = {arXiv},
       eprint = {2505.22742},
 primaryClass = {astro-ph.GA},
       adsurl = {https://ui.adsabs.harvard.edu/abs/2025ApJ...989..178P}
}

@ARTICLE{Proctor.etal.24,
       author = {{Proctor}, Katy L. and {Ludlow}, Aaron D. and {Lagos}, Claudia del P. and {Robotham}, Aaron S.~G.},
        title = "{Milky Way-mass disc galaxies with low-mass stellar haloes have diverse merger histories}",
      journal = {arXiv e-prints},
         year = 2024,
        month = jul,
          eid = {arXiv:2407.11444},
        pages = {arXiv:2407.11444},
          doi = {10.48550/arXiv.2407.11444},
archivePrefix = {arXiv},
       eprint = {2407.11444},
 primaryClass = {astro-ph.GA},
       adsurl = {https://ui.adsabs.harvard.edu/abs/2024arXiv240711444P}
}

@ARTICLE{Pu.etal.25,
       author = {{Pu}, Sy-Yun and {Cooper}, Andrew P. and {Grand}, Robert J.~J. and {G{\'o}mez}, Facundo A. and {Monachesi}, Antonela},
        title = "{Progenitor Diversity in the Accreted Stellar Halos of Milky Way{\textendash}like Galaxies}",
      journal = {\apj},
         year = 2025,
        month = feb,
       volume = {980},
       number = {1},
          eid = {63},
        pages = {63},
          doi = {10.3847/1538-4357/ada382},
archivePrefix = {arXiv},
       eprint = {2410.13491},
 primaryClass = {astro-ph.GA},
       adsurl = {https://ui.adsabs.harvard.edu/abs/2025ApJ...980...63P}
}

@ARTICLE{Rey.and.Starkenburg.2022,
       author = {{Rey}, Martin P. and {Starkenburg}, Tjitske K.},
        title = "{How cosmological merger histories shape the diversity of stellar haloes}",
      journal = {\mnras},
         year = 2022,
        month = mar,
       volume = {510},
       number = {3},
        pages = {4208-4224},
          doi = {10.1093/mnras/stab3709},
archivePrefix = {arXiv},
       eprint = {2106.09729},
 primaryClass = {astro-ph.GA},
       adsurl = {https://ui.adsabs.harvard.edu/abs/2022MNRAS.510.4208R}
}

@ARTICLE{Riley.etal.24,
       author = {{Riley}, Alexander H. and {Shipp}, Nora and {Simpson}, Christine M. and {Bieri}, Rebekka and {Fattahi}, Azadeh and {Brown}, Shaun T. and {Oman}, Kyle A. and {Fragkoudi}, Francesca and {G{\'o}mez}, Facundo A. and {Grand}, Robert J.~J. and {Marinacci}, Federico},
        title = "{Auriga Streams I: disrupting satellites surrounding Milky Way-mass haloes at multiple resolutions}",
      journal = {arXiv e-prints},
         year = 2024,
        month = oct,
          eid = {arXiv:2410.09144},
        pages = {arXiv:2410.09144},
          doi = {10.48550/arXiv.2410.09144},
archivePrefix = {arXiv},
       eprint = {2410.09144},
 primaryClass = {astro-ph.GA},
       adsurl = {https://ui.adsabs.harvard.edu/abs/2024arXiv241009144R}
}

@ARTICLE{RodriguezGomez.etal.16,
       author = {{Rodriguez-Gomez}, Vicente and {Pillepich}, Annalisa and {Sales}, Laura V. and {Genel}, Shy and {Vogelsberger}, Mark and {Zhu}, Qirong and {Wellons}, Sarah and {Nelson}, Dylan and {Torrey}, Paul and {Springel}, Volker and {Ma}, Chung-Pei and {Hernquist}, Lars},
        title = "{The stellar mass assembly of galaxies in the Illustris simulation: growth by mergers and the spatial distribution of accreted stars}",
      journal = {\mnras},
         year = 2016,
        month = may,
       volume = {458},
       number = {3},
        pages = {2371-2390},
          doi = {10.1093/mnras/stw456},
archivePrefix = {arXiv},
       eprint = {1511.08804},
 primaryClass = {astro-ph.GA},
       adsurl = {https://ui.adsabs.harvard.edu/abs/2016MNRAS.458.2371R}
}

@ARTICLE{RodriguezPuebla.etal.17,
       author = {{Rodr{\'\i}guez-Puebla}, Aldo and {Primack}, Joel R. and {Avila-Reese}, Vladimir and {Faber}, S.~M.},
        title = "{Constraining the galaxy-halo connection over the last 13.3 Gyr: star formation histories, galaxy mergers and structural properties}",
      journal = {\mnras},
         year = 2017,
        month = sep,
       volume = {470},
       number = {1},
        pages = {651-687},
          doi = {10.1093/mnras/stx1172},
archivePrefix = {arXiv},
       eprint = {1703.04542},
 primaryClass = {astro-ph.GA},
       adsurl = {https://ui.adsabs.harvard.edu/abs/2017MNRAS.470..651R}
}

@ARTICLE{RodriguezCardoso.etal.25,
       author = {{Rodr{\'\i}guez-Cardoso}, Ram{\'o}n and {Roca-F{\`a}brega}, Santi and {Jung}, Minyong and {Nguyễn}, Thịnh H. and {Kim}, Ji-Hoon and {Primack}, Joel and {Agertz}, Oscar and {Barrow}, Kirk S.~S. and {Gallego}, Jesus and {Nagamine}, Kentaro and {Powell}, Johnny W. and {Revaz}, Yves and {Vel{\'a}zquez}, Hector and {Genina}, Anna and {Kim}, Hyeonyong and {Lupi}, Alessandro and {Abel}, Tom and {Cen}, Renyue and {Ceverino}, Daniel and {Dekel}, Avishai and {Oh}, Boon Kiat and {Quinn}, Thomas R. and {The Agora Collaboration}},
        title = "{The AGORA High-Resolution Galaxy Simulations Comparison Project: VII. Satellite quenching in zoom-in simulation of a Milky Way-mass halo}",
      journal = {\aap},
         year = 2025,
        month = jun,
       volume = {698},
          eid = {A303},
        pages = {A303},
          doi = {10.1051/0004-6361/202453639},
archivePrefix = {arXiv},
       eprint = {2505.05844},
 primaryClass = {astro-ph.GA},
       adsurl = {https://ui.adsabs.harvard.edu/abs/2025A&A...698A.303R}
}

@ARTICLE{Sales.etal.22,
       author = {{Sales}, Laura V. and {Wetzel}, Andrew and {Fattahi}, Azadeh},
        title = "{Baryonic solutions and challenges for cosmological models of dwarf galaxies}",
      journal = {Nature Astronomy},
         year = 2022,
        month = jun,
       volume = {6},
        pages = {897-910},
          doi = {10.1038/s41550-022-01689-w},
archivePrefix = {arXiv},
       eprint = {2206.05295},
 primaryClass = {astro-ph.GA},
       adsurl = {https://ui.adsabs.harvard.edu/abs/2022NatAs...6..897S}
}

@ARTICLE{Sanderson.etal.18,
       author = {{Sanderson}, Robyn E. and {Garrison-Kimmel}, Shea and {Wetzel}, Andrew and {Keung Chan}, Tsang and {Hopkins}, Philip F. and {Kere{\v{s}}}, Du{\v{s}}an and {Escala}, Ivanna and {Faucher-Gigu{\`e}re}, Claude-Andr{\'e} and {Ma}, Xiangcheng},
        title = "{Reconciling Observed and Simulated Stellar Halo Masses}",
      journal = {\apj},
         year = 2018,
        month = dec,
       volume = {869},
       number = {1},
          eid = {12},
        pages = {12},
          doi = {10.3847/1538-4357/aaeb33},
archivePrefix = {arXiv},
       eprint = {1712.05808},
 primaryClass = {astro-ph.GA},
       adsurl = {https://ui.adsabs.harvard.edu/abs/2018ApJ...869...12S}
}

@ARTICLE{Santos.etal.22,
       author = {{Santos-Santos}, Isabel M.~E. and {Sales}, Laura V. and {Fattahi}, Azadeh and {Navarro}, Julio F.},
        title = "{Satellite mass functions and the faint end of the galaxy mass-halo mass relation in LCDM}",
      journal = {\mnras},
         year = 2022,
        month = sep,
       volume = {515},
       number = {3},
        pages = {3685-3697},
          doi = {10.1093/mnras/stac2057},
archivePrefix = {arXiv},
       eprint = {2111.01158},
 primaryClass = {astro-ph.GA},
       adsurl = {https://ui.adsabs.harvard.edu/abs/2022MNRAS.515.3685S}
}

@ARTICLE{Santistevan.etal.20,
       author = {{Santistevan}, Isaiah B. and {Wetzel}, Andrew and {El-Badry}, Kareem and {Bland-Hawthorn}, Joss and {Boylan-Kolchin}, Michael and {Bailin}, Jeremy and {Faucher-Gigu{\`e}re}, Claude-Andr{\'e} and {Benincasa}, Samantha},
        title = "{The formation times and building blocks of Milky Way-mass galaxies in the FIRE simulations}",
      journal = {\mnras},
         year = 2020,
        month = sep,
       volume = {497},
       number = {1},
        pages = {747-764},
          doi = {10.1093/mnras/staa1923},
archivePrefix = {arXiv},
       eprint = {2001.03178},
 primaryClass = {astro-ph.GA},
       adsurl = {https://ui.adsabs.harvard.edu/abs/2020MNRAS.497..747S}
}

@article{Sawicki.etal.2019,
  author       = {Marcin Sawicki and S. Arnouts and J. Huang and et~al.},
  title        = {The CFHT Large Area U-band Deep Survey (CLAUDS)},
  journal      = {Monthly Notices of the Royal Astronomical Society},
  year         = {2019},
  volume       = {489},
  pages        = {5202--5222},
  doi          = {10.1093/mnras/stz2270},
  url          = {https://arxiv.org/abs/1909.05898}
}

@ARTICLE{Sharpe.etal.24,
       author = {{Sharpe}, Katherine and {Naidu}, Rohan P. and {Conroy}, Charlie},
        title = "{What Is Missing from the Local Stellar Halo?}",
      journal = {\apj},
         year = 2024,
        month = mar,
       volume = {963},
       number = {2},
          eid = {162},
        pages = {162},
          doi = {10.3847/1538-4357/ad19ca},
archivePrefix = {arXiv},
       eprint = {2211.04562},
 primaryClass = {astro-ph.GA},
       adsurl = {https://ui.adsabs.harvard.edu/abs/2024ApJ...963..162S}
}

@ARTICLE{Shipp.etal.24,
       author = {{Shipp}, Nora and {Riley}, Alexander H. and {Simpson}, Christine M. and {Bieri}, Rebekka and {Necib}, Lina and {Arora}, Arpit and {Fragkoudi}, Francesca and {G{\'o}mez}, Facundo A. and {Grand}, Robert J.~J. and {Marinacci}, Federico},
        title = "{Auriga Streams II: orbital properties of tidally disrupting satellites of Milky Way-mass galaxies}",
      journal = {arXiv e-prints},
         year = 2024,
        month = oct,
          eid = {arXiv:2410.09143},
        pages = {arXiv:2410.09143},
          doi = {10.48550/arXiv.2410.09143},
archivePrefix = {arXiv},
       eprint = {2410.09143},
 primaryClass = {astro-ph.GA},
       adsurl = {https://ui.adsabs.harvard.edu/abs/2024arXiv241009143S}
}

@ARTICLE{Spergel.Steinhardt.00,
       author = {{Spergel}, David N. and {Steinhardt}, Paul J.},
        title = "{Observational Evidence for Self-Interacting Cold Dark Matter}",
      journal = {\prl},
         year = 2000,
        month = apr,
       volume = {84},
       number = {17},
        pages = {3760-3763},
          doi = {10.1103/PhysRevLett.84.3760},
archivePrefix = {arXiv},
       eprint = {astro-ph/9909386},
 primaryClass = {astro-ph},
       adsurl = {https://ui.adsabs.harvard.edu/abs/2000PhRvL..84.3760S}
}

@ARTICLE{Taffoni.etal.03,
   author = {{Taffoni}, G. and {Mayer}, L. and {Colpi}, M. and {Governato}, F.
	},
    title = "{On the life and death of satellite haloes}",
  journal = {\mnras},
   eprint = {astro-ph/0301271},
     year = 2003,
    month = may,
   volume = 341,
    pages = {434-448},
      doi = {10.1046/j.1365-8711.2003.06395.x},
   adsurl = {http://adsabs.harvard.edu/abs/2003MNRAS.341..434T}
}

@ARTICLE{Tau.etal.25,
       author = {{Tau}, Elisa A. and {Monachesi}, Antonela and {G{\'o}mez}, Facundo A. and {Grand}, Robert J.~J. and {Pakmor}, R{\"u}diger and {van de Voort}, Freeke and {Marinacci}, Federico and {Bieri}, Rebekka},
        title = "{Age and metallicity of low-mass galaxies: from their centres to their stellar halos}",
      journal = {arXiv e-prints},
         year = 2025,
        month = nov,
          eid = {arXiv:2511.20806},
        pages = {arXiv:2511.20806},
          doi = {10.48550/arXiv.2511.20806},
archivePrefix = {arXiv},
       eprint = {2511.20806},
 primaryClass = {astro-ph.GA},
       adsurl = {https://ui.adsabs.harvard.edu/abs/2025arXiv251120806T}
}

@ARTICLE{Taylor.Babul.01,
   author = {{Taylor}, J.~E. and {Babul}, A.},
    title = "{The Dynamics of Sinking Satellites around Disk Galaxies: A Poor Man's Alternative to High-Resolution Numerical Simulations}",
  journal = {\apj},
   eprint = {astro-ph/0012305},
     year = 2001,
    month = oct,
   volume = 559,
    pages = {716-735},
      doi = {10.1086/322276},
   adsurl = {http://adsabs.harvard.edu/abs/2001ApJ...559..716T}
}

@ARTICLE{Tissera.etal.13,
       author = {{Tissera}, Patricia B. and {Scannapieco}, Cecilia and {Beers}, Timothy C. and {Carollo}, Daniela},
        title = "{Stellar haloes of simulated Milky-Way-like galaxies: chemical and kinematic properties}",
      journal = {\mnras},
         year = 2013,
        month = jul,
       volume = {432},
       number = {4},
        pages = {3391-3400},
          doi = {10.1093/mnras/stt691},
archivePrefix = {arXiv},
       eprint = {1301.1301},
 primaryClass = {astro-ph.GA},
       adsurl = {https://ui.adsabs.harvard.edu/abs/2013MNRAS.432.3391T}
}

@ARTICLE{Tollet.etal.17,
   author = {{Tollet}, {\'E}. and {Cattaneo}, A. and {Mamon}, G. and {Moutard}, T. and 
	{van den Bosch}, F.},
    title = "{On stellar mass loss from galaxies in groups and clusters}",
  journal = {ArXiv e-prints},
archivePrefix = "arXiv",
   eprint = {1707.06264},
     year = 2017,
    month = jul,
   adsurl = {http://adsabs.harvard.edu/abs/2017arXiv170706264T}
}

@ARTICLE{vdBosch.02,
       author = {{van den Bosch}, Frank C.},
        title = "{The universal mass accretion history of cold dark matter haloes}",
      journal = {\mnras},
         year = 2002,
        month = mar,
       volume = {331},
       number = {1},
        pages = {98-110},
          doi = {10.1046/j.1365-8711.2002.05171.x},
archivePrefix = {arXiv},
       eprint = {astro-ph/0105158},
 primaryClass = {astro-ph},
       adsurl = {https://ui.adsabs.harvard.edu/abs/2002MNRAS.331...98V}
}

@ARTICLE{vdBosch.etal.08a,
   author = {{van den Bosch}, F.~C. and {Aquino}, D. and {Yang}, X. and {Mo}, H.~J. and 
	{Pasquali}, A. and {McIntosh}, D.~H. and {Weinmann}, S.~M. and 
	{Kang}, X.},
    title = "{The importance of satellite quenching for the build-up of the red sequence of present-day galaxies}",
  journal = {\mnras},
archivePrefix = "arXiv",
   eprint = {0710.3164},
     year = 2008,
    month = jun,
   volume = 387,
    pages = {79-91},
      doi = {10.1111/j.1365-2966.2008.13230.x},
   adsurl = {http://adsabs.harvard.edu/abs/2008MNRAS.387...79V}
}

@ARTICLE{vdBosch.etal.14,
   author = {{van den Bosch}, F.~C. and {Jiang}, F. and {Hearin}, A. and 
	{Campbell}, D. and {Watson}, D. and {Padmanabhan}, N.},
    title = "{Coming of age in the dark sector: how dark matter haloes grow their gravitational potential wells}",
  journal = {\mnras},
archivePrefix = "arXiv",
   eprint = {1409.2750},
     year = 2014,
    month = dec,
   volume = 445,
    pages = {1713-1730},
      doi = {10.1093/mnras/stu1872},
   adsurl = {http://adsabs.harvard.edu/abs/2014MNRAS.445.1713V}
}

@ARTICLE{vdBosch.Jiang.16,
   author = {{van den Bosch}, F.~C. and {Jiang}, F.},
    title = "{Statistics of dark matter substructure - II. Comparison of model with simulation results}",
  journal = {\mnras},
     year = 2016,
    month = may,
   volume = 458,
    pages = {2870-2884},
      doi = {10.1093/mnras/stw440},
   adsurl = {http://adsabs.harvard.edu/abs/2016MNRAS.458.2870V}
}

@ARTICLE{vdBosch.etal.18a,
   author = {{van den Bosch}, F.~C. and {Ogiya}, G. and {Hahn}, O. and {Burkert}, A.
	},
    title = "{Disruption of dark matter substructure: fact or fiction?}",
  journal = {\mnras},
archivePrefix = "arXiv",
   eprint = {1711.05276},
     year = 2018,
    month = mar,
   volume = 474,
    pages = {3043-3066},
      doi = {10.1093/mnras/stx2956},
   adsurl = {http://adsabs.harvard.edu/abs/2018MNRAS.474.3043V}
}

@ARTICLE{vdBosch.etal.18b,
   author = {{van den Bosch}, F.~C. and {Ogiya}, G.},
    title = "{Dark matter substructure in numerical simulations: a tale of discreteness noise, runaway instabilities, and artificial disruption}",
  journal = {\mnras},
archivePrefix = "arXiv",
   eprint = {1801.05427},
     year = 2018,
    month = apr,
   volume = 475,
    pages = {4066-4087},
      doi = {10.1093/mnras/sty084},
   adsurl = {http://adsabs.harvard.edu/abs/2018MNRAS.475.4066V}
}

@ARTICLE{Wang.etal.22,
       author = {{Wang}, Kuan and {Mao}, Yao-Yuan and {Zentner}, Andrew R. and {Guo}, Hong and {Lange}, Johannes U. and {van den Bosch}, Frank C. and {Mezini}, Lorena},
        title = "{Evidence of Galaxy Assembly Bias in SDSS DR7 Galaxy Samples from Count Statistics}",
      journal = {arXiv e-prints},
         year = 2022,
        month = apr,
          eid = {arXiv:2204.05332},
        pages = {arXiv:2204.05332},
archivePrefix = {arXiv},
       eprint = {2204.05332},
 primaryClass = {astro-ph.GA},
       adsurl = {https://ui.adsabs.harvard.edu/abs/2022arXiv220405332W}
}

@ARTICLE{Wechsler.Tinker.18,
   author = {{Wechsler}, R.~H. and {Tinker}, J.~L.},
    title = "{The Connection Between Galaxies and Their Dark Matter Halos}",
  journal = {\araa},
archivePrefix = "arXiv",
   eprint = {1804.03097},
     year = 2018,
    month = sep,
   volume = 56,
    pages = {435-487},
      doi = {10.1146/annurev-astro-081817-051756},
   adsurl = {http://adsabs.harvard.edu/abs/2018ARA%26A..56..435W}
}

@ARTICLE{Weinmann.etal.06,
   author = {{Weinmann}, S.~M. and {van den Bosch}, F.~C. and {Yang}, X. and 
	{Mo}, H.~J.},
    title = "{Properties of galaxy groups in the Sloan Digital Sky Survey - I. The dependence of colour, star formation and morphology on halo mass}",
  journal = {\mnras},
   eprint = {astro-ph/0509147},
     year = 2006,
    month = feb,
   volume = 366,
    pages = {2-28},
      doi = {10.1111/j.1365-2966.2005.09865.x},
   adsurl = {http://adsabs.harvard.edu/abs/2006MNRAS.366....2W}
}

@ARTICLE{Wetzel.etal.13,
   author = {{Wetzel}, A.~R. and {Tinker}, J.~L. and {Conroy}, C. and {van den Bosch}, F.~C.
	},
    title = "{Galaxy evolution in groups and clusters: satellite star formation histories and quenching time-scales in a hierarchical Universe}",
  journal = {\mnras},
archivePrefix = "arXiv",
   eprint = {1206.3571},
     year = 2013,
    month = jun,
   volume = 432,
    pages = {336-358},
      doi = {10.1093/mnras/stt469},
   adsurl = {http://adsabs.harvard.edu/abs/2013MNRAS.432..336W}
}

@ARTICLE{Williams.etal.25,
       author = {{Williams}, Devin J. and {Damjanov}, Ivana and {Sawicki}, Marcin and {Souchereau}, Harrison and {Chen}, Lingjian and {Desprez}, Guillaume and {George}, Angelo and {Annunziatella}, Marianna and {Arnouts}, St{\'e}phane and {Gwyn}, Stephen and {Marchesini}, Danilo and {Sajina}, Anna},
        title = "{The Growth of Galaxy Stellar Haloes over 0.2 {\ensuremath{\leq}} z {\ensuremath{\leq}} 1.1}",
      journal = {\apj},
         year = 2025,
        month = aug,
       volume = {989},
       number = {1},
          eid = {107},
        pages = {107},
          doi = {10.3847/1538-4357/ade9a8},
archivePrefix = {arXiv},
       eprint = {2412.03662},
 primaryClass = {astro-ph.GA},
       adsurl = {https://ui.adsabs.harvard.edu/abs/2025ApJ...989..107W}
}

@ARTICLE{Wright.etal.24,
       author = {{Wright}, Anna C. and {Tumlinson}, Jason and {Peeples}, Molly S. and {O'Shea}, Brian W. and {Lochhaas}, Cassandra and {Corlies}, Lauren and {Smith}, Britton D. and {Binh}, Nguyen and {Augustin}, Ramona and {Simons}, Raymond C.},
        title = "{Figuring Out Gas and Galaxies in Enzo (FOGGIE). VII. The (Dis)assembly of Stellar Halos}",
      journal = {\apj},
         year = 2024,
        month = jul,
       volume = {970},
       number = {1},
          eid = {70},
        pages = {70},
          doi = {10.3847/1538-4357/ad49a3},
archivePrefix = {arXiv},
       eprint = {2309.10039},
 primaryClass = {astro-ph.GA},
       adsurl = {https://ui.adsabs.harvard.edu/abs/2024ApJ...970...70W}
}

@ARTICLE{Yang.etal.12,
   author = {{Yang}, X. and {Mo}, H.~J. and {van den Bosch}, F.~C. and {Zhang}, Y. and 
	{Han}, J.},
    title = "{Evolution of the Galaxy-Dark Matter Connection and the Assembly of Galaxies in Dark Matter Halos}",
  journal = {\apj},
archivePrefix = "arXiv",
   eprint = {1110.1420},
     year = 2012,
    month = jun,
   volume = 752,
      eid = {41},
    pages = {41},
      doi = {10.1088/0004-637X/752/1/41},
   adsurl = {http://adsabs.harvard.edu/abs/2012ApJ...752...41Y}
}

@ARTICLE{Yang.etal.13,
   author = {{Yang}, X. and {Mo}, H.~J. and {van den Bosch}, F.~C. and {Bonaca}, A. and 
	{Li}, S. and {Lu}, Y. and {Lu}, Y. and {Lu}, Z.},
    title = "{Constraining the Star Formation Histories in Dark Matter Halos. I. Central Galaxies}",
  journal = {\apj},
archivePrefix = "arXiv",
   eprint = {1302.1265},
     year = 2013,
    month = jun,
   volume = 770,
      eid = {115},
    pages = {115},
      doi = {10.1088/0004-637X/770/2/115},
   adsurl = {http://adsabs.harvard.edu/abs/2013ApJ...770..115Y}
}

@ARTICLE{Zentner.Bullock.03,
   author = {{Zentner}, A.~R. and {Bullock}, J.~S.},
    title = "{Halo Substructure and the Power Spectrum}",
  journal = {\apj},
   eprint = {astro-ph/0304292},
     year = 2003,
    month = nov,
   volume = 598,
    pages = {49-72},
      doi = {10.1086/378797},
   adsurl = {http://adsabs.harvard.edu/abs/2003ApJ...598...49Z}
}

@ARTICLE{Zentner.etal.14,
   author = {{Zentner}, A.~R. and {Hearin}, A.~P. and {van den Bosch}, F.~C.},
    title = "{Galaxy assembly bias: a significant source of systematic error in the galaxy-halo relationship}",
  journal = {\mnras},
archivePrefix = "arXiv",
   eprint = {1311.1818},
     year = 2014,
    month = oct,
   volume = 443,
    pages = {3044-3067},
      doi = {10.1093/mnras/stu1383},
   adsurl = {http://adsabs.harvard.edu/abs/2014MNRAS.443.3044Z}
}

@ARTICLE{Zhao.etal.09,
       author = {{Zhao}, D.~H. and {Jing}, Y.~P. and {Mo}, H.~J. and {B{\"o}rner}, G.},
        title = "{Accurate Universal Models for the Mass Accretion Histories and Concentrations of Dark Matter Halos}",
      journal = {\apj},
         year = 2009,
        month = dec,
       volume = {707},
       number = {1},
        pages = {354-369},
          doi = {10.1088/0004-637X/707/1/354},
archivePrefix = {arXiv},
       eprint = {0811.0828},
 primaryClass = {astro-ph},
       adsurl = {https://ui.adsabs.harvard.edu/abs/2009ApJ...707..354Z}
}

@ARTICLE{Zhu.etal.22,
       author = {{Zhu}, Ling and {Pillepich}, Annalisa and {van de Ven}, Glenn and {Leaman}, Ryan and {Hernquist}, Lars and {Nelson}, Dylan and {Pakmor}, Ruediger and {Vogelsberger}, Mark and {Zhang}, Le},
        title = "{Mass of the dynamically hot inner stellar halo predicts the ancient accreted stellar mass}",
      journal = {\aap},
         year = 2022,
        month = apr,
       volume = {660},
          eid = {A20},
        pages = {A20},
          doi = {10.1051/0004-6361/202142496},
archivePrefix = {arXiv},
       eprint = {2110.13172},
 primaryClass = {astro-ph.GA},
       adsurl = {https://ui.adsabs.harvard.edu/abs/2022A&A...660A..20Z}
}

@ARTICLE{Zolotov.etal.09,
       author = {{Zolotov}, Adi and {Willman}, Beth and {Brooks}, Alyson M. and {Governato}, Fabio and {Brook}, Chris B. and {Hogg}, David W. and {Quinn}, Tom and {Stinson}, Greg},
        title = "{The Dual Origin of Stellar Halos}",
      journal = {\apj},
         year = 2009,
        month = sep,
       volume = {702},
       number = {2},
        pages = {1058-1067},
          doi = {10.1088/0004-637X/702/2/1058},
archivePrefix = {arXiv},
       eprint = {0904.3333},
 primaryClass = {astro-ph.GA},
       adsurl = {https://ui.adsabs.harvard.edu/abs/2009ApJ...702.1058Z}
}

@ARTICLE{Zolotov.etal.12,
   author = {{Zolotov}, A. and {Brooks}, A.~M. and {Willman}, B. and {Governato}, F. and 
	{Pontzen}, A. and {Christensen}, C. and {Dekel}, A. and {Quinn}, T. and 
	{Shen}, S. and {Wadsley}, J.},
    title = "{Baryons Matter: Why Luminous Satellite Galaxies have Reduced Central Masses}",
  journal = {\apj},
archivePrefix = "arXiv",
   eprint = {1207.0007},
     year = 2012,
    month = dec,
   volume = 761,
      eid = {71},
    pages = {71},
      doi = {10.1088/0004-637X/761/1/71},
   adsurl = {http://adsabs.harvard.edu/abs/2012ApJ...761...71Z}
}

@ARTICLE{Zu.Mandelbaum.18,
   author = {{Zu}, Y. and {Mandelbaum}, R.},
    title = "{Mapping stellar content to dark matter haloes - III. Environmental dependence and conformity of galaxy colours}",
  journal = {\mnras},
archivePrefix = "arXiv",
   eprint = {1703.09219},
     year = 2018,
    month = may,
   volume = 476,
    pages = {1637-1653},
      doi = {10.1093/mnras/sty279},
   adsurl = {http://adsabs.harvard.edu/abs/2018MNRAS.476.1637Z}
}

\appendix

\section{Impact of Free Parameters}
\label{App:freeparam}

\begin{figure*}
    \centering
    \includegraphics{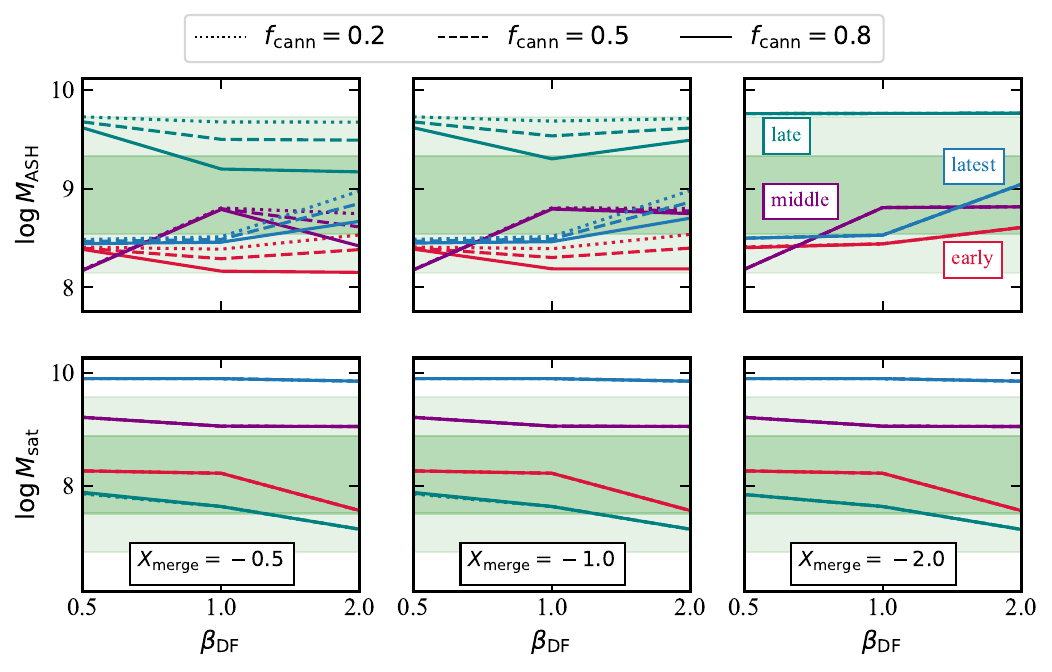}
    \caption{The impact of our three free parameters: dynamical friction strength ($\beta_{\rm DF}$), merging criterion ($X_{\rm merge}$), and cannibalization fraction ($\fcann$) on the four illustrative merger trees introduced in Section \ref{sec:examples}. For each tree, we explore $\beta_{\rm DF} = [0.5, 1, 2]$, $X_{\rm merge} = [-0.5, -1, -2]$, and $\fcann = [0.2, 0.5, 0.8]$, yielding 27 unique parameter combinations. The \textit{early} tree is shown in crimson, the \textit{middle} tree in purple, the \textit{late} in teal, and the \textit{latest} in blue. \textbf{Top Row of Panels}: The $\Mash$ component as a function of $\beta_{\rm DF}$.  \textbf{Bottom Row of Panels}: The $\Msat$ component as a function of $\beta_{\rm DF}$. The different columns correspond to different values of $X_{\rm merge}$, while line styles indicate $\fcann$. Shaded regions show the 16-84 and 5-95 percentile extents of the corresponding $S_0$ mass distributions. Despite substantial tree-to-tree variation in the absolute mass components, the relative response of $\Mash$ and $\Msat$ to changes in $\beta_{\rm DF}$, $X_{\rm merge}$, and $\fcann$ is modest, indicating that stochasticity in merger histories dominates over systematic effects from parameter choices.}
    \label{fig:free_params}
\end{figure*}

In this appendix, we discuss how the three free parameters [$\beta_{\rm DF}$, $X_{\rm merge}$, and $\fcann$] in our model impact our results using four illustrative merger trees from the $S_0$ sample introduced in Section~\ref{sec:stochasticity}. Specifically, for each tree we consider $\beta_{\rm DF} = [0.5, 1, 2]$, $X_{\rm merge} = [-0.5, -1, -2]$, and $\fcann = [0.2, 0.5, 0.8]$. Note that for each parameter combination, the subhalos are initialized with satellite galaxies of the same size and stellar mass, and evolved along the same orbit. Hence, the differences in the final mass components arise solely from changes in the free parameters. The 27 unique parameter combinations are shown in Figure \ref{fig:free_params} for the \textit{early} (crimson lines), \textit{middle} (purple lines), \textit{late} (teal lines), and \textit{latest} (blue lines) merger trees.

The top row of panels in Figure \ref{fig:free_params} shows the $\Mash$ mass component as a function of the assumed strength of dynamical friction, $\beta_{\rm DF}$ (see Section~\ref{sec:orbits}). Similarly, the bottom row of panels in Figure \ref{fig:free_params} shows the $\Msat$ mass component. We purposefully exclude the $M_{\rm cen}$ mass component from these figures because it is insensitive to our three free parameters, even when they are tuned to maximize the role of mergers in the assembly of the central galaxy. The three columns of panels distinguish between different assumed merging criteria ($X_{\rm merge}$), as indicated. Finally, the different line styles distinguish between the assumed cannibalization fractions ($\fcann$). In the background of all panels, we show the 16-84 and 5-95 percentile extents of the corresponding mass components in the $S_0$ distributions.

Notice that although the absolute values of the different mass components depend strongly on the particular merger tree used, the \textit{relative} changes in $\Mash$ and $\Msat$ in response to [$\beta_{\rm DF}$, $X_{\rm merge}$, and $\fcann$] are minor. Figure \ref{fig:free_params} therefore emphasizes that the inherent sources of stochasticity in the $\Mash$ and $\Msat$ components are far more dominant than systematic changes caused by our choice of free parameters. Recall that, using our fiducial model [$\beta_{\rm DF} = 1$, $X_{\rm merge} = -2.0$, and $\fcann = 0.8$], we have demonstrated that our framework is qualitatively consistent with the literature in terms of how MW-mass ASHs form in the context of $\Lambda$CDM \citep{Bullock.Johnston.05, Cooper.etal.10, Deason.etal.16, Monachesi.etal.19, Proctor.etal.24}.

\section{Input Feature / Target Variable Correlations}
\label{App:stack}

For completeness, Figure~\ref{fig:stack} shows contour plots of the three target variables ($\zfive$, $\Nnine$ and $\MMPdm$) vs. the three observable input features ($\Mcen$, $\Mash$ and $\Msat$) defined in the main text and used in the Random Forest Regression described in Section~\ref{sec:assembly_histories}. In each panel, contours indicate the 16-84, 5-95 and 1-99 percentiles of the distributions, with different colors corresponding to the three different merger tree samples, as indicated. The corresponding Spearman rank-order correlation coefficients are listed in Table~\ref{tab:correlations}. See the main text for a discussion of the various correlations.

\begin{figure*}
    \centering
    \includegraphics{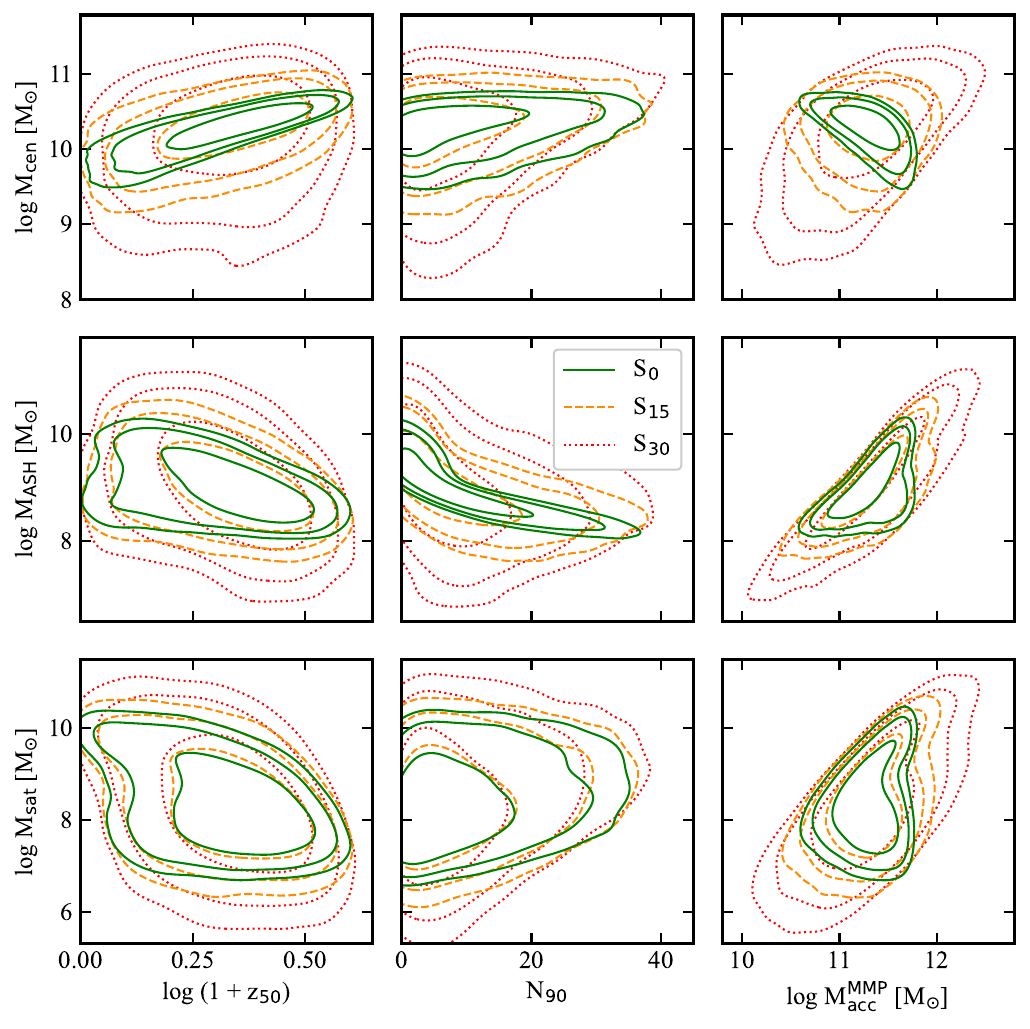}
    \caption{2D contour plots showing the correlations between input features and target variables. The different lines indicate the 16-84, 5-95 and 1-99 percentiles of the distributions. As always, the three different merger tree samples distinguished by color according to the legend. For clarity, all associated rank-order correlation coefficients ($\spear$) are listed in Table \ref{tab:correlations}. In this 3x3 grid, each column represents the mapping that the RFRa is attempting to learn. Notice that, in general, including host halo mass-mixing in the modeling weakens correlations.}
    \label{fig:stack}
\end{figure*}

\section{Random Forest Regression}
\label{App:RFR}

Random Forest Regression (RFR) is a supervised machine learning method designed to capture complex, non-linear relationships between a series of input features quantified by the vector $\bx$ and a single continuous target variable \citep{Breiman.84}. The algorithm operates by constructing an ensemble of decision trees, each of which recursively partitions the input feature space into nodes that are increasingly homogeneous in terms of the target variable. In the context of regression, the impurity of a node is measured using the mean squared error (MSE):
\begin{equation}
\mathrm{MSE}=\frac{1}{n} \sum_{i=1}^n\left(y_i - \bar{y}\right)^2
\end{equation}
where $n$ is the number of training objects that pass the node, $y_i$ are the observed target values, and $\bar{y}$ is their mean. Each split is chosen to minimize the sum of impurities, weighted by $n$, the number of objects in each node, producing regions where the target values are as similar as possible. Each terminal node in a tree assigns a value for the target variable based on the training data that fall into that region, which forms the basic predictive unit used in random forest regression.

Single decision trees are prone to overfitting the training data and typically exhibit large variance in their predictions \citep{Breiman.01}. RFRa overcome this limitation by averaging over a large collection of decision trees, each trained on a bootstrap sample of the input data. This procedure, known as ``bootstrap aggregation'', makes the ensemble less sensitive to noise in the training set. Another important aspect is ``feature randomness'': at each node, the split is selected from a random subset of input features. This decorrelates the trees and further reduces the variance of their predictions.

The final RFRa prediction of a target variable for a new (as in, not part of the training data) input feature set $\bx$ is then obtained by averaging over the predictions of all $T$ trees in the ensemble:
\begin{equation}
f(\bx) = \frac{1}{T}\sum_{t=1}^{T} f_t(\bx),
\end{equation}
where $f_t(\bx)$ is the prediction of the $t$-th tree.

RFRa also naturally quantify ``feature importance'' by measuring the total reduction in impurity attributable to splits on a specific input feature, summed over all nodes and averaged across the ensemble of trees. In other words, input features that consistently produce large reductions in impurity carry more information about the target variable than those that do not. This interpretability is a primary motivation for our choice of RFR over other machine-learning methods.

Model performance is evaluated using the coefficient of determination, $R^2$, which measures the fraction of variance in the target variable explained by the ensemble prediction. For a set of test objects with observed target values $y_i$ and RFR predictions $\hat{y}_i = f(\bx_i)$, $R^2$ is defined as
\begin{equation}\label{R2eq}
\mathrm{R}^2 = 1 - \frac{\sum_i \left(y_i - \hat{y}_i\right)^2}{\sum_i \left(y_i - \bar{y}\right)^2},
\end{equation}
where $\bar{y}$ is the mean of the observed target values in the test sample. The coefficient of determination ranges from 0 to 1, with $R^2=1$ representing perfect predictive accuracy.

Attempting to capture the complex mapping between input features and target variables (see Section \ref{sec:correlations}) using simple parametric models would be inefficient and overly contrived. Our assessment of the algorithm’s performance therefore relies on relative feature importance, $R^2$ scores, and visual inspection of ``True vs. Predicted'' scatter plots. In this analysis, we make use of the {\tt scikit-learn} Python package \citep{Pedregosa.etal.11}. In practice, RFRa have only a few hyperparameters that require tuning, such as the number of trees in the forest and the maximum depth of each tree. Because the final results are relatively insensitive to the exact choices of these hyperparameters, we empirically tuned them by selecting the combination that yielded the best $R^2$ scores without substantially increasing computational cost.

\end{document}